\documentclass[article,nojss]{jss}
\usepackage{thumbpdf,lmodern} 
\graphicspath{{Figures/}}

\author{N. Benjamin Erichson \\
  University of St Andrews 
  \And        
  Sergey Voronin \\
  Tufts University 
  \AND
  Steven L. Brunton \\
  University of Washington
  \And
  J. Nathan Kutz \\ 
  University of Washington
}

\title{Randomized Matrix Decompositions Using \proglang{R}}

\Plainauthor{N. Benjamin Erichson, Sergey Voronin, Steven L. Brunton, J. Nathan Kutz} 
\Plaintitle{Randomized Matrix Dompositions Using R} 

\Abstract{ Matrix decompositions are fundamental tools in the area of
  applied mathematics, statistical computing, and machine learning. In
  particular, low-rank matrix decompositions are vital, and widely
  used for data analysis, dimensionality reduction, and data
  compression.
  Massive datasets, however, pose a computational challenge for
  traditional algorithms, placing significant constraints on both
  memory and processing power.
  Recently, the powerful concept of randomness has been introduced as
  a strategy to ease the computational load.
  The essential idea of probabilistic algorithms is to employ some
  amount of randomness in order to derive a smaller matrix from a
  high-dimensional data matrix. The smaller matrix is then used to
  compute the desired low-rank approximation.
  Such algorithms are shown to be computationally efficient for
  approximating matrices with low-rank structure.
  We present the \proglang{R} package \pkg{rsvd}, and provide a
  tutorial introduction to randomized matrix
  decompositions. Specifically, randomized routines for the singular
  value decomposition, (robust) principal component analysis,
  interpolative decomposition, and CUR decomposition are discussed.
  Several examples demonstrate the routines, and show the
  computational advantage over other methods implemented in
  \proglang{R}.
}


\Plainkeywords{dimension reduction, randomized algorithm, low-rank
  approximations, singular value decomposition, principal component
  analysis, CUR decomposition, R} 

\Address{
	N. Benjamin Erichson\\
	School of Mathematics and Statistics\\
	University of St Andrews \\
	St Andrews, UK\\
	E-mail: \email{erichson@uw.edu} \\
}

\usepackage{booktabs}       
\usepackage{amsfonts}       
\usepackage{nicefrac}       
\usepackage{microtype}      

\usepackage{amsmath}
\usepackage{amssymb}
\usepackage{graphicx}
\usepackage{epstopdf}
\usepackage[justification=centering]{caption}
\usepackage{subcaption}
\usepackage{multirow}
\usepackage{float}
\usepackage{setspace}
\usepackage{rotating}
\usepackage{wrapfig,lipsum}
\usepackage{listings}
\usepackage{paralist}
\usepackage{bm}
\usepackage{overpic}

\usepackage{tabularx}
\usepackage{tabulary}

\usepackage[]{algorithm2e}

\usepackage{amsthm}

\theoremstyle{definition}

\theoremstyle{remark}
\newtheorem{remark}{Remark}

\usepackage{tikz}
\usetikzlibrary{matrix,positioning,shapes,shadows,arrows,calc,decorations.pathreplacing}

\tikzstyle{block} = [draw,line width=1.2pt, fill=white!30, rectangle, 
minimum height=3em, minimum width=3em, rounded corners]         
\tikzstyle{blocky} = [draw, line width=1.2pt,fill=yellow!40, rectangle, 
minimum height=2.5em, minimum width=2.5em, rounded corners]    
\tikzstyle{blockp} = [draw,line width=1.2pt, fill=purple!30, rectangle, 
minimum height=3em, minimum width=3em, rounded corners]       
\tikzstyle{blockb} = [draw,line width=1.2pt, fill=blue!20, rectangle, 
minimum height=2.5em, minimum width=2.5em, rounded corners]     
\tikzstyle{blockr} = [draw, line width=1.2pt,fill=red!30, rectangle, 
minimum height=2.5em, minimum width=2.5em, rounded corners]                 
\tikzstyle{blockgr} = [draw, line width=1.2pt,fill=black!10, rectangle, 
minimum height=2.5em, minimum width=2.5em, rounded corners]                  
\tikzstyle{line}=[-, line width=1.2pt]

\tikzstyle{blocky4} = [draw, line width=1.2pt,fill=white, rectangle, 
text centered,minimum height=4.5em, text width=6.5em, rounded corners]        

\tikzstyle{blocky5} = [draw, line width=1.2pt,fill=red!10, rectangle, 
text centered,minimum height=3em, text width=6.5em, rounded corners]  

\tikzstyle{blockyY} = [draw, line width=1.2pt,fill=red!10, rectangle, 
text centered,minimum height=2.5em, text width=6.5em, rounded corners]    

\tikzstyle{blockyZ} = [draw, line width=1.2pt,fill=red!10, rectangle, 
text centered,minimum height=2.5em, text width=4.0em, rounded corners] 

\tikzstyle{blockyC} = [draw, line width=1.2pt,fill=red!10, rectangle, 
text centered,minimum height=2.5em, text width=4.0em, rounded corners] 

\definecolor{darkred1}{RGB}{228,26,28}
\definecolor{darkblue1}{RGB}{55,126,184}
\definecolor{darkgreen1}{RGB}{77,175,74}

\newcommand{\bA}{\mathbf{A}}

\newcommand{\bQ}{\mathbf{Q}}
\newcommand{\bB}{\mathbf{B}}

\DeclareMathOperator*{\argmax}{arg\rm{}max}
\DeclareMathOperator*{\argmin}{arg\rm{}min}

\usepackage{kbordermatrix}

\begin{document}
\section[Introduction]{Introduction}

In the era of ``big data'', vast amounts of data are being collected and
curated in the form of arrays across the social, physical,
engineering, biological, and ecological sciences.
Analysis of the data relies on a variety of matrix decomposition
methods which seek to exploit low-rank features exhibited by the
high-dimensional data.
Indeed, matrix decompositions are often the workhorse algorithms for
scientific computing applications in the areas of applied mathematics,
statistical computing, and machine learning.
Despite our ever-increasing computational power, the emergence of
large-scale datasets has severely challenged our ability to analyze
data using traditional matrix algorithms.
Moreover, the growth of data collection is far outstripping
computational performance gains.
The computationally expensive singular value decomposition (SVD) is
the most ubiquitous method for dimensionality reduction, data
processing and compression.
The concept of randomness has recently been demonstrated as an
effective strategy to easing the computational demands of low-rank
approximations from matrix decompositions such as the SVD, thus
allowing for a scalable architecture for modern ``big data''
applications.
Throughout this paper, we make the following assumption: the data
matrix to be approximated has low-rank structure, i.e., the rank is
smaller than the ambient dimension of the measurement space.

\subsection{Randomness as a computational strategy}

\begin{figure}[!b]
	\centering
	\scalebox{0.75}{
		\begin{tikzpicture}[auto,node distance = 2cm,>=latex']
		\node [blocky4,name=fdata] { $\mathbf{A}$};       
		\node [blocky4,name=fdmd, right of=fdata,node distance=8.5cm]  {factors};
		\node [blocky5,name=cdata,below of=fdata,node distance=3.6cm] {$\mathbf{B}$};
		\node [blocky5,name=cdmd,below of=fdmd,node distance=3.6cm] {approximated factors};
		
		\path [draw, ->, line width=1.4pt] (fdata) -- node[name=toparrow, above of=fdata, node distance=.45cm]{deterministic algorithm}(fdmd);
		\path [draw, ->, line width=1.4pt, red] (cdata) -- node[name=bottomarrow,above of=cdata,node distance=.45cm]{deterministic algorithm}(cdmd);
		\path [draw, ->, line width=1.4pt, red] (fdata) -- node[name=leftarrow, left of=fdata, node distance = 2.3cm, text width=4.0cm]{probabilistic strategy to find a small matrix}(cdata);
		\path [draw, <-, line width=1.4pt, red] (fdmd) -- node[name=rightarrow, right of=fdmd, node distance = 1.9cm, text width=3.0cm]{recover near-optimal factors}(cdmd);
		\node [name=data,above of=fdata,node distance=1.35cm] {\textbf{Data}};
		\node [name=modes,above of=fdmd,node distance=1.35cm] {\textbf{Decomposition}};
		\node [name=full,left of=fdata,node distance=1.75cm]{\begin{sideways}\textbf{``Big''}\end{sideways}};
		\node [name=full,left of=cdata,node distance=1.75cm]{\begin{sideways}\textbf{``Small''}\end{sideways}};
		\end{tikzpicture}}
	\caption{First, randomness is used as a computational strategy to derive a smaller matrix $\mathbf{B}$ from $\mathbf{A}$. Then, the low-dimensional matrix is used to compute an approximate matrix decomposition. Finally, the near-optimal (high-dimensional) factors may be reconstructed.}
	\label{Fig:randLA}
\end{figure}
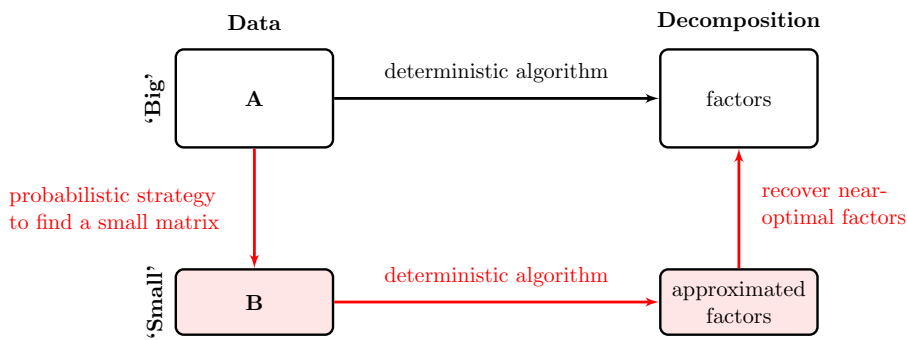

Randomness is a fascinating and powerful concept in science and nature. 
Probabilistic concepts can be used as an effective strategy for
designing better algorithms.  By the deliberate introduction of
randomness into computations~\citep{Motwani}, randomized algorithms
have not only been shown to outperform some of the best deterministic
methods, but they have also enabled the computation of previously
infeasible problems. The Monte Carlo method, invented by Stan Ulam,
Nick Metropolis and John von Neumann, is among the most prominent
randomized methods in computational statistics as well as one of the
``best'' algorithms of the 20th century~\citep{cipra2000best}.

Over the past two decades, probabilistic algorithms have been
established to compute matrix approximations, forming the field of
randomized numerical linear algebra~\citep{RandNLA}.
%
%
While randomness is quite controversial, and is often seen as an obstacle and a nuisance, modern randomized matrix algorithms are reliable and numerically stable.
The basic idea of probabilistic matrix algorithms is to employ a degree of randomness in order to derive a smaller matrix from a high-dimensional matrix, which captures the essential information.  
Thus, none of the ``randomness'' should obscure the dominant spectral information of the data as long as the input matrix features some low-rank structure. 
Then, a deterministic matrix factorization algorithm is applied to the smaller matrix to compute a near-optimal low-rank approximation. The principal concept is sketched in Figure~\ref{Fig:randLA}.

Several probabilistic strategies have been proposed to find a ``good'' smaller matrix, and we refer the reader to the surveys by~\cite{Mahoney2011},~\cite{liberty2013simple}, and ~\cite{halko2011rand} for an in-depth discussion, and theoretical results.
In addition to computing the singular value decomposition~\citep{sarlos2006improved,Martinsson201147} and principal component analysis~\citep{rokhlin2009randomized,halko2011algorithm}, it has been demonstrated that this probabilistic framework can also be used to compute the pivoted QR decomposition~\citep{doi:10.1137/15M1044680}, the pivoted LU decomposition~\citep{shabat2016randomized}, and the dynamic mode decomposition~\citep{erichson2017randomized}.

\subsection[The rsvd package: Motivation and contributions]{The \pkg{rsvd} package: Motivation and contributions}

The computational costs of applying deterministic matrix algorithms to massive data matrices can render the problem intractable. 
Randomized matrix algorithms are becoming increasingly popular as an
alternative, and implementations are available in a variety of
programming languages and machine learning libraries. For instance,
\cite{voronin2015rsvdpack} provide high performance, multi-core and
GPU accelerated randomized routines in \proglang{C}.

\begin{figure}[!b]
  \centering
  \DeclareGraphicsExtensions{.pdf}
  \includegraphics[width=0.9\textwidth]{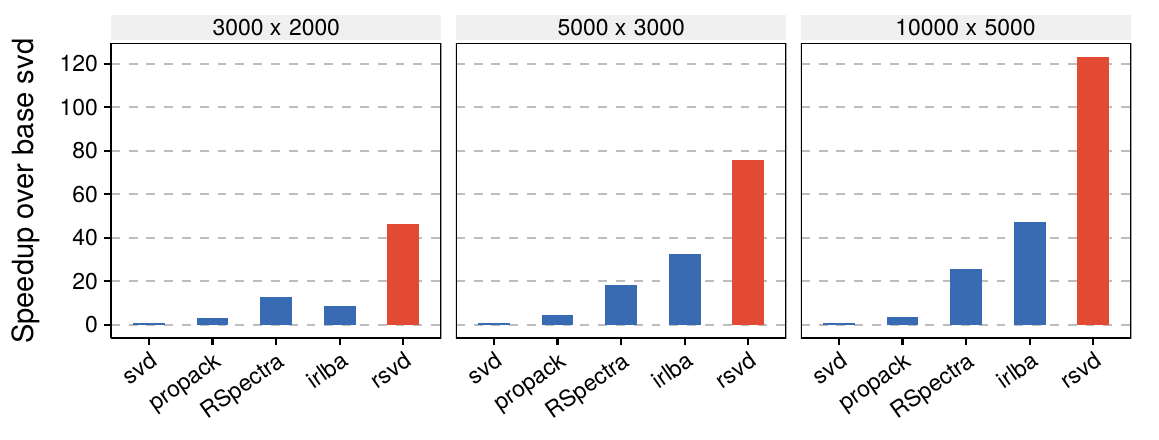}
  \caption{Runtime speedups (relative performance) of fast SVD algorithms compared to the base \code{svd()} routine in \proglang{R}. Here, the dominant $k=20$ singular values and vectors are computed for random low-rank matrices with varying dimension $m\times n$. Note, here we are using Microsoft Open 3.5.1 which provides the Intel MKL for parallel mathematical computing (using 4 cores).
    \label{fig:benchmark1}}
\end{figure}

The \pkg{rsvd} package aims to fill the gap in
\proglang{R}, providing the following randomized routines:
%
		\begin{compactitem}  
		\item Randomized singular value decomposition: \code{rsvd()}. 
		\item Randomized principal component analysis: \code{rpca()}.
		\item Randomized robust principal component analysis: \code{rrpca()}.
		\item Randomized interpolative decomposition: \code{rid()}.
		\item Randomized CUR decomposition: \code{rcur()}.
		\end{compactitem} 
The routines are, in particular, efficient for matrices with rapidly decaying singular values.
%
%
Figure~\ref{fig:benchmark1} compares the computational performance of the \code{rsvd()} function to other existing SVD routines in \proglang{R}, which are discussed in more detail in Section~\ref{sec:svdinr}. Specifically, for computing the dominant $k$ singular values and vectors, the randomized singular value decomposition function \code{rsvd()} results in significant speedups over other existing SVD routines in \proglang{R}. See Section~\ref{sec:compute_performance} for a more detailed performance evaluation. 

%

The computational benefits of the randomized SVD translates directly to principal component analysis (PCA), since both methods are closely related. Further, the randomized SVD can be used to accelerate the computation of robust principal component analysis (RPCA). More generally, the concept of randomness allows also one to efficiently compute modern matrix decompositions such as the interpolative decomposition (ID) and CUR decomposition.
While the performance of the randomized algorithms depends on the actual shape of the matrix, we can state (as a rule of thumb) that significant computational speedups are achieved if the target rank $k$ is about $3$-$6$ times smaller than the smallest dimension $\text{min}\{m,n\}$ of the matrix. The speedup for tall and thin matrices is in general less impressive than for ``big'' fat matrices.

The \pkg{rsvd} package is available from the Comprehensive
\proglang{R} Archive Network (CRAN) at
\url{https://CRAN.R-project.org/package=rsvd}.  Thus, to install and
load within \proglang{R} simply use:
\begin{CodeChunk}
\begin{CodeInput}
R> install.packages("rsvd") 
R> library("rsvd")
\end{CodeInput}
\end{CodeChunk}
Alternatively, the package can be obtained via github:
\url{https://github.com/erichson/rSVD}.

\subsection{Organization}
The remainder of this paper is organized as follows. 
First, Section~\ref{sec:framework} outlines the advocated probabilistic framework for low-rank matrix approximations.
Section~\ref{sec:rSVD} briefly reviews the singular value decomposition and the randomized SVD algorithm. Then, the \code{rsvd()} function and its computational performance are demonstrated.
Section~\ref{sec:rPCA} first describes the principal component analysis. Then, the randomized PCA algorithm is outlined, followed by the demonstration of the corresponding \code{rpca()} function. 
Section~\ref{sec:rrpca} outlines robust principal component analysis, and describes the randomized robust PCA algorithm as well as the \code{rrpca()} function. 
Section~\ref{sec:rid} gives a high-level overview of the interpolative and CUR decomposition.
Finally, concluding remarks and a roadmap for future developments are presented in Section~\ref{sec:conclusion}.  

\subsection{Notation}

In the following we give a brief overview of some notation used throughout this manuscript. 

Scalars are denoted by lower case letters $x$, and vectors both in
$\mathbb{R}^{n}$ and $\mathbb{C}^{n}$ are denoted as bold lower case
letters $\mathbf{x}=[x_1,x_2,\ldots,x_n]^\top$.
Matrices are denoted by bold capital letters $\mathbf{A}$ and the entry at row
$i$ and column $j$ is denoted as $\mathbf{A}{(i,j)}$. This notation is convenient for matrix slicing, for instance, $\mathbf{A}{(1:i,:)}$ extracts the first $1,2,\ldots,i$ rows, and $\mathbf{A}{(:,1:j)}$ extracts the first $1,2,\ldots,j$ columns.
The transpose of a real matrix is denoted as $\mathbf{A}^\top$, and without loss of generality, we restrict most of the discussion in the following to real matrices.
Further, the column space (range) of $\mathbf{A}$ is denoted as  $\text{col}(\mathbf{A})$, and the row space as $\text{row}(\mathbf{A})$. 
%

The spectral or operator norm of a matrix is defined as the largest singular value $\sigma_{\max}$ of $\mathbf{A}$, i.e., the square root of the largest eigenvalue $\lambda_{\max}$ of the positive-semidefinite matrix $\mathbf{A}^\top\mathbf{A}$:
\begin{equation*}
\| \mathbf{A} \|_2 = \sqrt{\lambda_{\max}(\mathbf{A}^\top\mathbf{A})} = \sigma_{\max}(\mathbf{A}) = \max_{x\neq0}\dfrac{\| \mathbf{A}\mathbf{x} \|_2}{\| \mathbf{x} \|_2}.
\end{equation*} 
The Frobenius norm is defined as the square root of the sum of the
absolute squares of its elements, which is equal to the square root of
the matrix trace of $\mathbf{A}^\top\mathbf{A}$:
\begin{equation*}
\| \mathbf{A} \|_F = \sqrt{ \sum_{i=1}^m\sum_{j=1}^n|\mathbf{A}{(i,j)}|^2} = \sqrt{\textrm{trace}(\mathbf{A}^\top\mathbf{A})}.
\end{equation*}

\section{Probabilistic framework for low-rank approximations}\label{sec:framework}

Assume that a matrix $\mathbf{A} \in \mathbb{R}^{m\times n}$ has rank
$r$, where $r \le \text{min}\{m,n\}$. Then, in general, the objective
of a low-rank matrix approximation is to find two smaller matrices
such that:
\begin{equation}\label{eq:matrixfac}
\begin{array}{cccc}
\mathbf{A} & \approx & \mathbf{E} & \mathbf{F}, \\
m\times n &   &  m\times r & r\times n
\end{array} 
\end{equation}
where the columns of the matrix $\mathbf{E}\in \mathbb{R}^{m\times r}$ span the column space of $\mathbf{A}$, and the rows of the matrix $\mathbf{F}\in \mathbb{R}^{r\times n}$ span the row space of $\mathbf{A}$. The factors $\mathbf{E}$ and $\mathbf{F}$ can then be used to summarize or to reveal some interesting structure in the data.
Further, the factors can be used to efficiently store the large data matrix $\mathbf{A}$. Specifically, while $\mathbf{A}$ requires $mn$ words of storage, $\mathbf{E}$ and $\mathbf{F}$ require only $mr + nr$ words of storage. 

In practice, most data matrices do not feature a precise rank $r$. Rather we are commonly interested in finding a rank-$k$ matrix $\mathbf{A}_k$, which is as close as possible to an arbitrary input matrix $\mathbf{A}$ in the least-square sense. We refer to $k$ as the target rank in the following.

In particular,  modern data analysis and scientific computing largely rely on low-rank approximations, since low-rank matrices are ubiquitous throughout the sciences.
However, in the era of ``big data'', the emergence of massive data poses a significant computational challenge for traditional deterministic algorithms.

In the following, we advocate the probabilistic framework, formulated by~\cite{halko2011rand}, to compute a near-optimal low-rank approximation. Conceptually, this framework splits the computational task into two logical stages:
\begin{itemize}
	\item \textbf{Stage A:} Construct a low dimensional subspace that approximates the column space of $\mathbf{A}$. This means, the aim is to find a matrix $\mathbf{Q} \in \mathbb{R}^{m\times k}$ with orthonormal columns such that $\mathbf{A} \approx \mathbf{Q}\mathbf{Q}^\top\mathbf{A}$ is satisfied. 
	
	\item \textbf{Stage B:} Form a smaller matrix $\mathbf{B}:=\mathbf{Q^\top A}\in \mathbb{R}^{k\times n}$, i.e., restrict the high-dimensional input matrix to the low-dimensional space spanned by the near-optimal basis $\mathbf{Q}$. The smaller matrix $\mathbf{B}$ can then be used to compute a desired low-rank approximation.
\end{itemize}
The first computational stage is where randomness comes into the play, while the second stage is purely deterministic. 
In the following, the two stages are described in detail. 

\subsection{The generic randomized algorithm}

\subsubsection{Stage A: Computing the near-optimal basis} 

\begin{figure}[t!]
  \centering
    \scalebox{0.85}{
			\begin{tikzpicture}[auto,node distance = 2cm,>=latex']
			\coordinate (A) at (0,0);
			\coordinate (B) at (8,0);
			\coordinate (C) at (11,3);
			\coordinate (D) at (3,3);

			\draw [line width=1.4pt]  (A) -- (B);
			\draw [line width=1.4pt]  (B) -- (C);
			\draw [line width=1.4pt]  (C) -- (D);
			\draw [line width=1.4pt]  (A) -- (D);

			\coordinate (O) at (5.5,1.5);		
			\draw [fill=black] (O) circle (3pt) node [below] {origin};
			
			\coordinate (Px) at (8,2);
			\draw [-latex, red, line width=1.4pt] (O) -- (Px);				
			\draw [fill=black] (Px) circle (2pt) node [right] {$\hat{\mathbf{x}}=\mathbf{P}\mathbf{x}$};
			\draw [fill=black] (O) circle (3pt);
			
			\coordinate (x) at (8,5);				
			\draw [fill=black] (x) circle (2pt) node [right] {$\mathbf{x}$};
			\draw [-latex, black, line width=1.4pt] (O) -- (x);
			
			\draw [dashed, line width=1.4pt]  (x) -- (Px);

			\coordinate (x) at (1,0.5);				
			\draw [fill=red] (x) circle (0pt) node [right] {col($\mathbf{A}$)};			
			\end{tikzpicture}}
	\caption{Geometric illustration of the orthogonal projection operator $\mathbf{P}$. A vector $\mathbf{x} \in \mathbb{R}^{m}$ is restricted to the column space of $\mathbf{A}$, where $\mathbf{P}\mathbf{x} \in \text{col}(\mathbf{A})$.  }
	\label{Fig:orthProjector}
\end{figure}
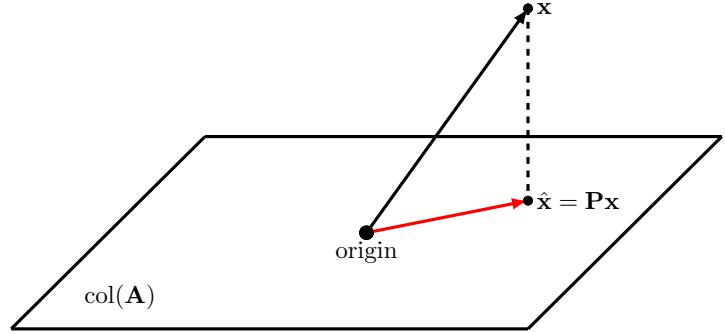

First, we aim to find a near-optimal basis $\mathbf{Q}$ for the matrix $\mathbf{A}$ such that
\begin{equation}\label{eq:AQQA}
\mathbf{A} \approx \mathbf{Q}\mathbf{Q}^\top\mathbf{A}
\end{equation}
is satisfied. The desired target rank $k$ is assumed to be
$k\ll \text{min}\{m,n\}$. Specifically,
$\mathbf{P}:=\mathbf{Q}\mathbf{Q}^\top$ is a linear orthogonal
projector.
A projection operator corresponds to a linear subspace, and transforms any vector to its orthogonal projection on the subspace. This is illustrated in Figure~\ref{Fig:orthProjector}, where a vector $\mathbf{x}$ is confined to the column space $\text{col}(\mathbf{A})$.  

The concept of random projections can be used to sample the range (column space) of the input matrix $\mathbf{A}$ in order to efficiently construct such a orthogonal projector. 
Random projections are data agnostic, and constructed by first drawing a set of $k$ random vectors $\{\bm{\omega}_i\}_{i=1}^k$, for instance, from the standard normal distribution. 
Probability theory guarantees that random vectors are linearly
independent with high probability. Then, a set of random projections
$\{\mathbf{y}_{i}\}_{i=1}^k$ is computed by mapping $\mathbf{A}$ to
low-dimensional space:
\begin{equation}\label{eq:samplevector}
\mathbf{y}_{i} := \mathbf{A}\bm{\omega}_{i} \quad  \textrm{for } i=1,2,\ldots,k.
\end{equation}
In other words, this process forms a set of independent randomly weighted linear combinations of the columns of $\mathbf{A}$, and reduces the number of columns from $n$ to $k$. While the input matrix is compressed, the Euclidean distances between the original data points are approximately preserved. Random projections are also well known as the Johnson-Lindenstrauss (JL) transform~\citep{johnson1984extensions}, and we refer to~\cite{ahfock2017statistical} for a recent statistical perspective.

Equation~\ref{eq:samplevector} can be efficiently executed in
parallel. Therefore, let us define the random test matrix
$\mathbf{\Omega} \in \mathbb{R}^{n\times k}$, which is again drawn
from the standard normal distribution, and the columns of which are
given by the vectors $\{\bm{\omega}_i\}$. The samples matrix
$\mathbf{Y} \in \mathbb{R}^{m\times k}$, also denoted as sketch, is
then obtained by post-multiplying the input matrix by the random test
matrix
\begin{equation}\label{eq:YAQ}
\mathbf{Y} := \mathbf{A}\mathbf{\Omega}.
\end{equation}
Once $\mathbf{Y}$ is obtained, it only remains to orthonormalize the
columns in order to form a natural basis
$\mathbf{Q} \in \mathbb{R}^{m\times k}$. This can be efficiently
achieved using the QR-decomposition
$\mathbf{Y}=:\mathbf{Q}\mathbf{R}$, and it follows that
Equation~\ref{eq:AQQA} is satisfied.

\subsubsection{Stage B: Compute the smaller matrix} 
\begin{figure}[t!]
  \centering
    \scalebox{0.95}{
			\begin{tikzpicture}[auto,node distance = 2cm,>=latex']

			\coordinate (O) at (0,0);
			\coordinate (X) at (2,0);
			\coordinate (Y) at (0,2);
			\coordinate (Z) at (-1,-1);

			\draw [-latex, black, line width=1.4pt]  (O) -- (X);
			\draw [-latex, black, line width=1.4pt]  (O) -- (Y);
			\draw [-latex, black, line width=1.4pt]  (O) -- (Z);
			
			\draw [fill=black] (0.5,0.5) circle (2pt);
			\draw [fill=black] (-0.5,0.5) circle (2pt);
			\draw [fill=red] (-0.6,-0.2) circle (2pt);
			\draw [fill=black] (-0.7,0.5) circle (2pt);
			\draw [fill=black] (0.2,-0.8) circle (2pt);
			\draw [fill=black] (0.4,0.9) circle (2pt);
			\draw [fill=black] (0.8,0.8) circle (2pt);
			\draw [fill=black] (0.6,-0.9) circle (2pt);
			\draw [fill=black] (0.7,0.2) circle (2pt);
			\draw [fill=red] (0.3,0.3) circle (2pt);
			\draw [fill=black] (0.2,-0.2) circle (2pt);
			\draw [fill=black] (0.4,-0.25) circle (2pt);
			\draw [dashed, line width=1.4pt, red]  (-0.6,-0.2) -- (0.3,0.3);		
			\draw [fill=black] (1.5,1.3) circle (0pt) node [right] {$\mathbb{R}^{m}$};																																
			\draw [-latex, black, line width=1.0pt]  (3,0) -- (4.5,0);			
			
			\draw [fill=black] (3.75,0) circle (0pt) node [above] {projection};

			\coordinate (A) at (5.5,-1.2);
			\coordinate (B) at (5.5+4,-1.2);
			\coordinate (C) at (5.5+5,1.7);
			\coordinate (D) at (5.5+1,1.7);
			\draw [line width=1.4pt]  (A) -- (B);
			\draw [line width=1.4pt]  (B) -- (C);
			\draw [line width=1.4pt]  (C) -- (D);
			\draw [line width=1.4pt]  (A) -- (D);	
			
			\draw [fill=black] (8+0.55,0.55) circle (2pt);
			\draw [fill=black] (8+-0.45,0.6) circle (2pt);
			\draw [fill=red] (8+-0.5,-0.24) circle (2pt);
			\draw [fill=black] (8+-0.65,0.54) circle (2pt);
			\draw [fill=black] (8+0.3,-0.7) circle (2pt);
			\draw [fill=black] (8+0.4,0.9) circle (2pt);
			\draw [fill=black] (8+0.8,0.8) circle (2pt);
			\draw [fill=black] (8+0.4,-0.98) circle (2pt);
			\draw [fill=black] (8+0.645,0.33) circle (2pt);
			\draw [fill=red] (8+0.25,0.36) circle (2pt);
			\draw [fill=black] (8+0.267,-0.245) circle (2pt);
			\draw [fill=black] (8+0.4,-0.35) circle (2pt);
			\draw [dashed, line width=1.4pt, red]  (8+-0.5,-0.24) -- (8+0.25,0.36);		
			\draw [fill=black] (8+1.5,1.3) circle (0pt) node [right] {$\mathbb{R}^{k}$};

			\end{tikzpicture}}
	\caption{Points in a high-dimensional space are projected into low-dimensional space, while the geometric structure is preserved in an Euclidean sense.  }
	\label{Fig:projectionPoints}
\end{figure}
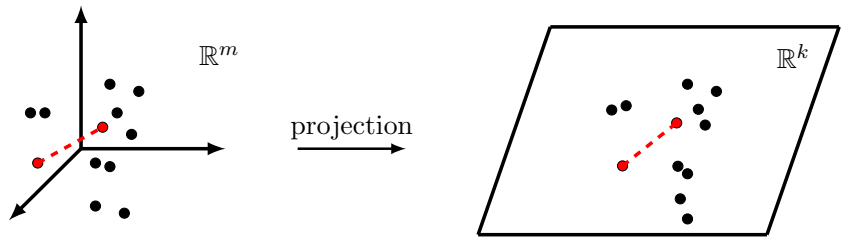

Now, given the near-optimal basis $\mathbf{Q}$, we aim to find a smaller matrix $\mathbf{B} \in \mathbb{R}^{k\times n}$. Therefore, we project the high-dimensional input matrix $\mathbf{A}$ to low-dimensional space
\begin{equation}\label{eq:BQA}
\mathbf{B} := \mathbf{Q}^\top\mathbf{A}.
\end{equation}
Geometrically, this is a projection (i.e., a linear transformation)
which takes points in a high-dimensional space into corresponding
points in a low-dimensional space, illustrated in
Figure~\ref{Fig:projectionPoints}.
This process preserves the geometric structure of the data in an Euclidean sense, i.e., the length of the projected vectors as well as the angles between the projected vectors are preserved. This is, due to the invariance of inner products~\citep{trefethen1997numerical}.
Substituting Equation~\ref{eq:BQA} into~\ref{eq:AQQA} yields then the following low-rank approximation
\begin{equation*}
\begin{array}{cccc}
\mathbf{A} & \approx & \mathbf{Q} & \mathbf{B}. \\
m\times n &   &  m\times k & k\times n
\end{array} 
\end{equation*}
This decomposition is referred to as the QB decomposition. Subsequently, the smaller matrix $\mathbf{B}$ can be used to compute a matrix decomposition using a traditional algorithm.

\subsection{Improved randomized algorithm}\label{sec:impRA}

The basis matrix $\mathbf{Q}$ often fails to provide a good approximation for the column space of the input matrix. This is because most real-world data matrices do not feature a precise rank $r$, and instead exhibit a gradually decaying singular value spectrum. The performance can be considerably improved using the concept of oversampling and the power iteration scheme.

\subsubsection{Oversampling} 
Most data matrices do not feature an exact rank, which means that the singular values $\{\sigma_i\}_{i=k+1}^n$ of the input matrix $\mathbf{A}$ are non-zero. 
As a consequence, the sketch $\mathbf{Y}$ does not exactly span the column space of the input matrix. Oversampling can be used to overcome this issue by using $l:=k+p$ random projections to form the sketch, instead of just $k$. Here, $p$ denotes the number of additional projections, and a small number $p=\{5,10\}$ is often sufficient to obtain a good basis that is comparable to the best possible basis~\citep{martinsson2016randomized}.

The intuition behind the oversampling scheme is the following. The sketch $\mathbf{Y}$ is a random variable, as it depends on the drawing of a random test matrix $\mathbf{\Omega}$.  Increasing the number of additional random projections allows one to decrease the variation in the singular value spectrum of the random test matrix, which subsequently improves the quality of the sketch. 

\subsubsection{Power iteration scheme}

The second method for improving the quality of the basis $\mathbf{Q}$ involves the concept of power sampling iterations~\citep{rokhlin2009randomized,halko2011rand,gu2015subspace}. 
Instead of obtaining the sketch $\mathbf{Y}$ directly, the data matrix $\mathbf{A}$ is first preprocessed as
\begin{equation}\label{eq:Aq}
\mathbf{A}^{(q)} := (\mathbf{A}\mathbf{A}^\top )^{q} \mathbf{A},
\end{equation}
where $q$ is an integer specifying the number of power iterations. This process enforces a more rapid decay of the singular values. Thus, we enable the algorithm to sample the relevant information related to the dominant singular values, while unwanted information is suppressed. 

\begin{figure}[t!]
  \centering
  \DeclareGraphicsExtensions{.pdf}
  \includegraphics[width=0.7\textwidth]{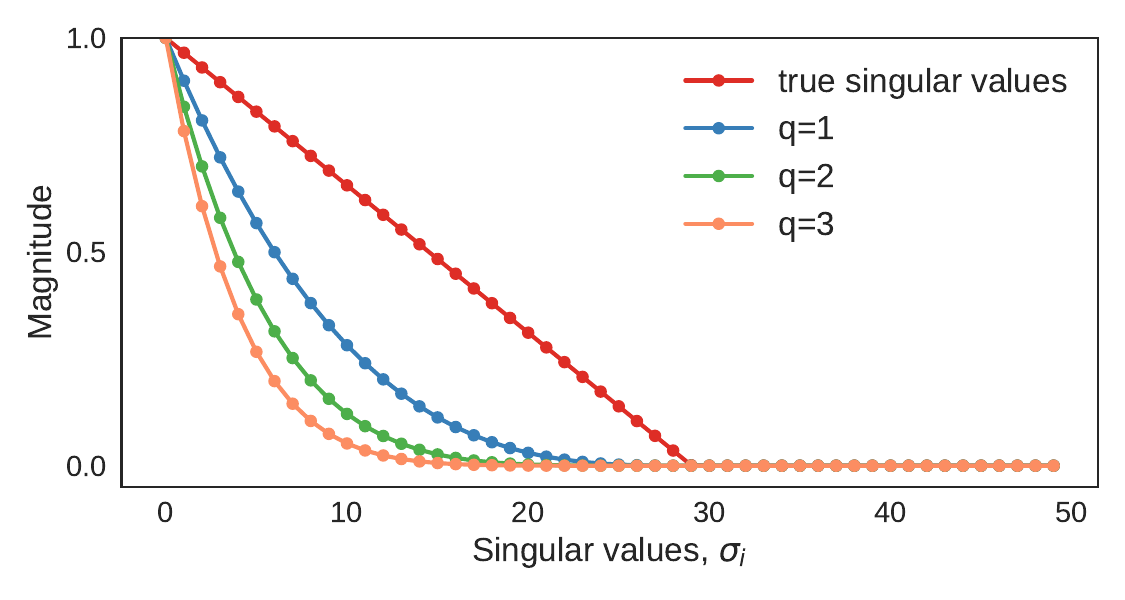}
  \caption{Singular value spectrum of a low-rank ($r=30$) matrix
    before and after preprocessing. The computation of power
    iterations enforce a more rapid decay of singular values.}
  \label{fig:poweritertions}
\end{figure}

Let $\mathbf{A}=\mathbf{U}\mathbf{\Sigma}\mathbf{V}^\top$ be the singular value decomposition. It is simple to show that $\mathbf{A}^{(q)} := ( \mathbf{A}\mathbf{A}^\top )^{q} \mathbf{A} = \mathbf{U} \mathbf{\Sigma}^{2q + 1} \mathbf{V}^\top$. Here, $\mathbf{U}$ and $\mathbf{V}$ are orthonormal matrices whose columns are the left and right singular vectors of $\mathbf{A}$, and $\mathbf{\Sigma}$ is a diagonal matrix containing the singular values in descending order. Hence, for $q > 0$, the modified matrix $\mathbf{A}^{(q)}$ has a relatively 
fast decay of singular values even when the decay in $\mathbf{A}$ is modest.
This is illustrated in Figure~\ref{fig:poweritertions}, showing the singular values of a $50 \times 50$ low-rank matrix before (red) and after computing $q=\{1,2,3\}$ power iterations.
Thus, substituting Equation~\ref{eq:Aq} into~\ref{eq:YAQ} yields an improved sketch
\begin{equation*}
\mathbf{Y} := \mathbf{A}^{(q)} \mathbf{\Omega}.
\end{equation*}
When the singular values of the data matrix decay slowly, as few as
$q=\{1,2,3\}$ power iterations can considerably improve the accuracy
of the approximation. The drawback of the power scheme is that $q$
additional passes over the input matrix are required.

Algorithm~\ref{alg:poweriter_direct} shows a direct implementation of
the power iteration scheme. Due to potential round-off errors,
however, this algorithm is not recommended in practice.

The numerical stability can be improved by orthogonalizing the sketch between each iteration. This scheme is shown in Algorithm~\ref{alg:subspace_iter}, and denoted as subspace iteration~\citep{halko2011rand,gu2015subspace}.
The pivoted LU decomposition can be used as an intermediate step instead of the QR decomposition as proposed by~\cite{rokhlin2009randomized}. Algorithm~\ref{alg:normalized_iter} is computationally more efficient, while slightly less accurate.

\begin{algorithm}[t!]
	\centering
	\scalebox{0.90}{\fbox{		
			\begin{minipage}{.3\textwidth}	
				\begin{tabbing}
					\hspace{2mm} \= \hspace{2mm} \= \hspace{2mm} \= \hspace{33mm} \=\kill
					\textbf{Input:} Input matrix $\mathbf{A}$, the sketch $\mathbf{Y}$, and parameter $q$.\\[1mm]
					\textbf{function} $\texttt{power\_iterations}(\mathbf{A, Y}, q)$\\[1mm]
					(1)  \> \> \textbf{for} $j = 1,\dots,q$ \> \> {\color{blue}$\textrm{perform q power iterations}$} \\[1mm]
					(2)  \> \> \> $\mathbf{Y} = \mathbf{A}^\top\mathbf{Y}$ \\[1mm]
					(3)  \> \> \> $\mathbf{Y} = \mathbf{A} \mathbf{Y}$ \\[1mm]
					\textbf{Return:} $\mathbf{Y}\in \mathbb{R}^{m\times k}$					
				\end{tabbing}
	\end{minipage}}}
	\vspace{+.10in}
	\caption{Direct implementation of the power scheme.}
	\label{alg:poweriter_direct}
\end{algorithm}

\begin{algorithm}[t!]
	\centering
	\scalebox{0.90}{\fbox{		
			\begin{minipage}{.5\textwidth}
				\begin{tabbing}
					\hspace{2mm} \= \hspace{2mm} \= \hspace{2mm} \= \hspace{33mm} \=\kill
					\textbf{Input:} Input matrix $\mathbf{A}$, the sketch $\mathbf{Y}$, and $q$.\\[1mm]
					\textbf{function} $\texttt{sub\_iterations}(\mathbf{A, Y}, q)$\\[1mm]
					(1)  \> \> \textbf{for} $j = 1,\dots,q$ \> \> {\color{blue}$\textrm{perform q iterations}$} \\[1mm]
					(2)  \> \> \> $\left[\mathbf{Q},\sim\right] = \texttt{qr}(\mathbf{Y})$ \> {\color{blue}$\textrm{economic QR}$}\\[1mm]
					(3)  \> \> \> $\left[\mathbf{Q},\sim\right] = \texttt{qr}(\mathbf{\mathbf{A}^\top \mathbf{Q}})$ \>  {\color{blue}$\textrm{economic QR}$}\\[1mm]
					(4)  \> \> \> $\mathbf{Y} = \mathbf{A} \mathbf{Q}$ \\
					\textbf{Return:} $\mathbf{Y}\in \mathbb{R}^{m\times k}$					
				\end{tabbing}
	\end{minipage}}}
	\vspace{+.10in}
	\caption{Subspace iterations.}
	\label{alg:subspace_iter}
\end{algorithm}
\begin{algorithm}[t!]
	\centering
	\scalebox{0.90}{\fbox{		
			\begin{minipage}{.5\textwidth}	
				\begin{tabbing}
					\hspace{2mm} \= \hspace{2mm} \= \hspace{2mm} \= \hspace{33mm} \=\kill
					\textbf{Input:} Input matrix $\mathbf{A}$, the sketch $\mathbf{Y}$, and $q$.\\[1mm]
					\textbf{function} $\texttt{norm\_iterations}(\mathbf{A, Y}, q)$\\[1mm]
					(1)  \> \> \textbf{for} $j = 1,\dots,q$ \> \>  {\color{blue}$\textrm{perform q iterations}$}\\[1mm]
					(2)  \> \> \> $\left[\mathbf{L},\sim\right] = \texttt{lu}(\mathbf{Y})$ \> {\color{blue}$\textrm{pivoted LU}$}\\[1mm]
					(3)  \> \> \> $\left[\mathbf{L},\sim\right] = \texttt{lu}(\mathbf{\mathbf{A}^\top \mathbf{L}})$ \>  {\color{blue}$\textrm{pivoted LU}$}\\[1mm]
					(4)  \> \> \> $\mathbf{Y} = \mathbf{A} \mathbf{L}$ \\ 
					\textbf{Return:} $\mathbf{Y}\in \mathbb{R}^{m\times k}$					
				\end{tabbing}
	\end{minipage}}}
	\vspace{+.10in}
	\caption{Normalized power iterations.}
	\label{alg:normalized_iter}
\end{algorithm}

\subsection{Random test matrices}

The probabilistic framework above essentially depends on the random test matrix $\mathbf{\Omega}$ used for constructing the sketch $\mathbf{Y}$. Specifically, we seek a matrix with independent identically distributed (i.i.d.) entries from some distribution, which ensures that its columns are linearly independent with high probability. Some popular choices for constructing the random test matrix are:

\begin{itemize}
 	\item \textbf{Gaussian.} The default choice to construct a random test matrix is to draw entries from the standard normal distribution, $\mathcal{N}(0,1)$.
	The normal distribution is known to have excellent performance for sketching in practice. Further, the theoretical properties of the normal distribution enable the derivation of accurate error bounds~\citep{halko2011rand}.

	\item \textbf{Uniform.} A simple alternative is to draw entries from the uniform distribution, $\mathcal{U}(-1,1)$. While the behavior is similar in practice, the generation of uniform random samples is computationally more efficient.

	\item \textbf{Rademacher.} Yet another approach to construct the random test matrix is to draw independent Rademacher entries. The Rademacher distribution is a discrete probability distribution, where the random variates take the values  $+1$ and $-1$ with equal probability. Rademacher entries are simple to generate, and they are cheaper to store than Gaussian and uniform random test matrices~\citep{tropp2016randomized}. 
\end{itemize}
 
Currently, the \pkg{rsvd} package only supports standard dense random
test matrices; however, dense matrix operations can become very
expensive for large-scale applications. This is because it takes
$O(mnk)$ time to apply an $n\times k$ dense random test matrix to any
$m\times n$ dense input matrix.
Structured random test matrices provide a computationally more efficient alternative~\citep{woolfe2008fast}, reducing the costs to $O(mn \text{ log}(k))$. 
Very sparse random test matrices are even simpler to
construct~\citep{li2006very}, yet slightly less accurate. These can be
applied to any $m\times n$ dense matrix in
$O(m \text{ nnz}(\mathbf{\Omega}))$ using sparse matrix multiplication
routines, where $\text{nnz}()$ denotes the non-zero entries in the
sparse random test matrix.

\section{Randomized singular value decompositions}\label{sec:rSVD}

The SVD provides a numerically stable matrix decomposition that can be
used to obtain low-rank approximations, to compute the pseudo-inverses
of non-square matrices, and to find the least-squares and minimum norm
solutions of a linear model. Further, the SVD is the workhorse
algorithm behind many machine learning concepts, for instance, matrix
completion, sparse coding, dictionary learning, PCA and robust PCA.
For a comprehensive technical overview of the SVD we refer to \cite{golub2012matrix}, and \cite{trefethen1997numerical}. 

\subsection{Brief historical overview} 
\begin{figure}[t!]
	\centering
	\DeclareGraphicsExtensions{.pdf}
	\includegraphics[width=0.86\textwidth]{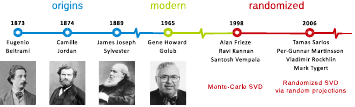}
	\caption{ A timeline of major singular value decomposition developments. }
	\label{Fig:timeline}
\end{figure}

While the origins of the SVD can be traced back to the late 19th
century, the field of randomized matrix algorithms is relatively
young. Figure~\ref{Fig:timeline} shows a short time-line of some major
developments of the singular value
decomposition. \cite{stewart1993early} gives an excellent historical
review of the five mathematicians who developed the fundamentals of
the SVD, namely Eugenio Beltrami (1835--1899), Camille Jordan
(1838--1921), James Joseph Sylvester (1814--1897), Erhard Schmidt
(1876--1959) and Hermann Weyl (1885--1955). The development and
fundamentals of modern high-performance algorithms to compute the SVD
are related to the seminal work of \cite{golub1965calculating} and
\cite{golub1970singular}, forming the basis for the \pkg{EISPACK}, and
\pkg{LAPACK} SVD routines.

Modern partial singular value decomposition algorithms are largely
based on Krylov methods, such as the Lanczos
algorithm~\citep{calvetti1994implicitly,larsen1998lanczos,ARPACK,baglama2005augmented}. These
methods are accurate and are particularly powerful for approximating
structured, and sparse matrices.

Randomized matrix algorithms for computing low-rank matrix
approximations have gained prominence over the past two
decades. \cite{frieze2004fast} introduced the ``Monte Carlo'' SVD, a
rigorous approach to efficiently compute the approximate low-rank SVD
based on non-uniform row and column
sampling. \cite{sarlos2006improved},~\cite{liberty2007randomized}
and~\cite{Martinsson201147} introduced a more robust approach based on
random projections. Specifically, the properties of random vectors are
exploited to efficiently build a subspace that captures the column
space of a matrix. \cite{woolfe2008fast} further improved the
computational performance by leveraging the properties of highly
structured matrices, which enable fast matrix
multiplications. Eventually, the seminal work by \cite{halko2011rand}
unified and expanded previous work on the randomized singular value
decomposition and introduced state-of-the-art prototype algorithms to
compute the near-optimal low-rank singular value decomposition.

\subsection{Conceptual overview}

Given a real matrix $\mathbf{A} \in \mathbb{R}^{m\times n}$ with $m \geq n$, the singular value decomposition takes the form 
\begin{equation*}
\mathbf{A} = \mathbf{U}\mathbf{\Sigma}\mathbf{V}^\top.
\end{equation*} 
The matrices $\mathbf{U} = [ \mathbf{u}_1, \ldots , \mathbf{u}_m ] \in \mathbb{R}^{m \times m}$ and $\mathbf{V} = [ \mathbf{v}_1, \ldots , \mathbf{v}_n ] \in \mathbb{R}^{n \times n}$ are orthonormal so that $\mathbf{U^\top U=I}$ and $ \mathbf{ V^\top V = I}$. 
The left singular vectors in $\mathbf{U}$ provide a basis for the range (column space), and the right singular vectors in $\mathbf{V}$ provide a basis for the domain (row space) of the matrix $\mathbf{A}$. 
The rectangular diagonal matrix $\mathbf{\Sigma} \in \mathbb{R}^{m \times n}$ contains the corresponding non-negative singular values $\sigma_{1} \geq \ldots \geq  \sigma_{n} \geq 0$, describing the spectrum of the data.

The so called ``economy'' or ``thin'' SVD computes only the left
singular vectors and singular values corresponding to the number
(i.e., $n$) of right singular vectors
\begin{equation*}
\mathbf{A} = \mathbf{U}\mathbf{\Sigma}\mathbf{V} = [ \mathbf{u}_1, \ldots , \mathbf{u}_n ] 
\textrm{diag}(\sigma_1,\dots,\sigma_n) [ \mathbf{v}_1, \ldots , \mathbf{v}_n ]^\top.
\end{equation*}
If the number of right singular vectors is small (i.e., $n\ll m$),
this is a more compact factorization than the full SVD. The ``economy''
SVD is the default form of the base \code{svd()} function in
\proglang{R}.

\begin{figure}[t!]
	\centering
	\DeclareGraphicsExtensions{.pdf}
	\includegraphics[width=0.9\textwidth]{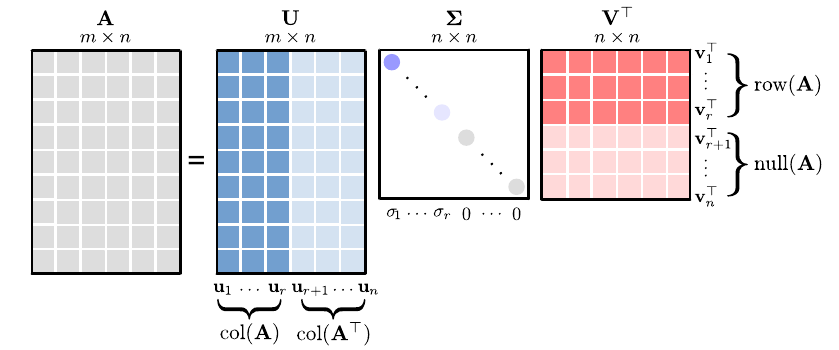}
	\caption{Schematic of the ``economy'' SVD for a rank-$r$ matrix, where $m\geq n$.}
	\label{Fig:SVD}
\end{figure}

Low-rank matrices feature a rank that is smaller than the dimension of
the ambient measurement space, i.e., $r$ is smaller than the number of
columns and rows. Hence, the singular values
$\{\sigma_{i}: i\geq r+1\}$ are zero, and the corresponding singular
vectors span the left and right null spaces. The concept of the
``economy'' SVD of a low-rank matrix is illustrated in
Figure~\ref{Fig:SVD}.

In practical applications matrices are often contaminated by errors, and the effective rank of a matrix can be smaller than its exact rank $r$. In this case, the matrix can be well approximated by including only those singular vectors which correspond to singular values of a significant magnitude. Hence, it is often desirable to compute a reduced version of the SVD
\begin{equation*}
\mathbf{A}_k := \mathbf{U}_k\mathbf{\Sigma}_k\mathbf{V}_k = [ \mathbf{u}_1, \ldots , \mathbf{u}_k ] 
\textrm{diag}(\sigma_1,\dots,\sigma_k) [ \mathbf{v}_1, \ldots , \mathbf{v}_k ]^\top,
\end{equation*}
where $k$ denotes the desired target rank of the approximation. In other words, this reduced form of the SVD allows one to express $\mathbf{A}$ approximately by the sum of $k$ rank-one matrices
\begin{equation*}
\mathbf{A}_k \approx \sum_{i=1}^{k} \sigma_i \mathbf{u}_i \mathbf{v}_i^\top.
\end{equation*}

Choosing an optimal target rank $k$ is highly dependent on the task. One can either be interested in a highly accurate reconstruction of the original data, or in a very low dimensional representation of dominant features in the data. In the former case $k$ should be chosen close to the effective rank, while in the latter case $k$ might be chosen to be much smaller. 

Truncating small singular values in the deterministic SVD gives an 
optimal approximation of the corresponding target rank $k$. Specifically, the Eckart-Young theorem~\citep{Eckart1936psych} states that the low-rank SVD provides the optimal 
rank-$k$ reconstruction of a matrix in the least-square sense
\begin{equation*} 
\mathbf{A}_k := \argmin_{\text{rank}(\mathbf{A}_k')=k} \| \mathbf{A} - \mathbf{A}_k' \|.
\end{equation*} 
The reconstruction error in both the spectral and Frobenius norms is given by
\begin{equation*}
\| \mathbf{A} - \mathbf{A}_k \|_2 = \sigma_{k+1}(\mathbf{A}) \quad \mbox{and} \quad \| \mathbf{A} - \mathbf{A}_k \|_F = \sqrt{\sum_{j = k+1}^{\min(m,n)} \sigma_j^2(\mathbf{A})}.
\end{equation*} 
For massive datasets, however, the truncated SVD is costly to compute. The cost to compute the full SVD of an $m \times n$ matrix is of the order $O(m n^2)$, from which the first $k$ components 
can then be extracted to form $\mathbf{A}_k$. 

\subsection{Randomized algorithm}

Randomized algorithms have been recently popularized, in large part due to their ``surprising'' reliability and computational efficiency~\citep{gu2015subspace}.
These techniques can be used to obtain an approximate rank-$k$ singular value decomposition at a cost of 
$O(m n k)$. When the dimensions of $\mathbf{A}$ are large, this is substantially more efficient than truncating the full SVD. 

We present details of the randomized low-rank SVD algorithm, which comes with favorable error bounds relative to the optimal truncated SVD, as presented in the seminal paper by~\cite{halko2011rand}, and further analyzed and implemented in~\cite{voronin2015rsvdpack}.

\begin{figure}[t!]
  \centering
  \begin{subfigure}[t]{0.82\textwidth}
    \centering
    \DeclareGraphicsExtensions{.pdf}
    \includegraphics[width=1\textwidth]{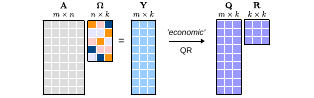}
    \vspace{+.03in}
  \end{subfigure}
  
  \begin{subfigure}[t]{0.82\textwidth}
    \centering
    \DeclareGraphicsExtensions{.pdf}
    \includegraphics[width=1\textwidth]{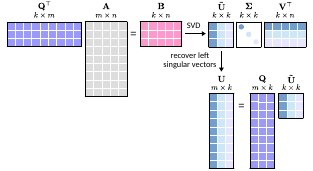}
    \label{fig:stageB}
  \end{subfigure}
  \caption{Conceptual architecture of the randomized singular value
    decomposition (rSVD). First, a natural basis $\mathbf{Q}$ is
    computed in order to derive the smaller matrix $\mathbf{B}$. Then,
    the SVD is efficiently computed using this smaller
    matrix. Finally, the left singular vectors $\mathbf{U}$ may be
    reconstructed from the approximate singular vectors
    $\tilde{\mathbf{U}}$ by the expression in Equation~\ref{eq:UQU}.}
  \label{Fig:rSVDillustration}
\end{figure}

Let $\mathbf{A}\in \mathbb{R}^{m\times n}$ be a low-rank matrix, and without loss of generality $m \geq n$. In the following, we seek the near-optimal low-rank approximation of the form
\begin{equation*}
\mathbf{A} \approx \mathbf{U}_k\mathbf{\Sigma}_k\mathbf{V}_k^\top,
\end{equation*} 
where $k$ denotes the target rank. Instead of computing the singular value decomposition directly, we embed the SVD into the probabilistic framework presented in Section~\ref{sec:framework}. The principal concept is sketched in Figure~\ref{Fig:rSVDillustration}.

Specifically, we first compute the near-optimal basis $\mathbf{Q} \in \mathbb{R}^{m\times l}$ using the randomized scheme as outlined in detail above. Note that we allow for both oversampling ($l=k+p$), and additional power iterations $q$, in order to obtain the near-optimal basis matrix.
The matrix $\mathbf{B} \in \mathbb{R}^{l\times n}$ is relatively small if $l \ll n$, and it is obtained by projecting the input matrix to low-dimensional space, i.e., $\mathbf{B}:=\mathbf{Q}^\top\mathbf{A}$.
The full SVD of $\mathbf{B}$ is then computed using a deterministic algorithm
\begin{equation*} 
\mathbf{B} = \tilde{\mathbf{U}} \mathbf{\Sigma}\mathbf{V}^\top.	
\end{equation*} 
Thus, we efficiently obtain the first $l$ right singular vectors $\mathbf{V} \in \mathbb{R}^{n\times l}$ as well as the corresponding singular values $\mathbf{\Sigma} \in \mathbb{R}^{l\times l}$. 
It remains to recover the left singular vectors $\mathbf{U} \in \mathbb{R}^{m\times l}$ from the approximate left singular vectors $\tilde{\mathbf{U}}\in \mathbb{R}^{l\times l}$ by pre-multiplying by $\mathbf{Q}$
\begin{equation} \label{eq:UQU}
\mathbf{U} \approx \mathbf{Q}\tilde{\mathbf{U}}	.
\end{equation}
The justification for the randomized SVD can be sketched as follows
\begin{equation*} 	
\mathbf{A} 	\approx \mathbf{Q}\mathbf{Q}^\top\mathbf{A}  
 =  \mathbf{Q}\mathbf{B} 
 =  \mathbf{Q}\tilde{\mathbf{U}} \mathbf{\Sigma}\mathbf{V}^\top 
 =  \mathbf{U} \mathbf{\Sigma}\mathbf{V}^\top. 	
\end{equation*} 
Algorithm~\ref{alg:RSVDalgorithm} presents an implementation using the randomized QB decomposition in Algorithm~\ref{alg:rqb}. The approximation quality can be controlled via oversampling and additional subspace iterations.
Note that if an oversampling parameter $p>0$ has been specified, the desired rank-$k$ approximation is simply obtained by truncating the left and right singular vectors and the singular values. 

The randomized singular value decomposition has several practical advantages: 

\begin{itemize}  
\item \textbf{Lower communication costs.} The randomized algorithm presented here requires few (at least two) passes over the input matrix. By passes we refer to the number of sequential reads of the entire input matrix. This aspect is crucial in the area of ``big data'' where communication costs play a significant role. For instance, the time to transfer the data from the hard-drive into fast memory can be substantially more expensive than the theoretical costs of the algorithm would suggest. Recently,~\cite{tropp2016randomized} have introduced an interesting set of new single pass algorithms, reducing the communication costs even further.
	
\item \textbf{Highly parallelizable.} Randomized algorithms are highly scalable by design. This is because the computationally expensive steps involve matrix-matrix operations, which are very efficient on parallel architectures. Hence, modern computational architectures such as multi-threading and distributed computing, can be fully exploited. 	
	
\item \textbf{General applicability.} 
	Randomized matrix algorithms work for matrices with arbitrary rates of singular value decay. The approximation for matrices with rapid singular value decay approaches that of the optimal truncated SVD, with high probability~\citep{martinsson2016randomized,gu2015subspace}.	
	
\end{itemize}

\begin{algorithm}[t!]
	\scalebox{0.9}{\fbox{		
			\begin{minipage}{210mm}
				\begin{tabbing}
					\hspace{2mm} \= \hspace{5mm} \= \hspace{2mm} \= \hspace{50mm} \=\kill
					\textbf{Input:} Input matrix $\mathbf{A}$ with dimensions $m\times n$, and target rank $k<\text{min}\{m,n\}$.\\[1mm]
					\textbf{Optional:} Parameters $p$ and $q$ to control oversampling, and the power scheme.\\[3mm] 
					
					\textbf{function} $\texttt{rqb}(\mathbf{A}, k, p, q)$\\[3mm]

					(1)  \> \> $l = k + p$ \> \> {\color{blue}$\textrm{slight oversampling}$} \\[1mm]
					(2)  \> \> $\mathbf{\Omega} = \texttt{rnorm}(n,l)$ \> \> {\color{blue}$\textrm{generate random test matrix}$}\\[1mm]
					(3)  \> \> $\mathbf{Y} = \mathbf{A} \mathbf{\Omega}$ \> \> {\color{blue}$\textrm{compute sketch}$}\\[1mm]
					
					(4)  \> \> $\mathbf{Y} = \texttt{sub\_iterations}(\mathbf{A,Y},q)$ \> \> {\color{blue}\textrm{optional: compute power scheme via Algorithm~\ref{alg:subspace_iter}}} \\[1mm]

					(9)  \> \> $\left[\mathbf{Q},\sim\right] = \texttt{qr}(\mathbf{Y})$ \> \> {\color{blue}\textrm{form orthonormal basis}}
					\\[1mm]
					(10)  \> \> $\mathbf{B} = \mathbf{Q}^\top \mathbf{A}$ \> \> {\color{blue}$\textrm{project to low-dimensional space}$} \\[3mm]

					\textbf{Return:} $\mathbf{Q}\in \mathbb{R}^{m\times l}$, $\mathbf{B}\in \mathbb{R}^{l\times n}$
				\end{tabbing}
			\end{minipage}}}
			\centering
			\vspace{+.15in}
			\caption{A randomized QB decomposition algorithm.}
			\label{alg:rqb}
		\end{algorithm}
\begin{algorithm}[t!]
  {
    \centering
    \scalebox{0.9}{\fbox{		
			\begin{minipage}{210mm}
				\begin{tabbing}
					\hspace{2mm} \= \hspace{5mm} \= \hspace{2mm} \= \hspace{50mm} \=\kill
					\textbf{Input:} Input matrix $\mathbf{A}$ with dimensions $m\times n$, and target rank $k<\text{min}\{m,n\}$.\\[1mm]
					\textbf{Optional:} Parameters $p$ and $q$ to control oversampling, and the power scheme.\\[3mm] 
					
					\textbf{function} $\texttt{rsvd}(\mathbf{A}, k, p, q)$\\[3mm]
					
					(1)  \> \> $[\mathbf{Q},\mathbf{B}] = \texttt{rqb}(\mathbf{A}, k, p, q)$ \> \> {\color{blue}\textrm{randomized QB decomposition via Algorithm~\ref{alg:rqb}}} \\[1mm]
					
					(2)  \> \> $[\tilde{\mathbf{U}},\mathbf{\Sigma},\mathbf{V}] = \texttt{svd}(\mathbf{B})$ \> \> {\color{blue}\textrm{compute economic SVD}} \\[1mm]
					(3)  \> \> $\mathbf{U} = \mathbf{Q} \tilde{\mathbf{U}}$  \> \> {\color{blue}\textrm{recover left singular vectors}} \\[3mm]
					
					\textbf{Return:} $\mathbf{U}(:,1:k)\in \mathbb{R}^{m\times k}$, $\mathbf{\Sigma}(1:k,1:k)\in \mathbb{R}^{k\times k}$ and $\mathbf{V}(:,1:k)\in \mathbb{R}^{n\times k}$
				\end{tabbing}
			\end{minipage}}}
			\vspace{+.15in}
			\caption{A randomized SVD algorithm.}
			\label{alg:RSVDalgorithm}
                      }
\begin{remark}
In general we achieve a good computational performance, if the target rank is much smaller than the ambient dimensions of the input matrix, e.g., $k < \text{min}\{m,n\}/ {4}$.
\end{remark}

\begin{remark}
  As default values for the oversampling and the power iteration
  scheme we recommend the parameters $p=10$ and $q=2$,
  respectively. 
\end{remark}
\end{algorithm}

\subsection{Theoretical performance}

Let us consider the low-rank matrix approximation $\mathbf{QB}$, where $\mathbf{B:=Q^\top A}$.
From the Eckart-Young theorem~\citep{Eckart1936psych} it follows that the smallest possible error achievable with the best possible basis matrix $\bQ$ is 
\begin{equation*}
\| \bA - \bQ \bB  \|_2  =  \sigma_{k+1}(\bA),
\end{equation*}
where $\sigma_{k+1}(\bA)$ denotes the $k+1$ largest singular value of the matrix $\mathbf{A}$. 

In Algorithm~\ref{alg:RSVDalgorithm} we 
compute the full SVD of $\mathbf{B}$, so it follows that 
$\| \mathbf{A} - \mathbf{A}_k \|_2 = \| \mathbf{A} - \mathbf{Q} \mathbf{B} \|_2 $, where $\mathbf{A}_k =\mathbf{U}_k \mathbf{\Sigma}_k \mathbf{V}^\top_k$. 
Following \cite{martinsson2016randomized}, the randomized algorithm for computing the low-rank matrix approximation has the following expected error:
\begin{equation}\label{eq:rsvd_expected_error}
\E\| \bA - \bA_k  \|_2  \leq  \Bigg[ 1 + \sqrt{\frac{k}{p-1}} + \frac{e\sqrt{k+p}}{p} \cdot \sqrt{\text{min}\{m,n\}-k}\Bigg]^\frac{1}{2q+1} \sigma_{k+1}(\bA).
\end{equation}
Here, the operator $\E$ denotes the expectation with respect to a Gaussian test matrix $\mathbf{\Omega}$, and Euler's number is denoted as $e$. Further, it is assumed that the oversampling parameter $p$ is greater or equal to two.  

From this error bound it follows that both the oversampling (parameter $p$) and the power iteration scheme (parameter $q$) can be used to control the approximation error. With increasing $p$ the second and third term on the right hand side tend towards zero, i.e., the bound approaches the theoretically optimal value of $\sigma_{k+1}(\bA)$. The parameter $q$ accelerates the rate of decay of the singular values of the sampled matrix, while maintaining the same eigenvectors. This yields better performance for matrices with otherwise modest decay.
Equation~\ref{eq:rsvd_expected_error} is a simplified version of one of the key theorems presented by \cite{halko2011rand}, who provide a detailed error analysis of the outlined probabilistic framework.
Further, \cite{witten2015randomized} provide sharp error bounds and interesting theoretical insights.

\pagebreak
\subsection[Existing functionality for SVD in R]{Existing functionality for SVD in \proglang{R}}\label{sec:svdinr}

The \code{svd()} function is the default option to compute the SVD in
\proglang{R}. This function provides an interface to the underlying
\pkg{LAPACK} SVD routines~\citep{anderson1999lapack}.
These routines are known to be numerical stable and highly accurate, i.e., full double precision. 

In many applications the full SVD is not necessary; only the truncated factorization is required.
The truncated SVD for an $m \times n$ matrix can be obtained by first computing the full SVD, and then extracting the $k$ dominant components to form $\mathbf{A}_k$. However, the computational time required to approximate large-scale data is tremendous using this approach.

Partial algorithms, largely based on Krylov subspace methods, are an efficient class of approximation methods to compute the dominant singular vectors and singular values. These algorithms are particularly powerful for approximating structured or sparse matrices. 
This is because Krylov subspace methods only require certain operations defined on the input matrix $\mathbf{A}$ such as matrix-vector multiplication. These basic operations can be computed very efficiently, if the input matrix features some structure like sparsity. 
The Lanczos algorithm and its variants are the most popular choice to compute the approximate SVD. Specifically, they first find the dominant $k$ eigenvalues and eigenvectors of the symmetric matrix $\mathbf{A^\top A}$ as
\begin{equation*}
\mathbf{A^\top A}\mathbf{V}_k = \mathbf{V}\mathbf{\Sigma}_k^2,
\end{equation*}
where $\mathbf{V}_k$ are the $k$ dominant right singular vectors, and $\mathbf{\Sigma}_k^2$ are the corresponding squared singular values. Depending on the dimensions of the input matrix, this operation can also be performed on $\mathbf{A A^\top}$. See, for instance,~\cite{demmel1997applied} and~\cite{martinsson2016randomized} for details on how the Lanczos algorithm builds the Krylov subspace, and subsequently approximates the eigenvalues and eigenvectors. 

The relationship between the singular value decomposition and eigendecomposition can then be used to approximate the left singular vectors as $\mathbf{U}_k = \mathbf{A} \mathbf{V}_k \mathbf{\Sigma}^{-1}$  (see also Section~\ref{sec:randomizedPCAalg}). 
However, computing the eigenvalues of the inner product $\mathbf{A^\top A}$ is not generally a good idea, because this process squares the condition number of $\mathbf{A}$.  
Further, the computational performance of Krylov
methods depends on factors such as the initial guess for the starting vector, and additional steps used to stabilize the algorithm~\citep{gu2015subspace}. 
While partial SVD algorithms have the same theoretical costs as randomized methods, i.e., they require $O(m n k)$ floating point operations, they have higher communication costs. This is because the matrix-vector operations do not permit data reuse between iterations.

The most competitive partial SVD routines in \proglang{R} are provided
by the \pkg{svd}~\citep{svdpackage}, \pkg{RSpectra}~\citep{RSpectra},
and the \pkg{irlba}~\citep{baglama2005augmented} packages.
The \pkg{svd} package provides a wrapper for the \pkg{PROPACK} SVD
algorithm~\citep{larsen1998lanczos}. The \pkg{RSpectra} package is
inspired by the software package \pkg{ARPACK}~\citep{ARPACK} and
provides fast partial SVD and eigendecompositon algorithms. The
\pkg{irlba} package implements implicitly restarted Lanczos
bidiagonalization methods for computing the dominant singular values
and vectors~\citep{baglama2005augmented}. The advantage of this
algorithm is that it avoids the implicit computation of the inner
product $\mathbf{A^\top A}$ or outer product $\mathbf{A
  A^\top}$. Thus, this algorithm is more numerically stable if
$\mathbf{A}$ is ill-conditioned.

\pagebreak
\subsection[The rsvd() function]{The \code{rsvd()} function}\label{sec:rsvdfunction}

The \pkg{rsvd} package provides an efficient routine to compute the low-rank SVD using Algorithm~\ref{alg:RSVDalgorithm}. The interface of the \code{rsvd()} function is similar to the base \code{svd()} function:
\begin{Code}
rsvd(A, k, nu = NULL, nv = NULL, p = 10, q = 2, sdist = "normal")
\end{Code}
%
%
The first mandatory argument \code{A} passes the $m\times n$ input data matrix. The second mandatory argument \code{k} defines the target rank, which is assumed to be chosen smaller than the ambient dimensions of the input matrix. 
The \code{rsvd()} function achieves significant speedups for target ranks chosen to be $k < \min\{m,n\}/4$. 
Similar to the \code{svd()} function, the arguments \code{nu} and \code{nv} can be used to specify the number of left and right singular vectors to be returned. 

The accuracy of the approximation can be controlled via the two tuning
parameters \code{p} and \code{q}. The former parameter is used to
oversample the basis, and is set by default to \code{p = 10}. This
setting guarantees a good basis with high probability in general. The
parameter \code{q} can be used to compute additional power iterations
(subspace iterations). By default this parameter is set to \code{q = 2},
which yields a good performance in our numerical experiments, i.e.,
the default values show an optimal trade-off between speed and
accuracy in standard situations. If the singular value spectrum of the
input matrix decays slowly, more power iterations are desirable.

Further, the \code{rsvd()} routine allows one to choose between a
standard normal, uniform and Rademacher random test matrices. The
different options can be selected via the argument \code{sdist =
  c("normal", "unif", "rademacher")}.

The resulting model object is itself a list.  It contains the
following components:
\begin{compactitem}  
\item \code{d}: $k$-dimensional vector containing the singular values.
\item \code{u}: $m\times k$ matrix containing the left singular vectors.
\item \code{v}: $n\times k$ matrix containing the right singular vectors. Note that \code{v} is not returned in its transposed form, as it is often returned in other programing languages.
\end{compactitem}
More details are provided in the corresponding documentation, see \code{?rsvd}.

\subsection{SVD example: Image compression} \label{sec:image_compression}
The singular value decomposition can be used to obtain a low-rank approximation of high-dimensional data. Image compression is a simple, yet illustrative example. The underlying structure of natural images can often be represented by a very sparse model. This means that images can be faithfully recovered from a relatively small set of basis functions. For demonstration, we use the following $1600 \times 1200$ grayscale image:
\begin{CodeChunk}
\begin{CodeInput}
R> data("tiger", package = "rsvd")
R> image(tiger, col = gray(0:255 / 255))
\end{CodeInput}
\end{CodeChunk}
A grayscale image may be thought of as a real-valued matrix
$\mathbf{A} \in \mathbb{R}^{m\times n}$, where $m$ and $n$ are the
number of pixels in the vertical and horizontal directions,
respectively. To compress the image we need to first decompose the
matrix $\mathbf{A}$. The singular vectors and values provide a
hierarchical representation of the image in terms of a new coordinate
system defined by dominant correlations within rows and columns of the
image. Thus, the number of singular vectors used for approximation
poses a trade-off between the compression rate (i.e., the number of
singular vectors to be stored) and the reconstruction fidelity. In the
following, we use the arbitrary choice $k=100$ as target rank. First,
the \proglang{R} base \code{svd()} function is used to compute the
truncated singular value decomposition:
\begin{CodeChunk}
\begin{CodeInput}
R> k <- 100 
R> tiger.svd <- svd(tiger, nu = k, nv = k)
\end{CodeInput}
\end{CodeChunk}
The \code{svd()} function returns three objects: \code{u}, \code{v}
and \code{d}. The first two objects are $m\times k$ and $n\times k$
arrays, namely the truncated left and right singular vectors. The
vector \code{d} is comprised of the $\min\{m,n\}$ singular values in
descending order. Now, the dominant $k=100$ singular values are
retained to approximate/reconstruct
($\mathbf{A}_k := \mathbf{U}_k\mathbf{D}_k\mathbf{V}_k^\top $) the
original image:
\begin{CodeChunk}
\begin{CodeInput}
R> tiger.re <- tiger.svd$u %*% diag(tiger.svd$d[1:k]) %*% t(tiger.svd$v) 
R> image(tiger.re, col = gray(0:255 / 255))
\end{CodeInput}
\end{CodeChunk}
The normalized root mean squared error (nrmse) is a common measure for the reconstruction quality of images, computed as: 
\begin{CodeChunk}
\begin{CodeInput}
R> nrmse <- sqrt(sum((tiger - tiger.re) ** 2 ) / sum(tiger ** 2))
\end{CodeInput}
\end{CodeChunk}
Using only $k=100$ singular values/vectors, a reconstruction error as low as $12.1\%$ is achieved. This illustrates the general fact that natural images feature a very compact representation. Note, that the singular value decomposition is also a numerically reliable tool for extracting a desired signal from noisy data. The central idea is that the small singular values mainly represent the noise, while the dominant singular values represent the desired signal. 

\begin{figure}[t!]
	\centering
	\begin{subfigure}[t]{0.4\textwidth}
		\centering
		\DeclareGraphicsExtensions{.pdf}
		\includegraphics[width=1\textwidth]{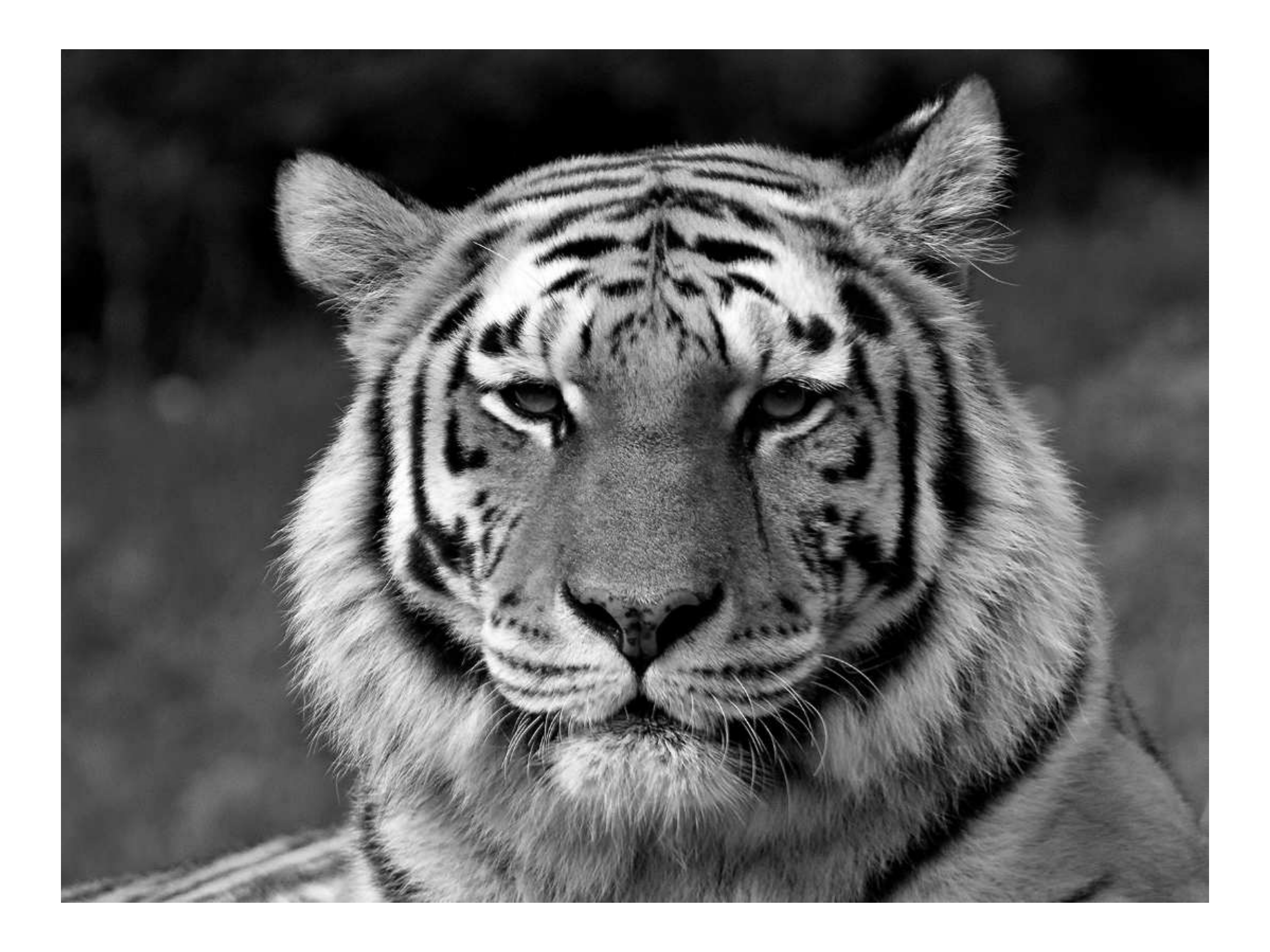}
		\caption{Original image. }
	\end{subfigure}
	\begin{subfigure}[t]{0.4\textwidth}
		\centering
		\DeclareGraphicsExtensions{.pdf}
		\includegraphics[width=1\textwidth]{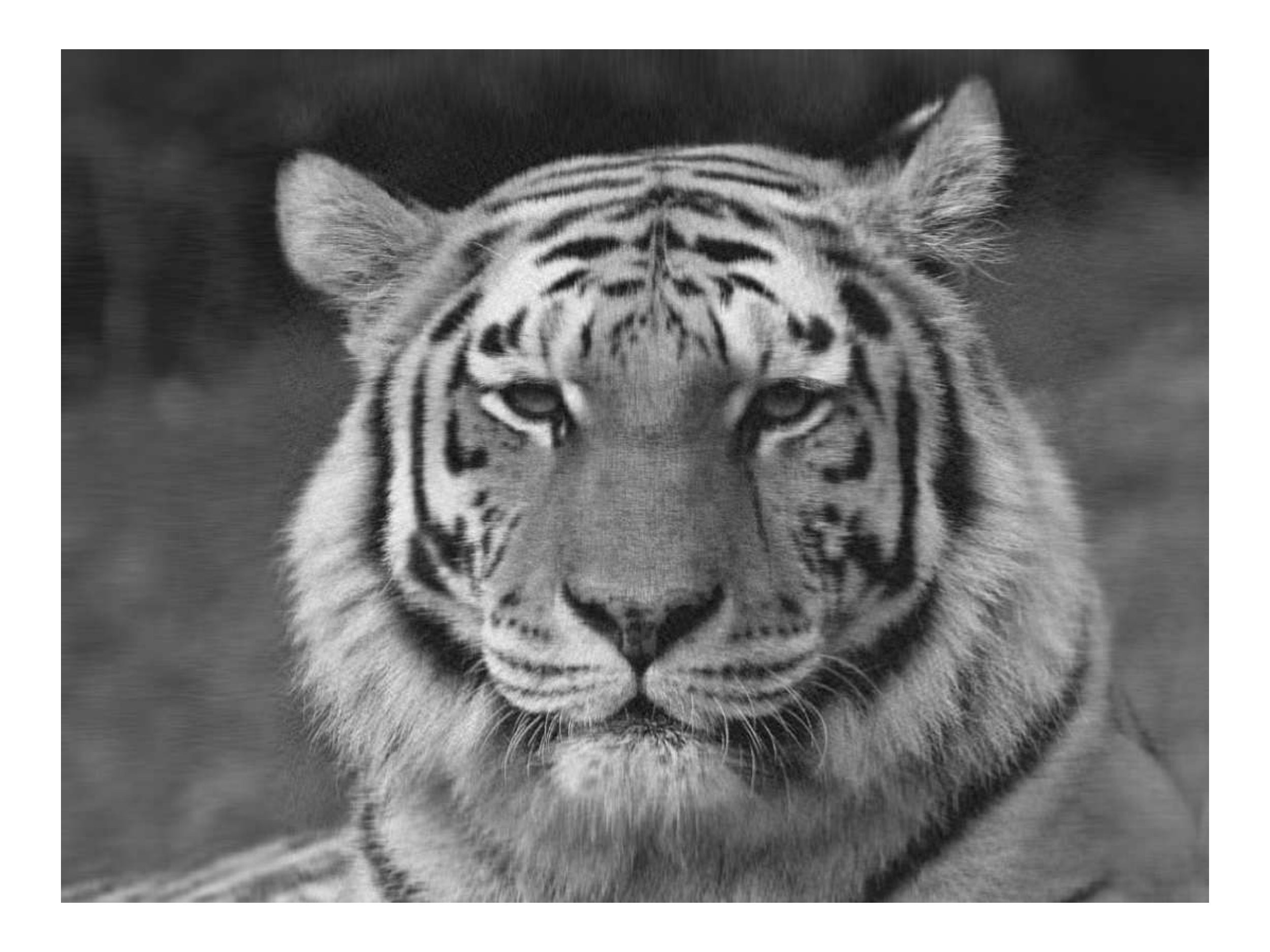}
		\caption{SVD (nrmse = $0.121$). }
	\end{subfigure}	
	
	\begin{subfigure}[t]{0.4\textwidth}
		\centering
		\DeclareGraphicsExtensions{.pdf}
		\includegraphics[width=1\textwidth]{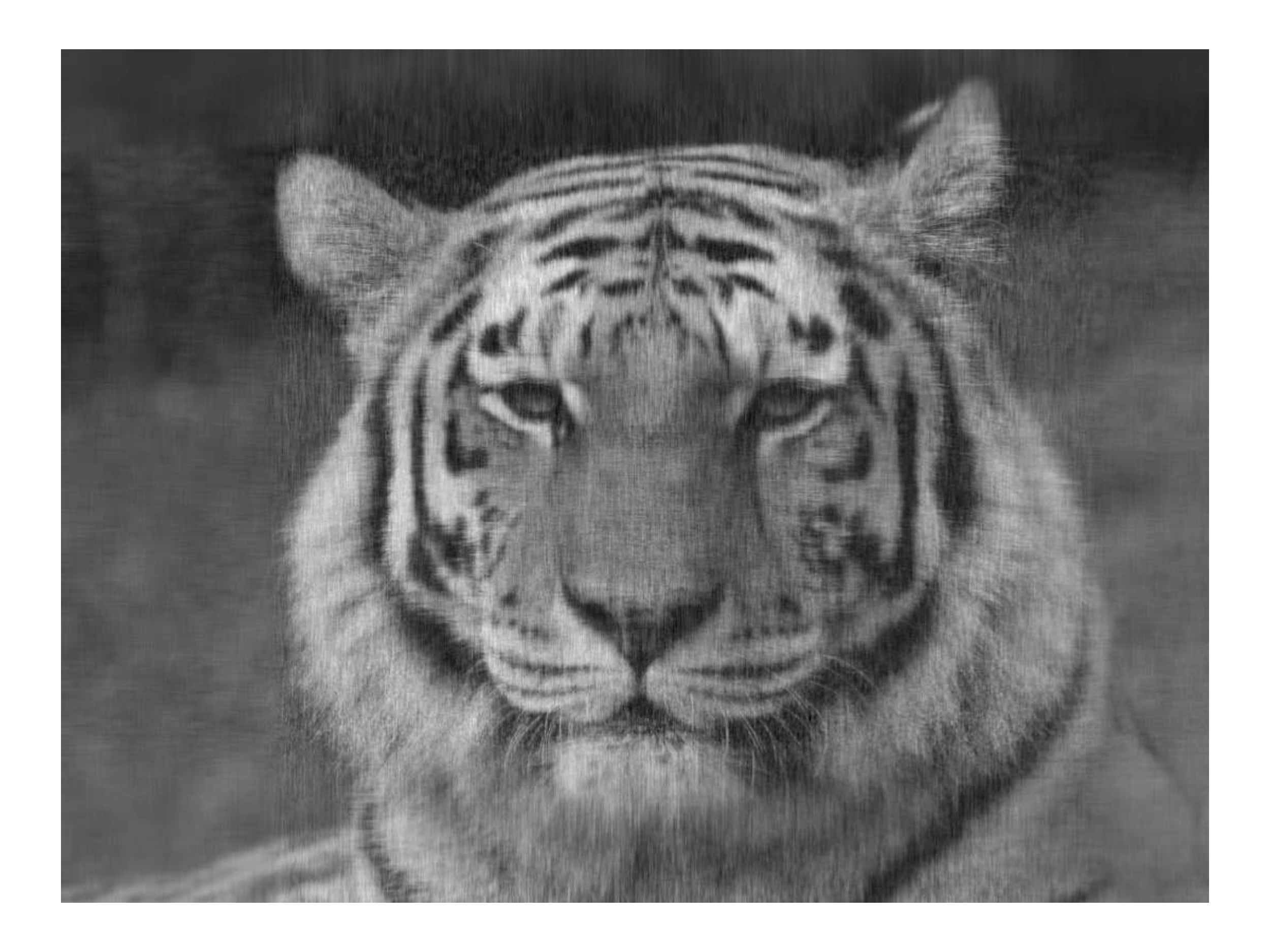}
		\caption{rSVD using $q=0$ (nrmse = $0.165$). }
	\end{subfigure}
	\begin{subfigure}[t]{0.4\textwidth}
		\centering
		\DeclareGraphicsExtensions{.pdf}
		\includegraphics[width=1\textwidth]{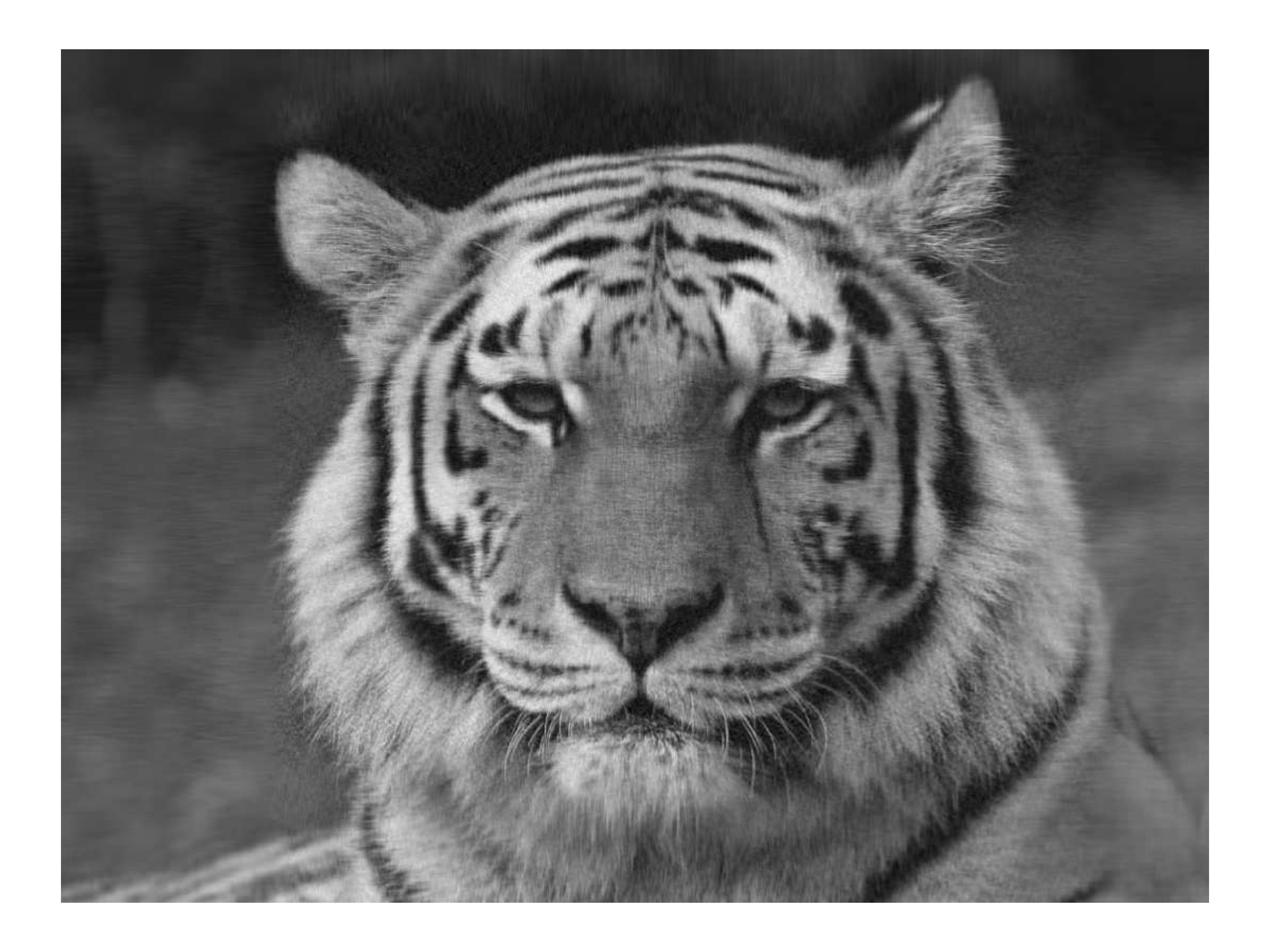}
		\caption{rSVD using $q=2$ (nrmse = $0.122$).}
	\end{subfigure}
	\caption{ Subplot (a) shows the original image, and subplots
          (b), (c) and (d) show the reconstructed images using the
          dominant
          $k=100$ components. The reconstruction quality of randomized
          SVD with power iterations in (d) is nearly as good as of the
          deterministic SVD.}
	\label{Fig:tiger}
\end{figure}

If the data matrix exhibits low-rank structure, the provided \code{rsvd()} function can be used as a plug-in function for the base \code{svd()} function, in order to compute the near-optimal low-rank singular value decomposition:
\begin{CodeChunk}
\begin{CodeInput}
R> tiger.rsvd <- rsvd(tiger, k = k)
\end{CodeInput}
\end{CodeChunk}
Similar to the base SVD function, the \code{rsvd()} function returns three objects: \code{u}, \code{v} and \code{d}. Again, \code{u} and \code{v} are $m\times k$ and $n\times k$ arrays containing the approximate left and right singular vectors and the vector \code{d} is comprised of the $k$ singular values in descending order.
Optionally, the approximation accuracy of the randomized SVD algorithm can be controlled by the two parameters \code{p} and \code{q}, as described in the previous section.  
Again, the approximated image and the reconstruction error can be computed as: 
\begin{CodeChunk}
\begin{CodeInput}
R> tiger.re <- tiger.rsvd$u %*% diag(tiger.rsvd$d) %*% t(tiger.rsvd$v) 
R> nrmse <- sqrt(sum((tiger - tiger.re) ** 2) / sum(tiger ** 2))
\end{CodeInput}
\end{CodeChunk}
The reconstruction error is about $0.122$, i.e., close to the optimal truncated SVD. Figure~\ref{Fig:tiger} presents the visual results using both the deterministic and randomized SVD algorithms. 
By visual inspection, no significant differences can be seen between (b) and (d).
However, the quality suffers by omitting subspace iterations in (c).  

Table~\ref{Tab:tiger2} shows the performance for different SVD
algorithms in \proglang{R}. The \code{rsvd()} functions achieves an
average speedup of about $4$--$7$ over the \code{svd()} function. The
\code{svds()} and \code{irlba()} functions achieve speedups of about
$1.5$. The computational gain of the randomized algorithm becomes more
pronounced with increased matrix dimension, e.g., images with higher
resolution. The trade-off between accuracy and speed of the partial
SVD algorithms depends on the precision parameter \code{tol}, and we
set the tolerance parameter for all algorithms to \code{tol = 1e-5}.

\begin{table}[t!]
	\centering
	\begin{tabular}{l l l c c c}
		\hline
		{\bf Package} & {\bf Function} & {\bf Parameters}  & {\bf Time (s)} & {\bf Speedup} & {\bf Error} \\ 
		\hline
		
		\pkg{base}    &  \code{svd()}    &  \code{nu = nv = 100}	    				   &	 \centering 0.37	& {\centering\arraybackslash * }  &  {\centering\arraybackslash 0.121 }  \\ 
		
		\pkg{svd}    & \code{propack.svd()}  & \code{neig = 100} & \centering 0.55	& 0.67  & {\centering\arraybackslash 0.121  }  \\   
		
		\pkg{RSpectra}    & \code{svds()}   & \code{k = 100}  & \centering 0.25	& 1.48  &  {\centering\arraybackslash 0.121 } \\   				

		\pkg{irlba}    & \code{irlba()}   & \code{nv = 100}  & \centering 0.24	& 1.54  &  {\centering\arraybackslash 0.121 } \\

		\pkg{rsvd}    & \code{rsvd()}   & \code{k = 100, q = 0}  & \centering 0.03	&   12.3  & {\centering\arraybackslash 0.165 } \\
		
		\pkg{rsvd} & \code{rsvd()} & \code{k = 100, q = 1}  & \centering 0.052 &   7.11  & {\centering\arraybackslash 0.125 } \\ 
	
		\pkg{rsvd} & \code{rsvd()} & \code{k = 100, q = 2}  & \centering 0.075 &   4.9 & {\centering\arraybackslash 0.122 }  \\ 
	
		\pkg{rsvd} & \code{rsvd()} & \code{k = 100, q = 3}  & \centering 0.097 &   3.8 & {\centering\arraybackslash 0.121 }  \\ 		
		\hline
	\end{tabular}
	\caption{Summary of algorithm runtimes (averaged over $20$ runs) and errors. The randomized routines achieve substantial speedups, while attaining similar reconstruction errors with $q\geq1$.}
	\label{Tab:tiger2}
\end{table}

\subsection[Computational performance]{Computational performance}\label{sec:compute_performance}

In the following we evaluate the performance of the randomized SVD routine and compare it to other SVD routines available in \proglang{R}.
To fully exploit the power of randomized algorithms we use the
enhanced \proglang{R} distribution {Microsoft \proglang{R} Open 3.4.3}. This
\proglang{R} distribution is linked with multi-threaded
\proglang{LAPACK} libraries, which use all available cores and
processors.
Compared to the standard CRAN \proglang{R} distribution, which uses
only a single thread (processor), the enhanced \proglang{R}
distribution shows significant speedups for matrix operations. For
benchmark results, see
\url{https://mran.microsoft.com/documents/rro/multithread/}.
However, we see also significant speedups when using the standard CRAN
\proglang{R} distribution.

A machine with Intel Core i7-7700K CPU Quad-Core 4.20GHz, 64GB fast memory, and operating-system Ubuntu 17.04 is used for all computations. The \pkg{microbenchmark} package is used for accurate timing \citep{microbenchmark}.

\begin{figure}[t!]
	\centering
	\DeclareGraphicsExtensions{.pdf}
	\includegraphics[width=0.75\textwidth]{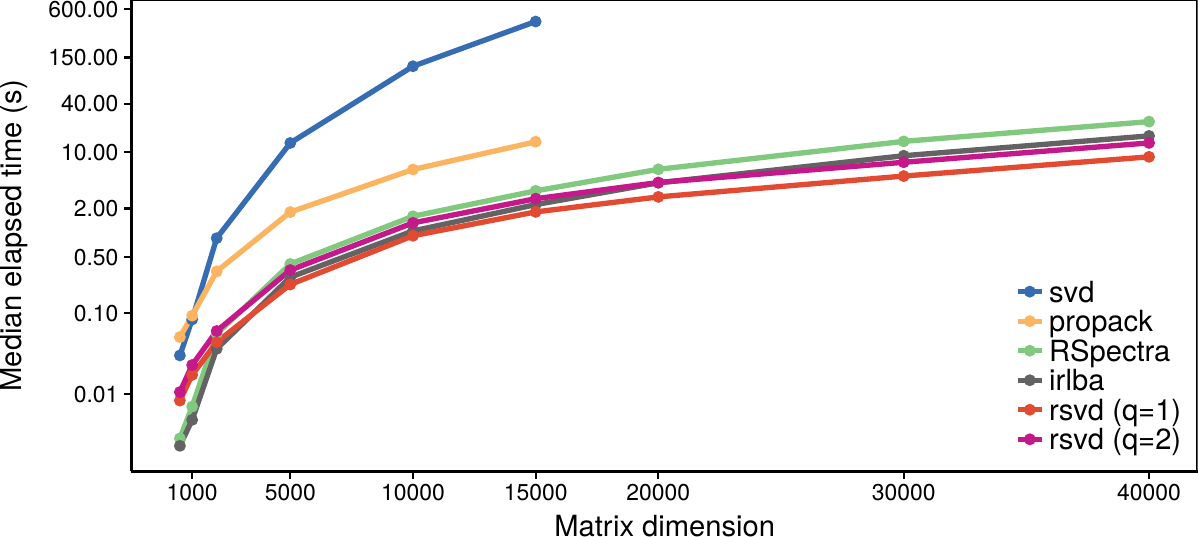}
	\caption{Runtimes for computing rank $k=20$ approximations for varying matrix dimensions.}	
	\label{fig:timing_dimensions}	
\end{figure}

\begin{figure}[t!]
	\centering
	\begin{subfigure}[t]{0.31\textwidth}
		\centering
		\DeclareGraphicsExtensions{.pdf}
		\includegraphics[width=1\textwidth]{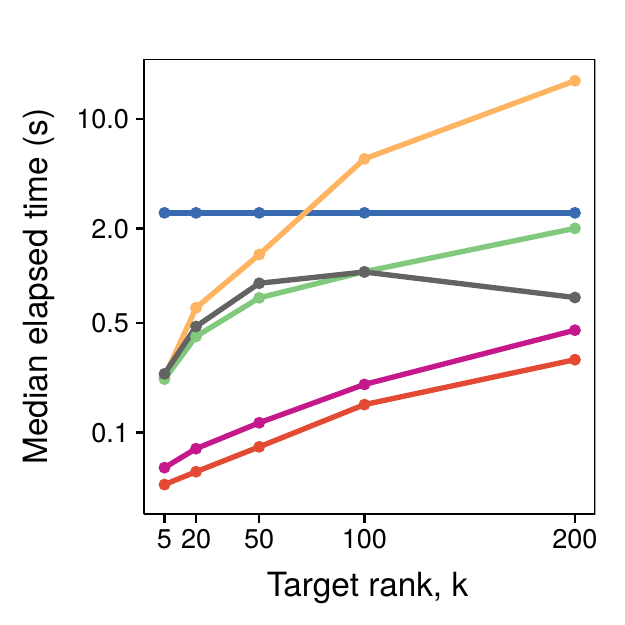}
		\vspace{-.25in}
		\caption{Runtime.}
	\end{subfigure}
	\begin{subfigure}[t]{0.31\textwidth}
		\centering
		\DeclareGraphicsExtensions{.pdf}
		\includegraphics[width=1\textwidth]{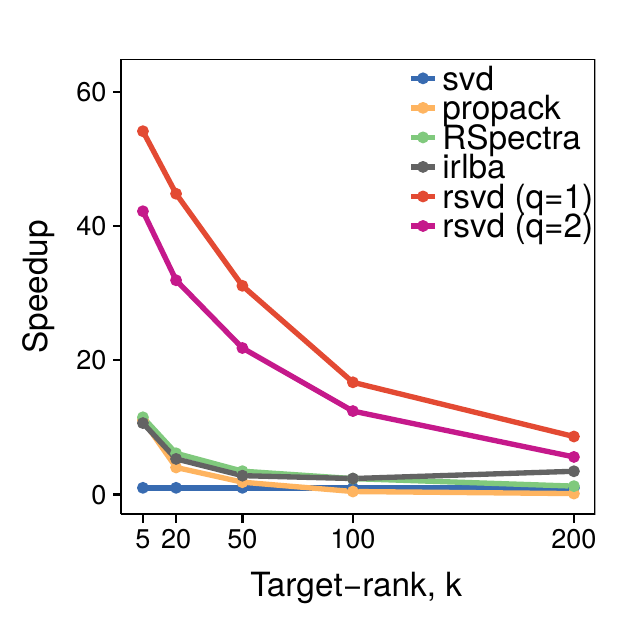}
		\vspace{-.25in}
		\caption{Speedups.}
	\end{subfigure}
	\begin{subfigure}[t]{0.31\textwidth}
		\centering
		\DeclareGraphicsExtensions{.pdf}
		\includegraphics[width=1\textwidth]{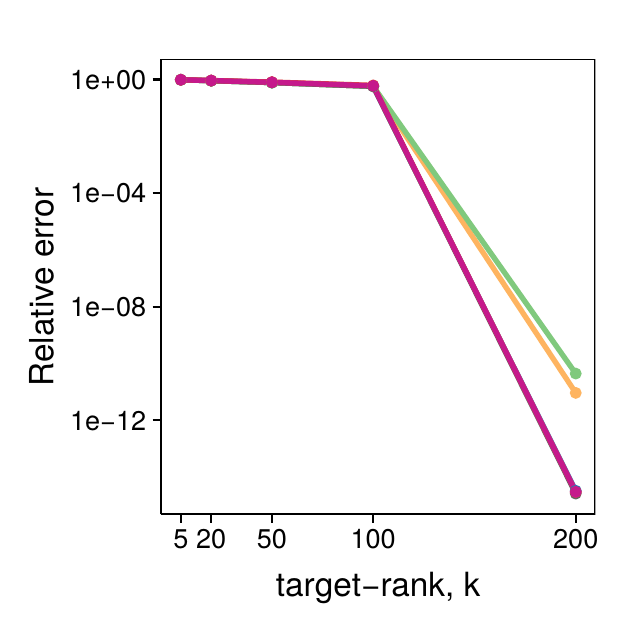}
		\vspace{-.25in}
		\caption{Relative errors.}
	\end{subfigure}	
	
	\caption{Computational performance for a dense $3000\times 2000$ low-rank matrix.}
	\label{fig:timing_3000x2000}	
\end{figure}

\begin{figure}[t!]
	\centering
	\begin{subfigure}[t]{0.31\textwidth}
		\centering
		\DeclareGraphicsExtensions{.pdf}
		\includegraphics[width=1\textwidth]{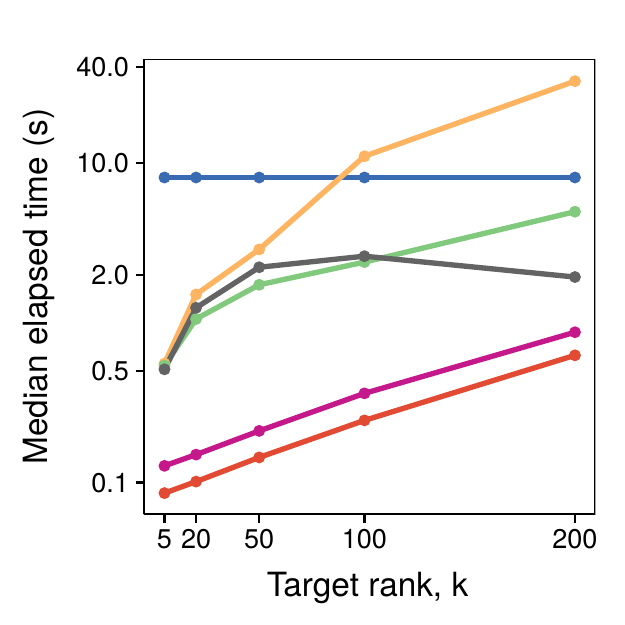}
		\caption{Runtime.}
	\end{subfigure}
	\begin{subfigure}[t]{0.31\textwidth}
		\centering
		\DeclareGraphicsExtensions{.pdf}
		\includegraphics[width=1\textwidth]{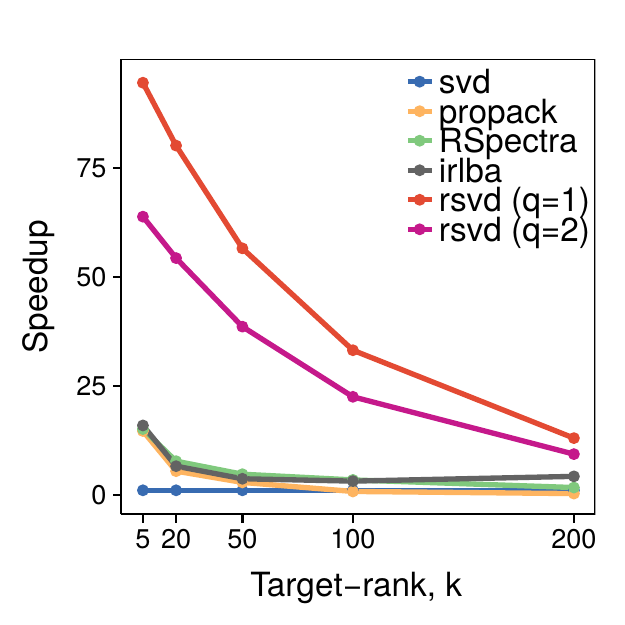}
		\caption{Speedups.}
	\end{subfigure}
	\begin{subfigure}[t]{0.31\textwidth}
		\centering
		\DeclareGraphicsExtensions{.pdf}
		\includegraphics[width=1\textwidth]{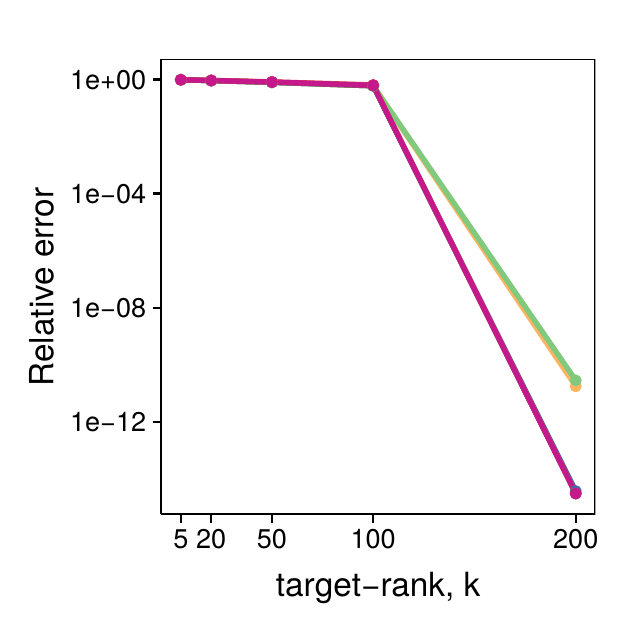}
		\caption{Relative errors.}
	\end{subfigure}	
	\caption{Computational performance for a dense $5000\times 3000$ low-rank matrix.}	
	\label{fig:timing_5000x3000}	
\end{figure}

\begin{figure}[t!]
	\centering
	\begin{subfigure}[t]{0.31\textwidth}
		\centering
		\DeclareGraphicsExtensions{.pdf}
		\includegraphics[width=1\textwidth]{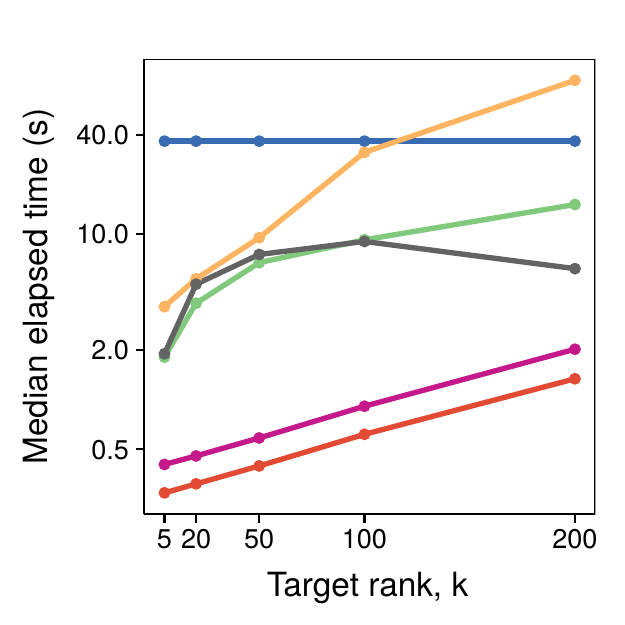}
		\caption{Runtime.}
	\end{subfigure}
	\begin{subfigure}[t]{0.31\textwidth}
		\centering
		\DeclareGraphicsExtensions{.pdf}
		\includegraphics[width=1\textwidth]{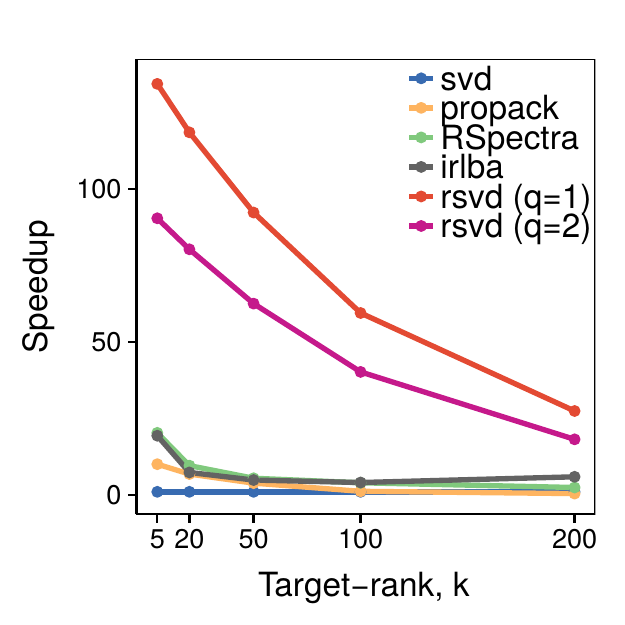}
		\caption{Speedups.}
	\end{subfigure}
	\begin{subfigure}[t]{0.31\textwidth}
		\centering
		\DeclareGraphicsExtensions{.pdf}
		\includegraphics[width=1\textwidth]{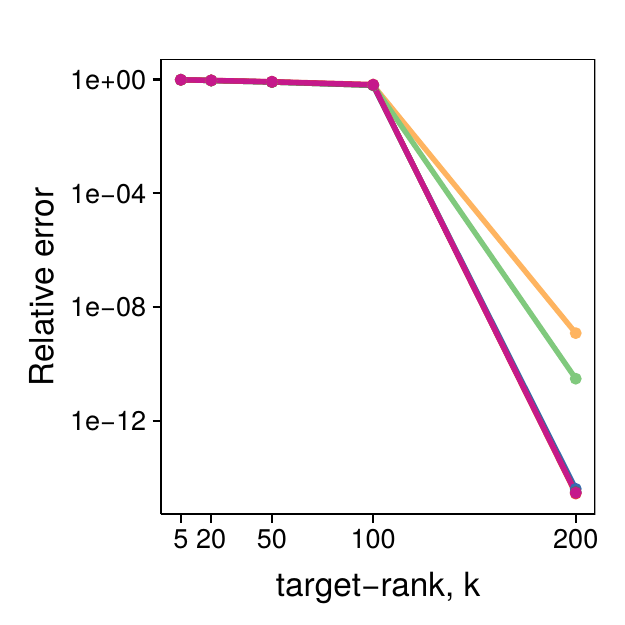}
		\caption{Relative errors.}
	\end{subfigure}	
	\caption{Computational performance for a dense $10000\times 5000$ low-rank matrix.}	
	\label{fig:timing_10000x5000}	
\end{figure}

To compare the computational performance of the SVD algorithms, we consider low-rank matrices with varying dimensions $m$ and $n$, and intrinsic rank $r=200$, generated as:
\begin{CodeChunk}
\begin{CodeInput}
R> A <- matrix(rnorm(m * r), m, r) %*% matrix(rnorm(r * n), r, n)  
\end{CodeInput}
\end{CodeChunk}
Figure~\ref{fig:timing_dimensions} shows the runtime for low-rank approximations (target-rank $k=20$) and varying matrix dimensions $m\times n$, where the second dimension is chosen to be $n := 0.75\cdot m$. While the routines of the \pkg{RSpectra} and \pkg{irlba} packages perform best for small dimensions, the computational advantage of the randomized SVD becomes pronounced with increasing dimensions.  
We now investigate the performance of the routines in more detail.
Figures~\ref{fig:timing_3000x2000},~\ref{fig:timing_5000x3000},
and~\ref{fig:timing_10000x5000} show the computational performance for
varying matrix dimensions and target ranks. The elapsed time is
computed as the median over $20$ runs. The speedups show the relative
performance compared to the base \code{svd()}, i.e., the average
runtime of the \code{svd()} function is divided by the runtime of the
other SVD algorithms. The relative reconstruction error is computed as
\begin{equation*}
\frac{\|\mathbf{A}-\mathbf{A}_k\|_F}{\|\mathbf{A}\|_F}, 
\end{equation*}
where $\mathbf{A}_k:=\mathbf{U}_k\mathbf{\Sigma}_k\mathbf{V}_k^\top$
is the rank-$k$ matrix approximation.

The \code{rsvd()} function achieves substantial speedups over the other SVD routines. 
Here, the oversampling parameter is fixed to $p=10$, but it can be
seen that additional power iterations improve the approximation
accuracy. This allows the user to control the trade-off between
computational time and accuracy, depending on the application. Note
that we have set the precision parameter of the \pkg{RSpectra},
\pkg{irlba} and \pkg{propack} routines to \code{tol = 1e-5}.

\begin{figure}[t!]
	\centering
	\begin{subfigure}[t]{0.31\textwidth}
		\centering
		\DeclareGraphicsExtensions{.pdf}
		\includegraphics[width=1\textwidth]{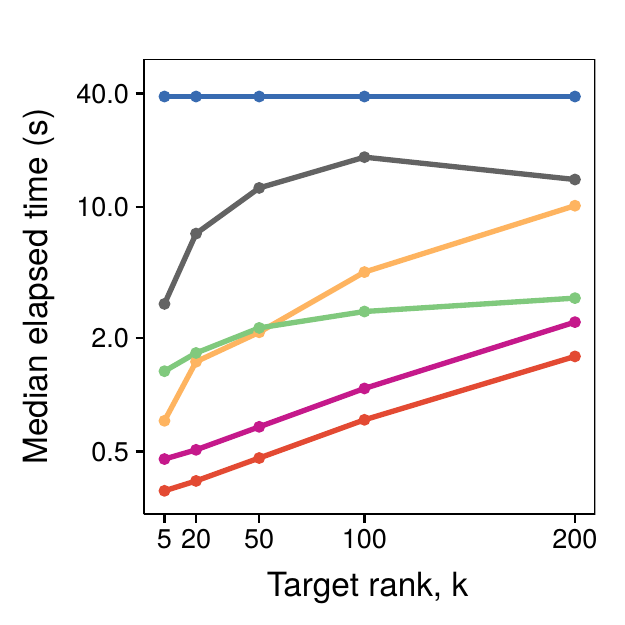}
		\caption{Runtime.}
	\end{subfigure}
	\begin{subfigure}[t]{0.31\textwidth}
		\centering
		\DeclareGraphicsExtensions{.pdf}
		\includegraphics[width=1\textwidth]{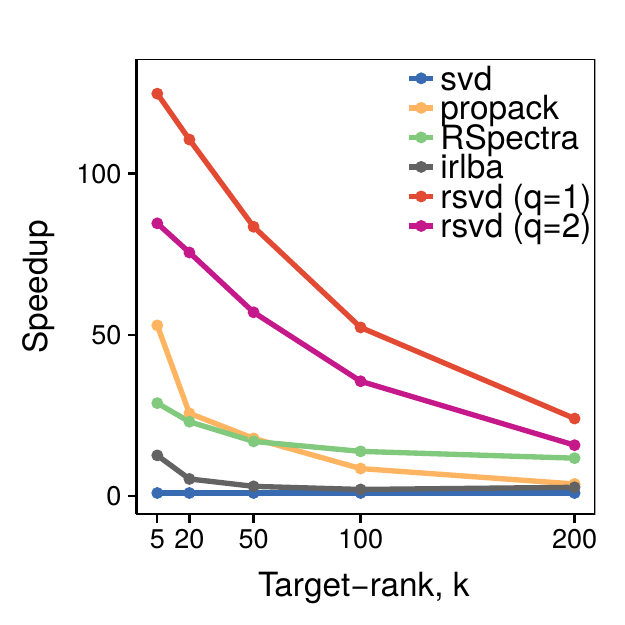}
		\caption{Speedups.}
	\end{subfigure}
	\begin{subfigure}[t]{0.31\textwidth}
		\centering
		\DeclareGraphicsExtensions{.pdf}
		\includegraphics[width=1\textwidth]{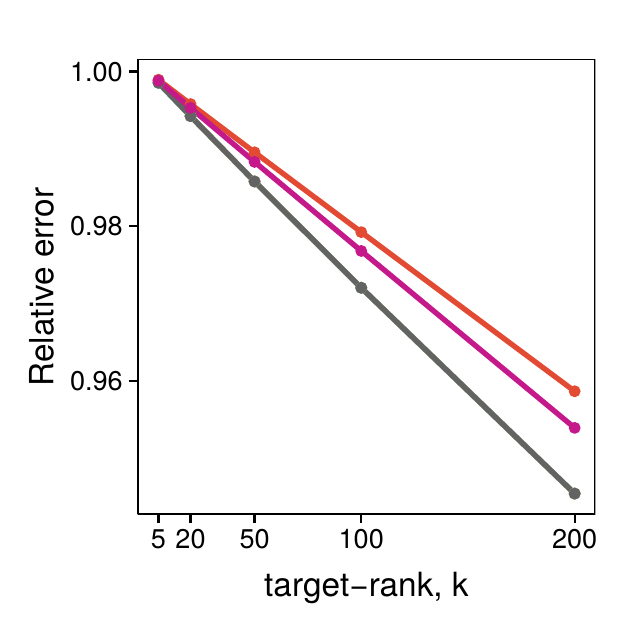}
		\caption{Relative errors.}
	\end{subfigure}	
	\caption{Computational performance for a sparse $10000\times 5000$ matrix.}	
	\label{fig:timing_s10000x5000}	
\end{figure}

Figure~\ref{fig:timing_s10000x5000} show the computational performance for sparse matrices with about $5\%$ non-zero elements. The \pkg{RSpectra}, and \pkg{propack} routines are specifically designed for sparse and structured matrices, and show considerably better computational performance. 

Note that the random sparse matrices do not feature low-rank structure; hence, the large relative error. Still, the randomized SVD shows a good trade-off between speedup and accuracy.

\section{Randomized principal component analysis} \label{sec:rPCA}

Dimensionality reduction is a fundamental concept in modern data analysis. The idea is to exploit relationships among points in high-dimensional space in order to construct some low-dimensional summaries. This process aims to eliminate redundancies, while preserving interesting characteristics of the data~\citep{burges2010dimension}. Dimensionality reduction is used to improve the computational tractability, to extract interesting features, and to visualize data which are comprised of many interrelated variables. The most important linear dimension reduction technique is principal component analysis (PCA), originally formulated by~\cite{PCApearson} and~\cite{hotelling1933analysis}. 
PCA plays an important role, in particular, due to its simple geometric interpretation.
\cite{jolliffe2002principal} provides a comprehensive introduction to PCA.

\subsection{Conceptual overview}
\begin{figure}[t!]
	\centering
	\begin{subfigure}{0.49\textwidth}
		\centering
		\DeclareGraphicsExtensions{.pdf}
		\includegraphics[]{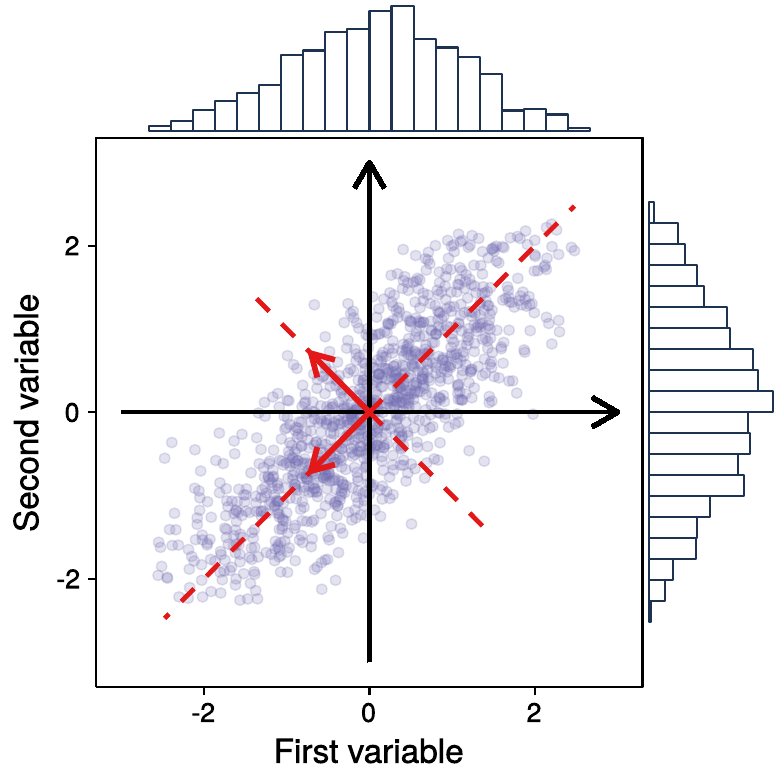}
		\caption{Original (standardized) data.}
		\label{fig:2dPCAa}
	\end{subfigure}%
	\hspace*{-1.0cm}
	\begin{subfigure}{0.49\textwidth}
		\centering
		\DeclareGraphicsExtensions{.pdf}
		\includegraphics[]{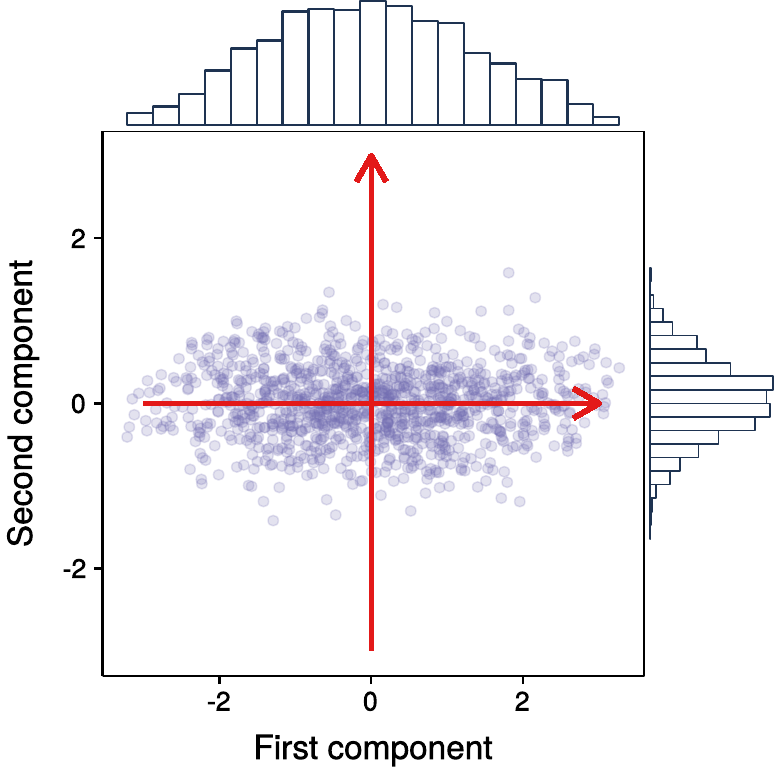}
		\caption{Rotated data.}
		\label{fig:2dPCAb}
	\end{subfigure}
	\caption{PCA seeks to find a new set of uncorrelated variables. Plot (a) shows some two-dimensional data, and its two principal directions. Plot (b) shows the new principal components. Geometrically, the PCs are simply a rotation and reflection of the original data, so that the first component accounts for most of the variation in the data, now. }
	\label{fig:2dPCA}
\end{figure}

Principal component analysis aims to find a new set of uncorrelated variables. 
The so called principal components (PCs) are constructed such that the first PC explains most of the variation in the data; the second PC most of the remaining variation and so on. 
This property ensures that the PCs sequentially capture most of the total variation (information) present in the data. 
In practice, we often aim to retain only a few number of PCs which
capture a ``good'' amount of the variation, where ``good'' depends on
the application.
The idea is depicted for two correlated variables in Figure~\ref{fig:2dPCA}.
Figure~\ref{fig:2dPCAa} illustrates the two principal directions of the data, which span a new coordinate system. The first principal direction is the vector pointing in the direction which accounts for most of the variability in data. The second principal direction is orthogonal (perpendicular) to the first one and captures the remaining variation in the data. 
Figure~\ref{fig:2dPCAb} shows the original data using the principal directions as a new coordinate system.
Compared to the original data, the histograms indicate that most of the variation is now captured by just the first principal component, while less by the second component.  

To be more formal, assume a data matrix $\mathbf{X} \in \mathbb{R}^{m \times n}$ with $m$ observations and $n$ variables (column-wise, mean-centered). Then, the principal components can be expressed as a weighted linear combination of the original variables
\begin{equation*}
\mathbf{z}_{i} := \mathbf{X}\mathbf{w}_{i},
\end{equation*}
where $\mathbf{z}_{i} \in \mathbb{R}^{m}$ denotes the $i${th} principal component. The vector $\mathbf{w}_{i} \in \mathbb{R}^{n}$ is the $i${th} principal direction, where the elements of $\mathbf{w}_{i}=[w_{1},\ldots,w_{n}]^\top$ are the principal component coefficients. 

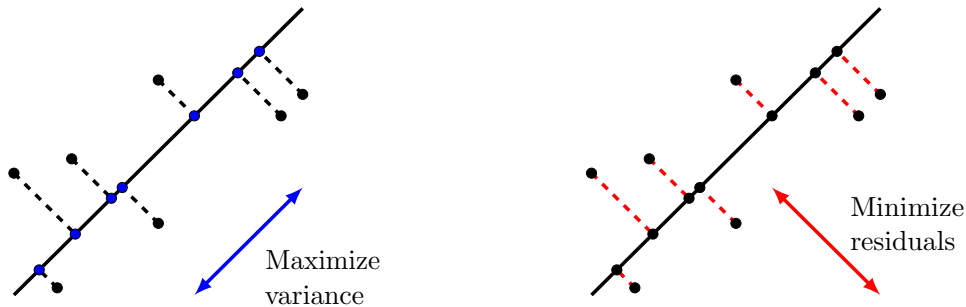
\begin{figure}[t!]
  \centering
    \scalebox{0.95}{
			\begin{tikzpicture}[auto,node distance = 2cm,>=latex']
			
			\coordinate (A) at (-2,-2);
			\coordinate (B) at (2,2);
			\draw [line width=1.4pt]  (A) -- (B);
			
			\draw [fill=black] (0,1) circle (2pt);			
			\draw [dashed, line width=1.4pt]  (0,1) -- (0.5,0.5);
			\draw [fill=blue] (0.5,0.5) circle (2pt);	

			\draw [fill=black] (1.7,0.5) circle (2pt);
			\draw [dashed, line width=1.4pt]  (1.7,0.5) -- (1.1,1.1);
			\draw [fill=blue] (1.1,1.1) circle (2pt);										
			\draw [fill=black] (2,0.8) circle (2pt);						
			\draw [dashed, line width=1.4pt]  (2,0.8) -- (1.4,1.4);	
			\draw [fill=blue] (1.4,1.4) circle (2pt);									
			\draw [fill=black] (-1.4,-1.9) circle (2pt);									
			\draw [dashed, line width=1.4pt]  (-1.4,-1.9) -- (-1.65, -1.65);
			\draw [fill=blue] (-1.65, -1.65) circle (2pt);

			\draw [fill=black] (0,-1) circle (2pt);
			\draw [dashed, line width=1.4pt]  (0,-1) -- ( -0.5, -0.5);			
			\draw [fill=blue] ( -0.5, -0.5) circle (2pt);

			\draw [fill=black] (-1.2,-0.1) circle (2pt);		
			\draw [dashed, line width=1.4pt]  (-1.2,-0.1) -- (-0.65, -0.65);			
			\draw [fill=blue] (-0.65, -0.65) circle (2pt);

			\draw [fill=black] (-2,-0.3) circle (2pt);
			\draw [dashed, line width=1.4pt]  (-2,-0.3) -- (-1.15, -1.15);			
			\draw [fill=blue] (-1.15, -1.15) circle (2pt);				
			
			\draw [latex-latex, blue, line width=1.4pt] (0.5,-2) -- (2,-0.5);
			\draw [fill=black] (1.35,-1.75) circle (0pt) node [right, text width=1.5cm] {Maximize variance};


			\coordinate (A2) at (8+-2,-2);
			\coordinate (B2) at (8+2,2);
			\draw [line width=1.4pt]  (A2) -- (B2);			
			
			\draw [dashed, red, line width=1.4pt]  (8+0,1) -- (8+0.5,0.5);
			\draw [fill=black] (8+0,1) circle (2pt);			
			\draw [fill=black] (8+0.5,0.5) circle (2pt);	

			\draw [dashed, red, line width=1.4pt]  (8+1.7,0.5) -- (8+1.1,1.1);			
			\draw [fill=black] (8+1.7,0.5) circle (2pt);
			\draw [fill=black] (8+1.1,1.1) circle (2pt);										
			\draw [dashed,  red, line width=1.4pt]  (8+2,0.8) -- (8+1.4,1.4);				
			\draw [fill=black] (8+2,0.8) circle (2pt);						
			\draw [fill=black] (8+1.4,1.4) circle (2pt);									
			\draw [dashed,  red, line width=1.4pt]  (8+-1.4,-1.9) -- (8+-1.65, -1.65);			
			\draw [fill=black] (8+-1.4,-1.9) circle (2pt);									
			\draw [fill=black] (8+-1.65, -1.65) circle (2pt);
			
			\draw [dashed,  red, line width=1.4pt]  (8+0,-1) -- (8+ -0.5, -0.5);						
			\draw [fill=black] (8+0,-1) circle (2pt);
			\draw [fill=black] (8+ -0.5, -0.5) circle (2pt);					
			
			\draw [dashed,  red, line width=1.4pt]  (8+-1.2,-0.1) -- (8+-0.65, -0.65);						
			\draw [fill=black] (8+-1.2,-0.1) circle (2pt);		
			\draw [fill=black] (8+-0.65, -0.65) circle (2pt);

			\draw [dashed,  red, line width=1.4pt]  (8+-2,-0.3) -- (8+-1.15, -1.15);						
			\draw [fill=black] (8+-2,-0.3) circle (2pt);
			\draw [fill=black] (8+-1.15, -1.15) circle (2pt);				
			
			\draw [latex-latex, red, line width=1.4pt] (8+0.5,-0.5) -- (8+2,-2);
			\draw [fill=black] (8.1+1.35,-1.0) circle (0pt) node [right, text width=1.8cm] {Minimize residuals};			
			
			\end{tikzpicture}}
	\caption{Principal component analysis can be formulated either as a variance maximization or as a least square minimization problem. Both views are equivalent.}
	\label{Fig:pcaviews}
\end{figure}

\begin{figure}[t!]
  \centering
    \scalebox{0.95}{
			\begin{tikzpicture}[auto,node distance = 2cm,>=latex']
			
			\coordinate (A) at (-4,0);
			\coordinate (B) at (4,0);
			\draw [line width=1.4pt]  (A) -- (B);

			\draw [-latex, blue, line width=1.4pt] (-2,0) -- (2,0);	
			\draw [-latex, black, line width=1.4pt] (-2,0) -- (2,2);	
			\draw [dashed, red, line width=1.4pt]  (2,0) -- (2,2);						
			\draw [fill=black] (-2,0) circle (2pt) node [below] {origin};								
			\draw [fill=black] (2,2) circle (2pt) node [above] {data point};	
			\draw [fill=black] (2,0) circle (2pt) node [below, text width=1.9cm] {projection};				
			
			\draw [fill=black] (0,1.5) circle (0pt) node [left, text width=1.5cm] {total variation};
			
			\draw [fill=black] (2,1) circle (0pt) node [right, red, text width=1.5cm] {unexplained variation};
			
			\draw [fill=black] (0,-0.5) circle (0pt) node [below, blue, text width=3.5cm] {explained variation};
			
			\draw [dashed, line width=1.4pt]  (4,0) -- (6,0);
			\draw [fill=black] (6,0) circle (0pt) node [below, black] {principal component};
			
			\end{tikzpicture}}
	\caption{The Pythagorean theorem provides a geometrical explanation for the relationship between the two views: The PCs can be obtained by either maximizing the variance or by minimizing the unexplained variation (squared residuals) of the data. }
	\label{Fig:pcaGeom}
\end{figure}
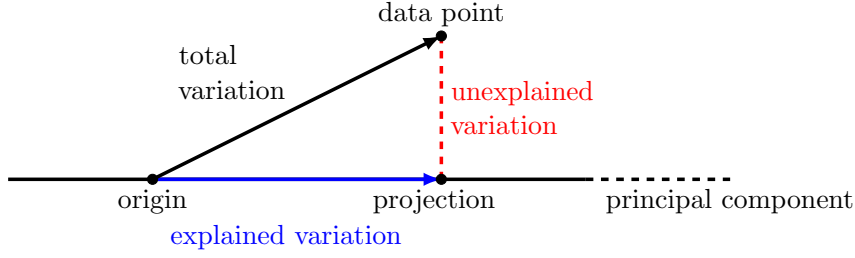

The problem is now to find a suitable vector $\mathbf{w}_{1}$ such that the first principal component $\mathbf{z}_{1}$ captures most of the variation in the data.
Mathematically, this problem can be formulated either as a least square problem or as a variance maximization problem~\citep{cunningham2015linear}. The two views are illustrated in Figure~\ref{Fig:pcaviews}.
This is, because the total variation equals the sum of the explained and unexplained variation~\citep{jolliffe2002principal}, illustrated in Figure~\ref{Fig:pcaGeom}. 

We follow the latter view, and maximize the variance of the first principal component $\mathbf{z}_{1} = \mathbf{X}\mathbf{w}_{1}$ subject to the normalization constraint $\|\mathbf{w}\|_2^2=1$
\begin{equation}\label{eq:varXw}		
\mathbf{w}_{1} : =\argmax_{\|\mathbf{w}\|_2^2=1}  \VAR(\mathbf{X}\mathbf{w}),
\end{equation}
where $\VAR$ denotes the variance operator. We can rewrite Equation~\ref{eq:varXw} as
\begin{equation}\label{eq:wXXw}	
\mathbf{w}_{1} :=  \argmax_{\|\mathbf{w}\|_2^2=1}  \frac{1}{m-1}  \| \mathbf{X}\mathbf{w} \|^{2}_2 = \argmax_{\|\mathbf{w}\|_2^2=1} \mathbf{w}^\top (\frac{1}{m-1}\mathbf{X}^\top \mathbf{X})\mathbf{w}.
\end{equation}
We note that the scaled inner product $\mathbf{X}^\top \mathbf{X}$ forms the sample covariance matrix
\begin{equation*}	
\mathbf{C} := \dfrac{1}{m-1}\mathbf{X}^\top \mathbf{X}.
\end{equation*}
$\mathbf{C}$ corresponds to the sample correlation matrix if the columns of $\mathbf{X}$ are both centered and scaled.
We substitute $\mathbf{C}$ into Equation~\ref{eq:wXXw}
\begin{equation*}
\mathbf{w}_{1} := \argmax_{\|\mathbf{w}\|_2^2=1} \mathbf{w}^\top \mathbf{C} \mathbf{w}.
\end{equation*}
Next, the method of Lagrange multipliers is used to solve the problem. First, we formulate the Lagrange function 
\begin{equation*}
\mathcal{L}(\mathbf{w}_{1},\lambda_{1}) = \mathbf{w}_{1}^\top \mathbf{C} \mathbf{w}_{1} - \lambda_{1}(\mathbf{w}_{1}^\top\mathbf{w}_{1}-1).
\end{equation*}
Then, we maximize the Lagrange function by differentiating with respect to $\mathbf{w}_{1}$
\begin{equation*}
\frac{\partial \mathcal{L}(\mathbf{w}_{1},\lambda_{1})}{\partial \mathbf{w}_{1}} = \mathbf{C} \mathbf{w}_{1} - \lambda_{1}\mathbf{w}_{1},
\end{equation*}
which leads to the well known eigenvalue problem.
Thus, the first principal direction for the mean centered matrix $\mathbf{X}$ is given by the dominant eigenvector $\mathbf{w}_1$ of the covariance matrix $\mathbf{C}$. The amount of variation explained by the first principal component is expressed by the corresponding eigenvalue $\lambda_1$. 
More generally, the subsequent principal component directions can be obtained by computing the eigendecompositon of the covariance or correlation matrix
\begin{equation*}
\mathbf{C}\mathbf{W} = \mathbf{W} \mathbf{\Lambda}.
\end{equation*}
The columns of $\mathbf{W} \in \mathbb{R}^{n \times n}$ are the eigenvectors (principal directions) which are orthonormal, i.e., $\mathbf{W}^\top\mathbf{W}=\mathbf{W}\mathbf{W}^\top=\mathbf{I}$. The diagonal elements of $\mathbf{\Lambda} \in \mathbb{R}^{n \times n}$ are the corresponding eigenvalues. 
%
The matrix $\mathbf{W}$ can also be interpreted as a projection matrix that maps the original observations to new coordinates in eigenspace. Hence, the $n$ principal components $\mathbf{Z} \in \mathbb{R}^{m \times n}$ can be more concisely expressed as 
\begin{equation*}
\mathbf{Z} := \mathbf{XW}.
\end{equation*}
Since the eigenvectors have unit norm, the projection should be purely rotational without any scaling; thus, $\mathbf{W}$ is also denoted as rotation matrix.

\subsubsection{PCA whitening}


In some situations, the scaled eigenvectors 
\begin{equation*}
\mathbf{L} := \mathbf{W}\mathbf{\Lambda}^{0.5}
\end{equation*}
provide a more insightful interpretation of the data.
$\mathbf{L} \in \mathbb{R}^{n \times n}$ is denoted as a loading matrix and provides a factorization of the covariance (correlation) matrix
\begin{equation*}	
\mathbf{C} = \mathbf{L}\mathbf{L}^\top =  \mathbf{W}\mathbf{\Lambda}\mathbf{W}^\top.
\end{equation*}
Thus, the loadings have the following two interesting properties: 
\begin{itemize}
	\item The squared column sums equal the eigenvalues.
	\item The squared row sums equal the variable's variance.
\end{itemize}

Further, the loading matrix $\mathbf{L}$ can be used to compute the $n$ whitened principal components
\begin{equation*}
\mathbf{Z}_\text{white} := \mathbf{XL}.
\end{equation*}

\begin{figure}[t!]
	\centering
	\begin{subfigure}{0.49\textwidth}
		\centering
		\DeclareGraphicsExtensions{.pdf}
		\includegraphics[]{pca_illustration2}
		\caption{Rotated data.}
		\label{fig:2dPCAa2}
	\end{subfigure}%
	\hspace*{-1.0cm}
	\begin{subfigure}{0.49\textwidth}
		\centering
		\DeclareGraphicsExtensions{.pdf}
		\includegraphics[]{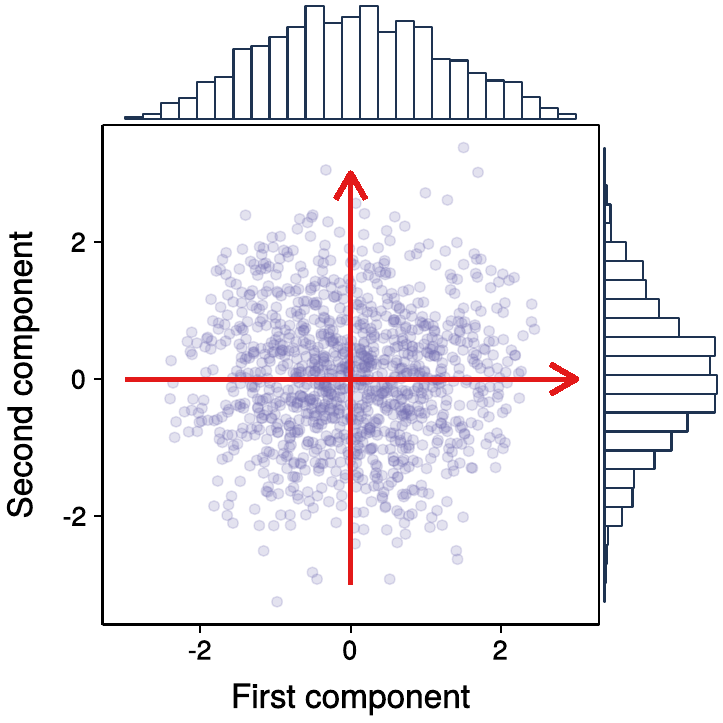}
		\caption{Rotated and whitened data.}
		\label{fig:2dPCAb2}
	\end{subfigure}
	\caption{Plot (a) shows the principal components and plot (b) shows the whitened principal components. The whitened components are uncorrelated and have unit variance. }
	\label{fig:2dPCA2}
\end{figure}

Essentially, whitening rescales the principal components so that they have unit variance. This process is also called sphering~\citep{Optimal_Whitening}. 
In other words, whitening scales the $i${th} principal component by the corresponding eigenvalue $1/\sqrt{\lambda_i}$ as
\begin{equation*}
\mathbf{z}_{\text{white}} := \dfrac{\mathbf{z}_i}{\sqrt{\lambda_i}}.
\end{equation*}
This is best illustrated by revisiting the above example shown in Figure~\ref{fig:2dPCA}.  In Figure~\ref{fig:2dPCA2} we show both the rotated and the whitened version of the data.

\subsubsection{Dimensionality reduction}

In practice, we often seek a useful low-dimensional representation to reveal the coherent structure of the data.
%
Choosing a ``good'' target-rank $k$, i.e., the number of PCs to retain, is a subtle issue and often domain specific. 
Little is gained by retaining too many components. Conversely,  a bad approximation is produced  if the number of retained components is too small.
Fortunately, the eigenvalues tell us the amount of variance captured by keeping only $k$ components given by
\begin{equation*}
\dfrac{\sum_{i=1}^k \lambda_i }{\sum_{i=1}^{n}\lambda_i}.
\end{equation*}
Thus, PCs corresponding to eigenvalues of small magnitude account only for a small amount of information in the data.

Many different heuristics, like the scree plot and Kaiser criterion, have been proposed to identify the optimal number of components~\citep{jolliffe2002principal}.
A computational intensive approach to determine the optimal number of components is via cross-validation approximations~\citep{josse2012selecting}, while a mathematically refined approach is the optimal hard threshold method for singular values, formulated by \cite{gavish2014optimal}. An interesting Bayesian approach to estimate the intrinsic dimensionality of
a high-dimensional dataset was recently proposed by~\cite{bouveyron2017exact}.

\subsection{Randomized algorithm}\label{sec:randomizedPCAalg}

The singular value decomposition provides a computationally efficient and numerically stable approach for computing the principal components. Specifically, the eigenvalue decomposition 
of the inner and outer dot product of $\mathbf{X}=\mathbf{U \Sigma V}^\top$ can be related to the 
SVD as
\begin{subequations}
	\begin{eqnarray}
	\mathbf{X}^\top \mathbf{X} = (\mathbf{V \Sigma U }^\top)(\mathbf{U \Sigma V }^\top)  =  \mathbf{V} \mathbf{\Sigma}^2 \mathbf{V}^\top,  \\
	\mathbf{X}\mathbf{X}^\top = (\mathbf{U \Sigma V}^\top)(\mathbf{V \Sigma U}^\top) = \mathbf{U} \mathbf{\Sigma}^2 \mathbf{U}^\top.
	\end{eqnarray} 
\end{subequations}
It follows that the eigenvalues are equal to the squared singular values. Thus, we recover the eigenvalues of the sample covariance matrix $\mathbf{C} := (m-1)^{-1} \mathbf{X}^\top \mathbf{X}$ as $\mathbf{\Lambda} = (m-1)^{-1} \mathbf{\Sigma}^2$. 

The left singular vectors $\mathbf{U}$ correspond to the eigenvectors of the outer product $\mathbf{X}\mathbf{X}^\top$, and the right singular vectors $\mathbf{V}$ correspond to the eigenvectors of the inner product $\mathbf{X}^\top\mathbf{X}$. 
This allows us to define the projection (rotation) matrix $\mathbf{W} := \mathbf{V}$.

Having established the connection between the singular value decomposition and principal component analysis, it is straight forward to show that the principal components can be computed as
\begin{equation*}
\mathbf{Z} := \mathbf{XW} = \mathbf{U \Sigma V}^\top \mathbf{W} = \mathbf{U \Sigma}. 
\end{equation*}

The randomized singular value decomposition can then be used to efficiently approximate the dominant $k$ principal components. 
This approach is denoted as randomized principal component analysis,
first introduced by~\cite{rokhlin2009randomized}, and later by
~\cite{halko2011algorithm}. \cite{szlam2014implementation} provide
some additional interesting implementation details for large-scale
applications.
Algorithm~\ref{alg:RPCAalgorithm} presents an implementation of the randomized PCA. The approximation accuracy can be controlled via oversampling and additional power iterations as described in Section~\ref{sec:impRA}.

\begin{algorithm}[t!]
  \centering
	\scalebox{0.9}{\fbox{		
			\begin{minipage}{210mm}
				\begin{tabbing}
					\hspace{2mm} \= \hspace{5mm} \= \hspace{2mm} \= \hspace{50mm} \=\kill
					\textbf{Input:} Centered/scaled input matrix $\mathbf{X}$ with dimensions $m\times n$, and target rank $k<\text{min}\{m,n\}$.\\[1mm]
					\textbf{Optional:} Parameters $p$ and $q$ to control oversampling, and the power scheme.\\[3mm] 
					\textbf{function} $\texttt{rpca}(\mathbf{X}, k, p, q)$\\[3mm]
				
					(1)  \> \> $[\mathbf{U},\mathbf{\Sigma},\mathbf{W}] = \texttt{rsvd}(\mathbf{X}, k, p, q)$ \> \> {\color{blue}\textrm{randomized SVD (Algorithm~\ref{alg:RSVDalgorithm})}} \\[1mm]
					
					(2)  \> \> $\mathbf{\Lambda} = \mathbf{\Sigma}^2 / (m-1)$  \> \> {\color{blue}\textrm{recover eigenvalues}} \\[1mm]

					(3)  \> \> $\mathbf{Z} = \mathbf{U}\mathbf{\Sigma}$ \> \> {\color{blue}{\textrm{optional: compute $k$ principal components}}} \\[3mm]

					\textbf{Return:} $\mathbf{W}\in \mathbb{R}^{n\times k}$, $\mathbf{\Lambda}\in \mathbb{R}^{k\times k}$ and $\mathbf{Z}\in \mathbb{R}^{m\times k}$
				\end{tabbing}
			\end{minipage}}}
                    \vspace{+.15in}
                    \caption{A randomized PCA algorithm.}
                    \label{alg:RPCAalgorithm}
\end{algorithm}

\subsection[Existing functionality for PCA in R]{Existing functionality for PCA in \proglang{R}}

The \code{prcomp()} and \code{princomp()} functions are the default options for performing PCA in \proglang{R}. The \code{prcomp()} routine uses the singular value decomposition and the \code{princomp()} function uses the eigenvalue decomposition to compute the principal components~\citep{venables2013modern}. 
%
Other options in \proglang{R} are the PCA routines of the \pkg{ade4}~\citep{ade4} and \pkg{FactoMineR}~\citep{FactoMineR} packages, which provide extended plot and summary capabilities. 
All these routines, however, are  based on computationally demanding algorithms. 

In many applications, only the dominant principal components are required. In this case, partial algorithms are an efficient alternative to constructing low-rank approximations, as discussed in Section~\ref{sec:rSVD}. 
For instance, the \pkg{irlba} package provides a computationally efficient routine for computing the dominant principal components using the implicitly restarted Lanczos method~\citep{irlba}. 

Another class of methods are incremental PCA algorithms, also denoted
as online PCA. These techniques are interesting if the data matrix is
not entirely available to start with, i.e., the algorithms allow one
to update the decomposition with each new arriving observation in
time. \citet{cardot2015online} give an overview of online PCA
algorithms and the corresponding routines are provided via
the~\pkg{onlinePCA} package~\citep{onlinePCA}. Similarly, the
\pkg{idm} package provides an incremental PCA algorithm \citep{idm}.
	
\subsection[The rpca() function]{The \code{rpca()} function}

The \code{rpca()} function provides an efficient routine for computing the dominant principal components using Algorithm~\ref{alg:RPCAalgorithm}. This routine is in particular relevant if the information in large-scale data matrices can be approximated by the first few principal components.

The interface of the \code{rpca()} function is similar to the \code{prcomp()} function:
\begin{Code}
rpca(A, k, center = TRUE, scale = TRUE, retx = TRUE, p = 10, q = 2)
\end{Code}
The first mandatory argument \code{A} passes the $m\times n$ input data matrix.
Note, that the analysis can be affected if the variables have different units of measurement. In this case, scaling is required to ensure a meaningful interpretation of the components. The \code{rpca()} function  centers and scales the input matrix by default, i.e., the analysis is based on the implicit correlation matrix.
However, if all of the variables have same units of measurement, there is the option to work with either the covariance or correlation matrix. In this case, the choice largely depends on the data and the aim of the analysis. 
The default options can be changed via the arguments \code{center} and \code{scale}. 
The second mandatory argument \code{k} sets the target rank, and it is assumed that \code{k} is smaller than the ambient dimensions of the input matrix.
%
%
The principal components are returned by default; otherwise the argument \code{retx} can be set to \code{FALSE} to not return the PCs. 
The parameters \code{p} and \code{q} are described in Section~\ref{sec:rsvdfunction}.

The resulting model object is a list and contains the following components:
\begin{compactitem}  
	\item \code{rotation}: $n\times k$ matrix containing the eigenvectors. 
	\item \code{eigvals}: $k$-dimensional vector containing the eigenvalues.
	\item \code{sdev}: $k$-dimensional vector containing the standard deviations of the principal components, i.e., the square root of the eigenvalues. 
	\item \code{x}: $m\times k$  matrix containing the principal components (rotated variables).
	\item \code{center, scale}: the numeric centering and scalings used (if any). 
\end{compactitem}

\subsubsection{Utility functions}
The \code{rpca()} routine comes with methods that can be used to
summarize and display the model information. These are similar to the
\code{prcomp()} function.
The \code{summary()} function provides information about the explained variance, standard deviations, proportion of variance as well as the cumulative proportion of the computed principal components. The \code{print()} function can be used to print the eigenvectors (principal directions).
The \code{plot()} function can be used to visualize the results, using the \pkg{ggplot2} package \citep{ggplot2}. 

\subsection{PCA example: Handwritten digits} \label{sec:eigenfaces}

\begin{figure}[t!]
	\centering
	\DeclareGraphicsExtensions{.pdf}
	\includegraphics[width=0.75\textwidth]{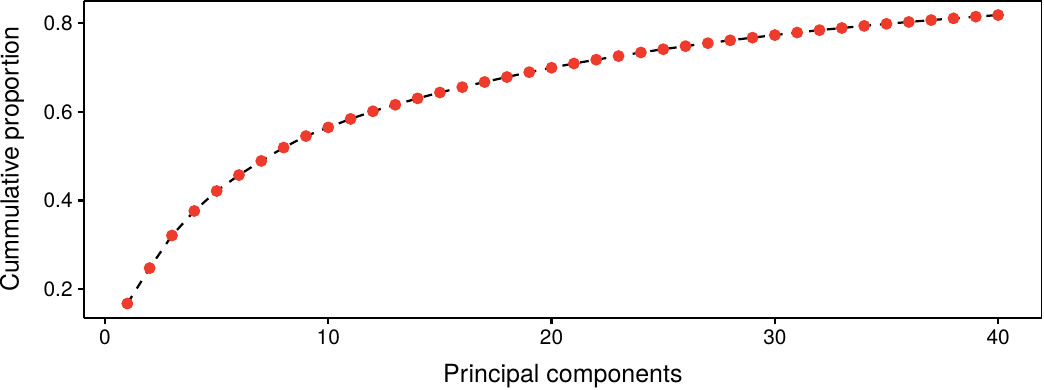}
	\caption{Cumulative proportion of the variance explained by the principal components. The first $40$ PCs explain about $82\%$ of the total variation in the data. }
	\label{fig:eigenPlots}
\end{figure}

\begin{figure}[t!]
	\centering
	\begin{subfigure}[t]{0.3\textwidth}
		\centering
		\DeclareGraphicsExtensions{.pdf}
		\includegraphics[width=1\textwidth]{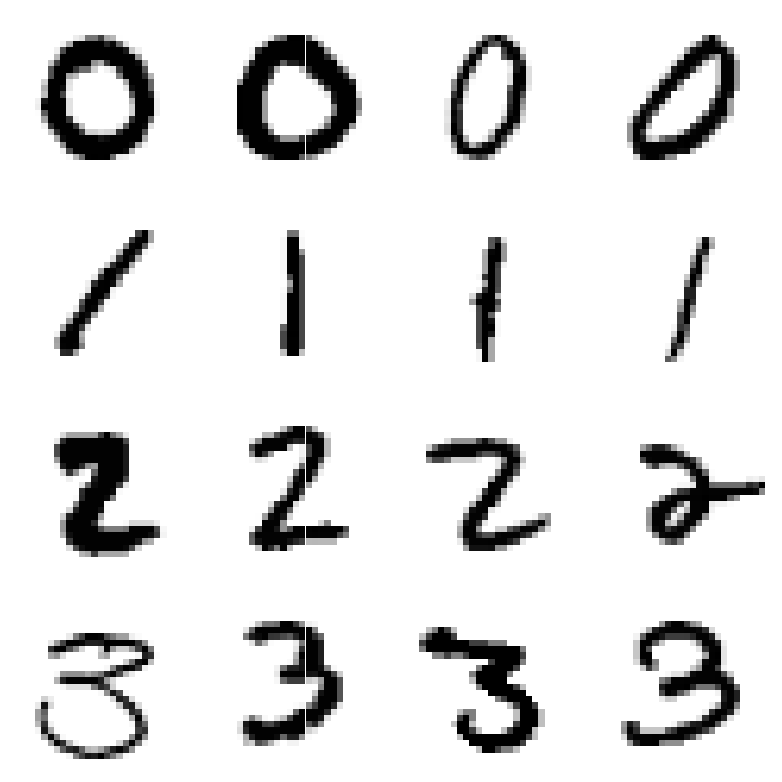}
		\caption{Handwritten digits. }
		\label{fig:digits_samples}
	\end{subfigure}
	~
	\begin{subfigure}[t]{0.3\textwidth}
		\centering
		\DeclareGraphicsExtensions{.pdf}
		\includegraphics[width=1\textwidth]{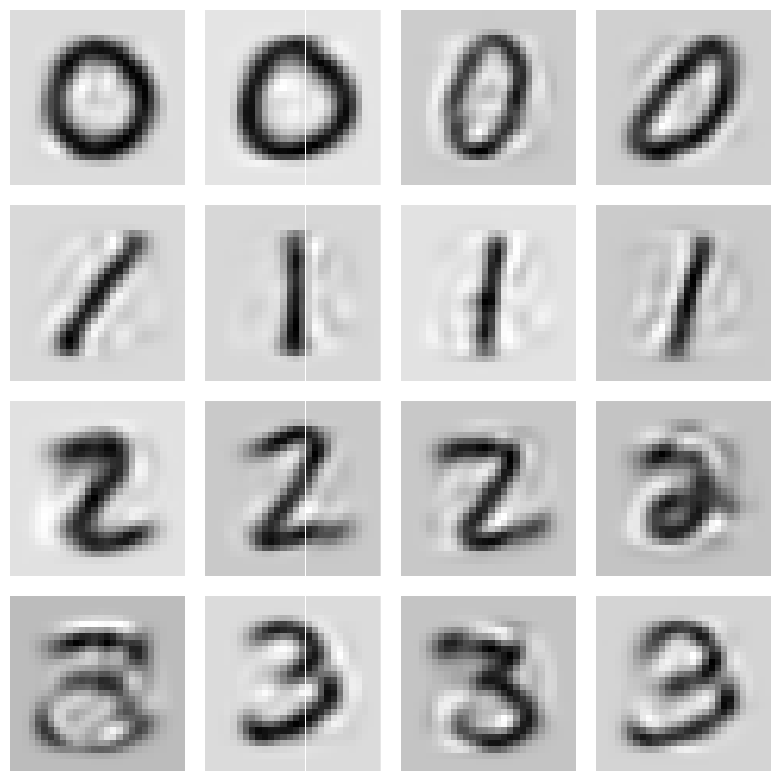}
		\caption{Deterministic.}
		\label{fig:digits_pca}
	\end{subfigure}	
	~
	\begin{subfigure}[t]{0.3\textwidth}
		\centering
		\DeclareGraphicsExtensions{.pdf}
		\includegraphics[width=1\textwidth]{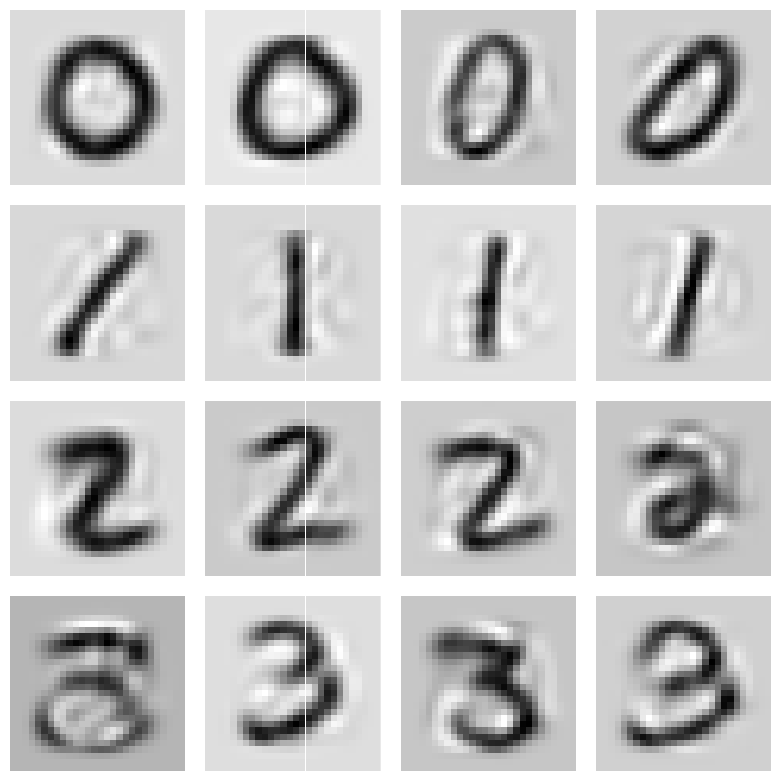}
		\caption{Randomized.}
		\label{fig:digits_rpca}
	\end{subfigure}

	\caption{ Handwritten digits, and its low-rank approximations using $k=40$ components.}
	\label{fig:digits}
\end{figure}

Handwritten digit recognition is a widely studied problem~\citep{MNIST}. 
In the following, we use a downsampled version of the MNIST (Modified National Institute of Standards and Technology) database of handwritten digits. The data are obtained from \url{http://yann.lecun.com/exdb/mnist} 
and can be loaded from the command line:
\begin{CodeChunk}
\begin{CodeInput}
R> data("digits", package = "rsvd")
R> label <- as.factor(digits[, 1])
R> digits <- digits[, 2:785]
\end{CodeInput}
\end{CodeChunk}
The data matrix is of dimension $12000\times 785$. Each row corresponds to a digit between $0$ and $3$. The first column is comprised of the class labels, while the following $784$ columns record the pixel intensities for the flattened $28\times 28$ image patches. Figure~\ref{fig:digits_samples} shows some of the digits.
In \proglang{R} the first digit can be displayed as:
\begin{CodeChunk}
\begin{CodeInput}
R> digit <- matrix(digits[1, ], nrow = 28, ncol = 28)
R> image(digit[, 28:1], col = gray(255:0 / 255))
\end{CodeInput}
\end{CodeChunk}
The aim of principal component analysis is to find a low-dimension representation which captures most of the variation in the data. PCA helps to understand the sources of variability in the data as well as to understand correlations between variables. The principal components can be used for visualization, or as features to train a classifier. A common choice is to retain the dominant $40$ principal components for classifying digits, using the $k$-nearest neighbor algorithm~\citep{MNIST}. 
Those can be efficiently approximated using the \code{rpca()} function:
\begin{CodeChunk}
\begin{CodeInput}
R> digits.rpca <- rpca(digits, k = 40, center = TRUE, scale = FALSE)
\end{CodeInput}
\end{CodeChunk}
The target rank is defined via the argument \code{k}. By default, the data are mean centered and standardized, i.e., the correlation matrix is implicitly computed. Here, we set \code{scale = FALSE}, since the variables have the same units of measurement, namely pixel intensities. 
The analysis can be summarized using the \code{summary()} function.
%
%
The screeplot function can be used to visualize the cumulative proportion of the variance captured by the principal components:%
\begin{CodeChunk}
\begin{CodeInput}
R> ggscreeplot(digits.rpca, type = "cum") 
\end{CodeInput}
\end{CodeChunk}
Figure~\ref{fig:eigenPlots} shows the corresponding plot.
Next, the PCs can be plotted in order to visualize the data in low-dimensional space:
\begin{CodeChunk}
\begin{CodeInput}
R> ggindplot(digits.rpca, groups = label, ellipse = TRUE, ind_labels = FALSE)
\end{CodeInput}
\end{CodeChunk}
\begin{figure}[t!]
	\centering
	\begin{subfigure}[t]{0.42\textwidth}
		\centering
		\DeclareGraphicsExtensions{.pdf}
		\includegraphics[width=1\textwidth]{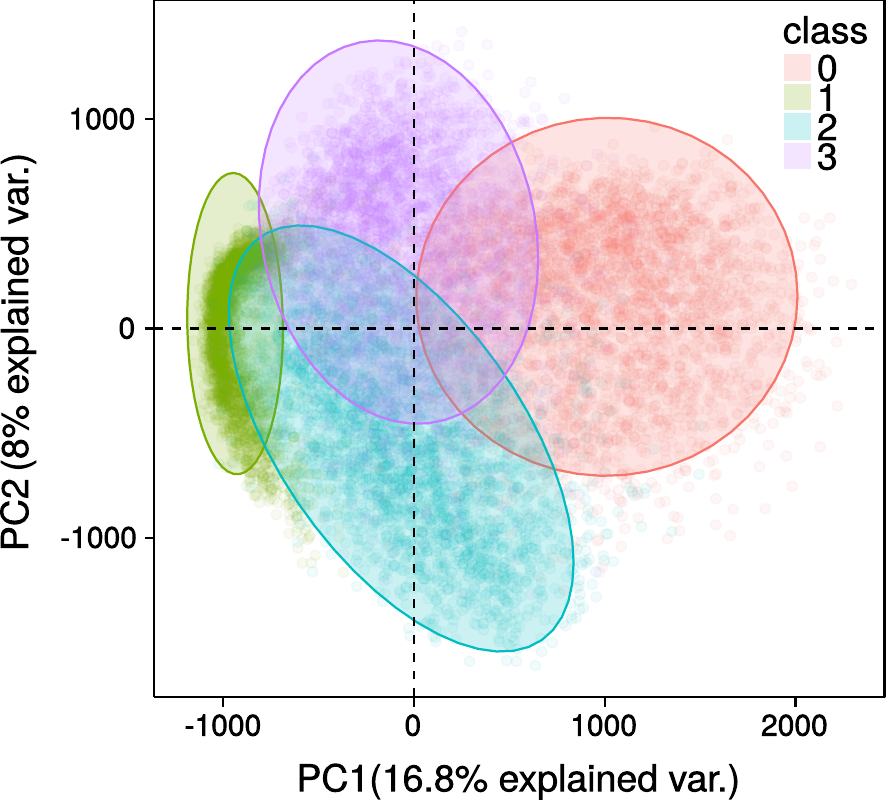}
		\caption{Individuals factor map. }
		\label{fig:ggindplot}
	\end{subfigure}
	~
	\begin{subfigure}[t]{0.42\textwidth}
		\centering
		\DeclareGraphicsExtensions{.pdf}
		\includegraphics[width=1\textwidth]{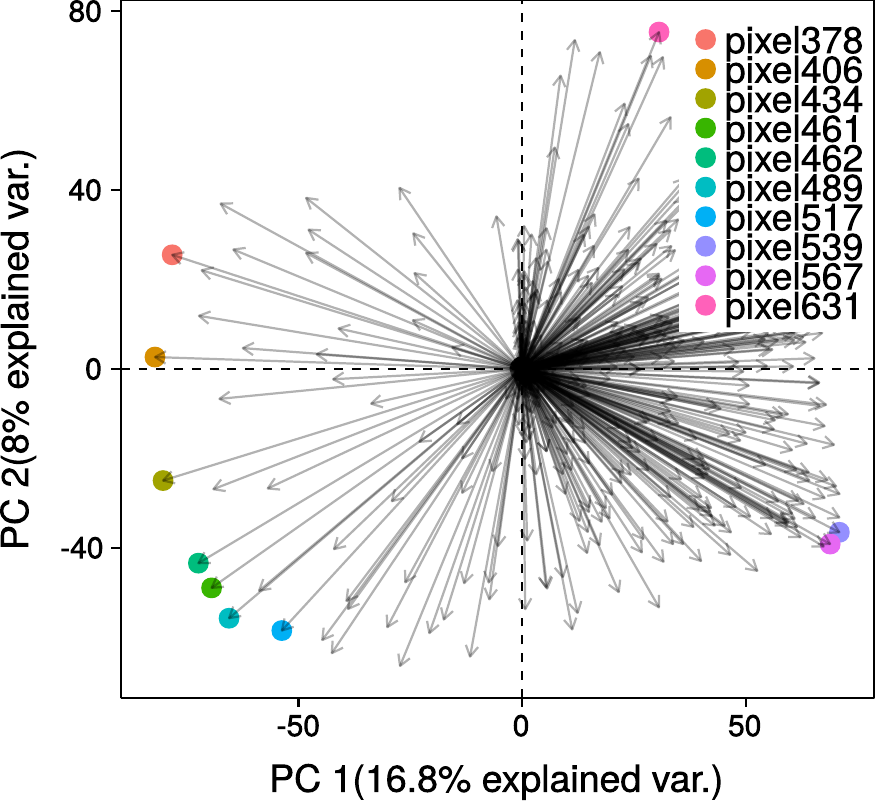}
		\caption{Variables factor map.}
		\label{fig:ggcorplot}
	\end{subfigure}		
	\caption{Plotting functionality to visualize the PCs: (a) shows the individuals factor map, overlaid with ellipses for each class; (b) shows the variables factor map.}
	\label{fig:ggplots}
\end{figure}

The so-called individual factor map, using the first and second principal component, is shown in Figure~\ref{fig:ggindplot}. The plot helps to reveal some interesting patterns in the data. For instance, $0$'s are distinct from $1$'s, while 3's share commonalities with all the other classes. 
Further, the correlation between the original variables and the PCs can be visualized:
\begin{CodeChunk}
\begin{CodeInput}
R> ggcorplot(digits.rpca, alpha = 0.3, top.n = 10)
\end{CodeInput}
\end{CodeChunk}
The correlation plot, also denoted as variables factor map, is shown in Figure~\ref{fig:ggcorplot}.  It shows the projected variables in eigenspace. This representation of the data gives some insights into the structural relationship (correlation) between the variables and the principal components. 

In order to quantify the quality of the dimensionality reduction, we can compute the relative error between the low-rank approximation and the original data. Recall, that the dominant $k$ principal component were defined as $\mathbf{Z}_k := \mathbf{XW}_k$. Hence, we can approximate the input matrix as $\mathbf{X} \approx \mathbf{Z}_k \mathbf{W}^\top_k$:
\begin{CodeChunk}
\begin{CodeInput}
R> digits.re <- digits.rpca$x %*% t(digits.rpca$rotation)  
\end{CodeInput}
\end{CodeChunk}
Since the procedure has centered the data, we need to add the mean pixel values back:
\begin{CodeChunk}
\begin{CodeInput}
R> digits.re <- sweep(digits.re, 2, digits.rpca$center, FUN = "+")
\end{CodeInput}
\end{CodeChunk}
The relative error can then be computed: 
\begin{CodeChunk}
\begin{CodeInput}
R> norm(digits - digits.re, "F") / norm(digits, "F")
\end{CodeInput}
\end{CodeChunk}
The relative error is approximately $32.8\%$. Figure~\ref{fig:digits_pca} and~\ref{fig:digits_rpca} show the samples of the reconstructed digits using both the \code{prcomp()} and \code{rpca()} function. By visual inspection, there is virtually no noticeable difference between the deterministic and the randomized approximation.

Runtimes and relative errors for different PCA functions in \proglang{R} are listed in Table~\ref{Tab:digits}.
The randomized algorithm is much faster than the \code{prcomp()} function, while attaining near-optimal results. 
Both the \code{dudi.pca()} and \code{PCA()} functions are slower than
the base \code{prcomp()} function. This is because we are using the
MKL (math kernel library) accelerated \proglang{R} distribution
{Microsoft \proglang{R} Open 3.4.1}. The timings can vary compared to using the
standard \proglang{R} distribution.
%

\begin{table}[t!]
	\centering
	\begin{tabular}{l l l c c c}
		\hline
		{\bf Package} & {\bf Function} & {\bf Parameters}  & {\bf Time (s)} & {\bf Speedup} & {\bf Error} \\ 
		\hline

          \pkg{base}    &  \code{prcomp()}    &  \code{rank. = 40} &	 \centering 0.56	& {\centering\arraybackslash * }  &  {\centering\arraybackslash 0.327 }  \\

		\pkg{FactoMineR}    & \code{PCA()}  & \code{ncp = 40}    & \centering 0.97	& 0.57  & {\centering\arraybackslash 0.327  }  \\

		\pkg{ade4}    & \code{dudi.pca()}   &  \code{nf = 40} 		   & \centering 0.91 & 0.61  &  {\centering\arraybackslash 0.327 } \\ 

		\pkg{irlba}    & \code{prcomp_irlba()}   & \code{n = 40}  & \centering {\centering\arraybackslash 0.47 }	&   {\centering\arraybackslash 1.2 }  & {\centering\arraybackslash {\centering\arraybackslash 0.327 } } \\			
		  				
		\pkg{rsvd}    & \code{rpca()}   & \code{k = 40}  & \centering 0.37 &  1.5  & {\centering\arraybackslash 0.328 } \\		
		\hline
	\end{tabular}
	\caption{Summary of the computational performance of different PCA functions.}
	\label{Tab:digits}
\end{table}

\subsubsection{Handwritten digit recognition}

The principal component scores can be used as features to efficiently
train a classifier. This is because PCA assumes that the interesting
information in the data are reflected by the dominant principal
components. This assumption is not always valid, i.e., in some
applications it can be the case that the variance corresponds to noise
rather than to the underlying signal. The question is, how good are
the randomized principal components suited for this task? In the
following, we use a simple $k$-nearest neighbor (kNN) algorithm to
classify handwritten digits in order to compare the performance. The
idea of kNN is to find the closest point (or set of points) to a given
target point~\citep{hastie2009elements}. There are two reasons to use
PCA for dimensionality reduction: (a) kNN is known to perform poorly
in high-dimensional space, due to the ``curse of
dimensionality''~\citep{donoho2000high}; (b) kNN is computational
expensive when high-dimensional data points are used for training.

\begin{figure}[t!]
	\centering
	\DeclareGraphicsExtensions{.pdf}
	\includegraphics[width=0.84\textwidth]{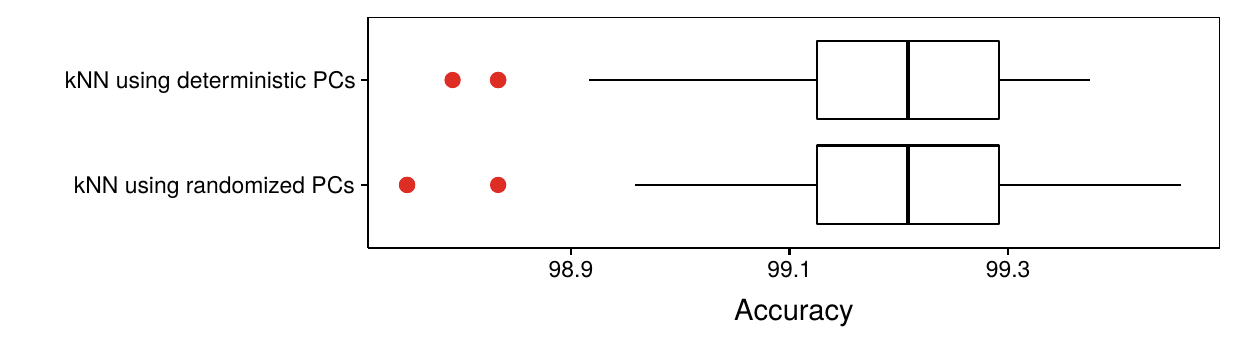}
	\caption{Performance of digits classification over $50$ random splits. There is no significant difference in terms of the accuracy between using the randomized and deterministic PCs. 
	}
	\label{fig:knn}
\end{figure}

First, we split the dataset into a training and a test set using the \pkg{caret} package~\citep{kuhn2008caret}. We aim to create a balanced split of the dataset, using about 80\% of the data for training:
\begin{CodeChunk}
\begin{CodeInput}
R> library("caret")
R> trainIndex <- createDataPartition(label, p = 0.8, list = FALSE)
\end{CodeInput}
\end{CodeChunk}
We then compute the dominant $k=40$ randomized principal components of the training set:
\begin{CodeChunk}
\begin{CodeInput}
R> train.rpca <- rpca(digits[trainIndex, ], k = 40, scale = FALSE)
\end{CodeInput}
\end{CodeChunk}
We can use the \code{predict()} function to rotate the test set into low-dimensional space:
\begin{CodeChunk}
\begin{CodeInput}
R> test.x <- predict(train.rpca, digits[-trainIndex, ]) 
\end{CodeInput}
\end{CodeChunk}
The base \pkg{class} package provides a kNN algorithm, which we use for classification:
\begin{CodeChunk}
\begin{CodeInput}
R> library("class")
R> knn.1 <- knn(train.rpca$x, test.x, label[trainIndex], k = 1)
\end{CodeInput}
\end{CodeChunk}
The test images are simply assigned to the class of the single nearest neighbor. The performance can be quantified by computing the accuracy, i.e., the number of correctly classified digits divided by the total number of predictions made:
\begin{CodeChunk}
\begin{CodeInput}
R> 100 * sum(label[-trainIndex] == knn.1) / length(label[-trainIndex]) 
\end{CodeInput}
\end{CodeChunk}
For comparison, the above steps can be repeated with the additional argument \code{rand = FALSE} or using the \code{prcomp()} function.  
Both the randomized PCA and the deterministic PCA algorithms achieve an accuracy of about $99.18\%$. 
%
%
%
%
Figure~\ref{fig:knn} shows the performance.

\section{Randomized robust principal component analysis}\label{sec:rrpca}

Thus far, we have viewed matrix approximations as a factorization of a given matrix into a product of smaller (low-rank) matrices, as formulated in Equation~\ref{eq:matrixfac}. However, there is another interesting class of matrix decompositions, which aim to separate a given matrix into low-rank, sparse and noise components.  
Such decompositions are motivated by the need for robust methods which can more effectively account for corrupt or missing data.  Indeed, outlier rejection is critical in many applications as data is rarely free of corrupt elements.  Robustification methods decompose data matrices as follows: 
\begin{equation*}
\begin{array}{cccccccc}
\mathbf{A} & \approx & \mathbf{L} & + & \mathbf{S} & + &  \mathbf{E}, \\
m\times n  &         &  m\times n &   & m\times n  &   & m\times n
\end{array} 
\end{equation*}
where ${\mathbf{L} \in \mathbb{R}^{m \times n}}$ denotes the low-rank matrix, ${\mathbf{S} \in \mathbb{R}^{m \times n}}$ the sparse matrix, and ${\mathbf{E} \in \mathbb{R}^{m \times n}}$ the noise matrix. 
Note that the sparse matrix $\mathbf{S}$ represents the corrupted entries (outliers) of the data matrix $\mathbf{A}$. 
In the following we consider only the special case $\mathbf{A}=\mathbf{L} + \mathbf{S}$, i.e., the decomposition of a matrix into its
sparse and low-rank components. This form of additive decomposition is also denoted as robust principal component analysis (RPCA), and its remarkable ability to separate high-dimensional matrices into low-rank and sparse component makes RPCA an invaluable tool for data science.
The additive decomposition is, however, different from classical robust PCA methods known in the statistical literature. These techniques are concerned with computing robust estimators for the empirical covariance matrix, for instance, see the seminal work by~\cite{hubert2005robpca} and~\cite{croux2005high}.

While traditional principal component analysis minimizes the spectral norm of the reconstruction error, robust PCA aims to recover the underlying low-rank matrix of a heavily corrupted input matrix.
\cite{candes2011robust} proved that it is possible to exactly separate such a data matrix $\mathbf{A} \in \mathbb{R}^{m \times n}$ into both its low-rank and sparse components, under rather broad assumptions. 
This is achieved by solving a convenient convex optimization problem called {\em principal component pursuit} (PCP). 
The objective is to minimize a weighted combination of the nuclear norm $\|\cdot \|_* := \sum_{i}\sigma_{i}$ and the $\ell_{1}$ norm $\|\cdot \|_1 := \sum_{ij}|m_{ij}|$ as 
\begin{equation*}
\min_{{\bf L}, {\bf S}}\|{\bf L}\|_* + \lambda \|{\bf S}\|_1 \,\,\text{ subject to }{\bf A}-{\bf L}-{\bf S}=0,
\end{equation*}
where $\lambda$ is an arbitrary balance parameter which puts some weight on the sparse error term in the cost function. Typically $\lambda$ is chosen to be $\lambda =\text{max}\{n,m\}^{-0.5}$. The PCP concept is mathematically sound, and has been applied successfully to many applications like video surveillance and face recognition~\citep{wright2009robust}.  Robust PCA is particular relevant if the underlying model of the data naturally features a low-rank subspace that is polluted with sparse components.  The concept of matrix recovery can also be extended to the important problem of matrix completion.

The biggest challenge for robust PCA is computational efficiency,
especially given the iterative nature of the optimization required.
\cite{bouwmans2015decomposition} have identified more than $30$
related algorithms to the original PCP approach, aiming to overcome
the computational complexity, and to generalize the original
algorithm.

\subsection{The inexact augmented Lagrange multiplier method}

A popular choice to compute RPCA, due to its favorable computational properties, is the inexact augmented Lagrange multiplier (IALM) method~\citep{ExactALM}. This method formulates the following Lagrangian function 
\begin{equation}\label{eq:IALM}
\mathcal{L}({\bf L},{\bf S},{\bf Z},{\bf \mu})=\|{\bf L}\|_* + \lambda  \|{\bf S}\|_1 + \langle {\bf Z}, {\bf A}-{\bf L}-{\bf S} \rangle  + \frac{\mu}{2} \|{\bf A}-{\bf L}-{\bf S} \|_F^2, 
\end{equation}
where ${\bf \mu}$ and ${\bf \lambda}$ are positive scalars, and {\bf Z} the Lagrange multiplier.
Further, $\langle \cdot, \cdot \rangle$ is defined as $\langle {\bf A}, {\bf B} \rangle := \textrm{trace}({\bf A}^\top{\bf B})$.
The method of augmented Lagrange multipliers can be used to solve the optimization problem~\citep{bertsekas1999nonlinear}. \cite{ExactALM} have proposed both an exact and inexact algorithm to solve Equation~\ref{eq:IALM}. Here, we advocate the latter approach. 
Specifically, the inexact algorithm avoids solving the problem
\begin{equation*}
{\bf L}_{i+1}, {\bf E}_{i+1} = \argmin_{{\bf L},{\bf E}} \mathcal{L}({\bf L}, {\bf E}, {\bf Z}_i, {\mu_k}),
\end{equation*}
by alternately solving the following two sub-problems at step $i$:
\begin{equation*}
{\bf L}_{i+1} = \argmin_{{\bf L}} \mathcal{L}({\bf L}, {\bf E}_i, {\bf Z}_i, {\mu_k}),
\end{equation*}
and
\begin{equation*}
{\bf E}_{i+1} = \argmin_{{\bf E}} \mathcal{L}({\bf L}_{i+1}, {\bf E}, {\bf Z}_i, {\mu_k}).
\end{equation*}
For details, we refer the reader to \cite{ExactALM}.  

\subsection{Randomized algorithm}

The singular value decomposition is the workhorse algorithm behind the IALM method. Thus, the computational costs can become intractable for ``big'' datasets.
However, randomized SVD can be used to substantially ease the computational burden of the IALM method. Algorithm~\ref{alg:rpca} outlines the randomized implementation.

\begin{algorithm}[t!]
  {\centering
    \scalebox{0.93}{\fbox{		
			\begin{minipage}{210mm}
				\begin{tabbing}
					\hspace{2mm} \= \hspace{5mm} \= \hspace{2mm} \= \hspace{75mm} \=\kill
					\textbf{Input:} Input matrix $\mathbf{A}$ with dimensions $m\times n$, and $\lambda$ to put weight on the sparse error term.\\[1mm]
					\textbf{Optional:} Parameters $p$ and $q$ to control oversampling, and the power scheme.\\[3mm] 
					\textbf{function} $\texttt{rrpca}(\mathbf{A}, \lambda, p, q)$\\[3mm]

					(1)  \> \> $k = 2$ \> \> {\color{blue}$\textrm{initialize target rank}$} \\[1mm]	
		
					(2)  \> \> $\mu = 1.25 \cdot \|\mathbf{A}\|_2$ \> \> {\color{blue}$\textrm{initialize }  \mu$} \\[1mm]	
			
					(3)  \> \> $\mathbf{Z} = \mathbf{A} \cdot \texttt{dual\_norm}(\mathbf{A})^{-1} $ \> \> {\color{blue}$\textrm{initialize Lagrange multiplier}$} \\[1mm]					

					(4)  \> \> $\mathbf{S} = \texttt{matrix}(0, m,n) $ \> \> {\color{blue}$\textrm{initialize sparse matrix}$} \\[2mm]						
					(5)  \> \> \textbf{repeat}  \> \> {\color{blue}} \\[1mm]

					(6)  \> \> \> $\left[\mathbf{U},\mathbf{\Sigma},\mathbf{V}\right] = \texttt{rsvd}(\mathbf{A}-\mathbf{S} + \mathbf{Z} \cdot \mu^{-1}, k, p , q)$ \> {\color{blue} randomized SVD using Algorithm~\ref{alg:RSVDalgorithm}}\\[1mm]

					(7)  \> \> \> $k, l = \texttt{predict\_rank}(\mathbf{\Sigma}, \mu^{-1})$ \> {\color{blue} predicted rank, and updated target rank}\\[1mm]

					(8)  \> \> \> $\mathbf{\Sigma}_l = \texttt{soft\_thres}( \texttt{diag}(\mathbf{\Sigma})(1:l), \mu^{-1})$ \> {\color{blue} soft threshold top $l$ singular values}\\[1mm]

					(9)  \> \> \> $\mathbf{L} = \mathbf{U}(:,1:l)\mathbf{\Sigma}_l\mathbf{V}(:,1:l)^\top$ \> {\color{blue} update low-rank matrix}\\[1mm]						
					
					(10)  \> \> \> $\mathbf{S} = \texttt{soft\_thres}(\mathbf{A}-\mathbf{L}+\mathbf{Z} \cdot \mu^{-1}, \lambda \cdot \mu^{-1})$\> {\color{blue} update sparse matrix via soft thresholding}\\[1mm]

					(11)  \> \> \> $\mathbf{Z} = \mathbf{Z} + (\mathbf{A}-\mathbf{L} -\mathbf{S}) \cdot \mu$\> {\color{blue} update Lagrange multiplier}\\[1mm]					
					
					(12)  \> \> \> update $\mu$ \\[1mm]

					(13)  \> \> \textbf{until} some convergence criterion is reached\\[3mm]
					
					\textbf{Return:} $\mathbf{L}\in \mathbb{R}^{m\times n}$, and $\mathbf{S}\in \mathbb{R}^{m\times n}$
				\end{tabbing}
			\end{minipage}}}
			\vspace{+.15in}
			\caption{A randomized robust PCA algorithm.}
			\label{alg:rpca}}

		\begin{remark}
                  \cite{ExactALM} provide details on how to predict the rank in Step (6). 
		\end{remark}

		\begin{remark}
                  Our randomized RPCA algorithm automatically switches
                  from the randomized SVD to the deterministic SVD, if
                  the target rank is predicted to be
                  $k > \min\{m,n\}/4$.
		\end{remark}			
			
		\end{algorithm}

\subsection[Existing functionality for robust PCA in R]{Existing functionality for robust PCA in \proglang{R}}

Only few \proglang{R} packages provide robust PCA routines. For comparison, we consider the \pkg{rpca} package~\citep{robustpcapkg}. The provided RPCA function implements the algorithm described by~\cite{candes2011robust}. This algorithm is  highly accurate.  However, a large number of iterations is required for convergence. 

\subsection[The rrpca() function]{The \code{rrpca()} function}
The \code{rrpca()} function implements the inexact augmented Lagrange multiplier method.
The interface of the \code{rrpca()} function takes the form of:
\begin{Code}
rrpca(A, lambda = NULL, maxiter = 50, tol = 1.0e-5, p = 10, q = 2, 
  trace = FALSE, rand = TRUE)
\end{Code}
The first mandatory argument \code{A} passes the $m\times n$ input data matrix.
The second argument \code{lambda} is used to put some weight on the sparse error term in the cost function. By default $\lambda$ is set to $\lambda =\text{max}\{n,m\}^{-0.5}$. 
The next two parameters \code{maxiter}, and \code{tol} are used to control the stopping criterion of the algorithm. The routine stops either if a specified maximal number of iterations, or if a certain tolerance level is reached. 
The parameters \code{p} and \code{q} are described in Section~\ref{sec:framework}.
The argument \code{rand} can be used to switch between the deterministic and randomized algorithms. By default the randomized algorithm is selected (i.e., the randomized SVD is used). Setting this argument \code{rand = FALSE} selects the deterministic algorithm (i.e., the deterministic SVD is used). 
To print out progress information, the argument \code{trace} can be set \code{TRUE}.

The resulting model object is a list and contains the following components:
\begin{compactitem}  
	\item \code{L}: $m\times n$ matrix containing the low-rank component. 
	\item \code{S}: $m\times n$ matrix containing the sparse component. 	
\end{compactitem}

\subsection{Robust PCA example: Grossly corrupted handwritten digits}

\begin{figure}[t!]
	\centering
	\begin{subfigure}[t]{0.23\textwidth}
		\centering
		\DeclareGraphicsExtensions{.pdf}
		\includegraphics[width=1\textwidth]{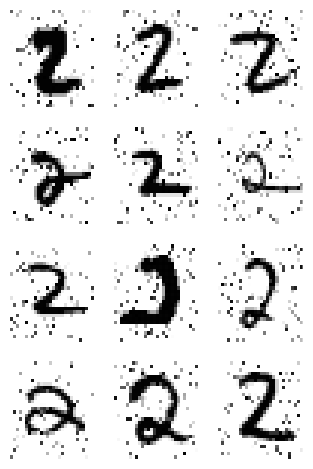}
		\caption{Noisy digits. }
		\label{fig:digits_noisy_samples}
	\end{subfigure}
	~
	\begin{subfigure}[t]{0.23\textwidth}
		\centering
		\DeclareGraphicsExtensions{.pdf}
		\includegraphics[width=1\textwidth]{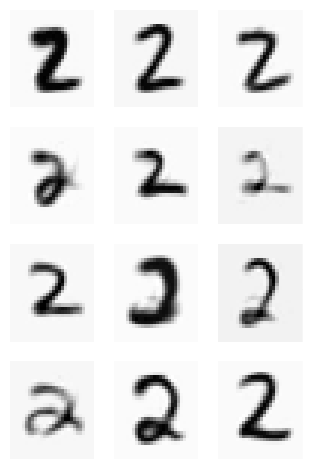}
		\caption{Deterministic $\mathbf{L}$.}
		\label{fig:digits_noisy_rrpca_det}
	\end{subfigure}		
	~
	\begin{subfigure}[t]{0.23\textwidth}
		\centering
		\DeclareGraphicsExtensions{.pdf}
		\includegraphics[width=1\textwidth]{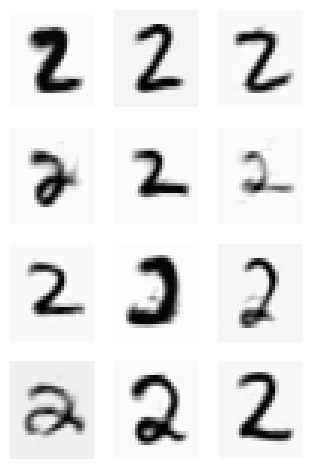}
		\caption{Randomized $\mathbf{L}$.}
		\label{fig:digits_noisy_rrpca}
	\end{subfigure}
	~
	\begin{subfigure}[t]{0.23\textwidth}
		\centering
		\DeclareGraphicsExtensions{.pdf}
		\includegraphics[width=1\textwidth]{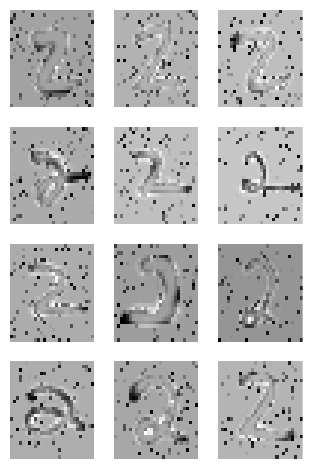}
		\caption{Randomized $\mathbf{S}$.}
		\label{fig:digits_noisy_rrpca_S}
	\end{subfigure}

	\caption{Separation of noisy handwritten digits into a low-rank and a sparse component.}
	\label{fig:digits_noisy}
\end{figure}

To demonstrate the randomized robust PCA algorithm, we consider a grossly corrupted subset of the handwritten digits dataset. We first extract a subset comprising only twos:
\begin{CodeChunk}
\begin{CodeInput}
R> data("digits", package = "rsvd")
R> two <- subset(digits[, 2:785], digits[, 1] == 2)
\end{CodeInput}
\end{CodeChunk}
Then, we corrupt the data using salt and pepper noise, i.e., we draw
i.i.d.~uniform entries in the interval $[0,255]$ and sparsify the
matrix so that about 10\% nonzero elements are retained:
\begin{CodeChunk}
\begin{CodeInput}
R> m <- nrow(two); n <- ncol(two)
R> S <- matrix(runif(m * n, 0, 255), nrow = m, ncol = n)
R> S <- S * matrix(rbinom(m*n, size = 1, prob = 0.1), nrow = m, ncol = n) 
\end{CodeInput}
\end{CodeChunk}
The digits are then corrupted as follows: 
\begin{CodeChunk}
\begin{CodeInput}
R> two_noisy <- two + S
R> two_noisy <- ifelse(two_noisy > 255, 255, two_noisy)
\end{CodeInput}
\end{CodeChunk}
Note, the last line ensures that the pixel intensities remain in the interval $[0,255]$. Samples of the corrupted digits are shown in Figure~\ref{fig:digits_noisy_samples}. 
Robust PCA is now used for matrix recovery (denoising) by separating the data into a low-rank and sparse component:
\begin{CodeChunk}
\begin{CodeInput}
R> two.rrpca <- rrpca(two_noisy, trace = TRUE, rand = TRUE)
\end{CodeInput}
\end{CodeChunk}
Figure~\ref{fig:digits_noisy_rrpca} and~\ref{fig:digits_noisy_rrpca_S} shows samples of the low-rank component $\mathbf{L}$ and the sparse component $\mathbf{S}$.
For comparison, Figure~\ref{fig:digits_noisy_rrpca_det} shows samples
of the low-rank component which are computed using the deterministic
routine (\code{rand = FALSE}).

Table~\ref{Tab:digits_noisy} summarizes the computational results. 
The randomized routine does not show a computational advantage in this case. The reasons are twofold. First, the dataset requires a relatively large target rank to approximate the data accurately, i.e., in this example the predicted rank-$k$ of the IALM algorithm for the final iteration is $232$. Secondly, the data matrix is tall and thin (i.e., the ratio of rows to columns is large). For both of theses reasons, the performance of the deterministic algorithm remains competitive. The advantage of the randomized algorithms becomes pronounced for higher-dimensional problems which feature low-rank structure.   

\begin{figure}[t!]
	\centering
	\DeclareGraphicsExtensions{.pdf}
	\includegraphics[width=0.85\textwidth]{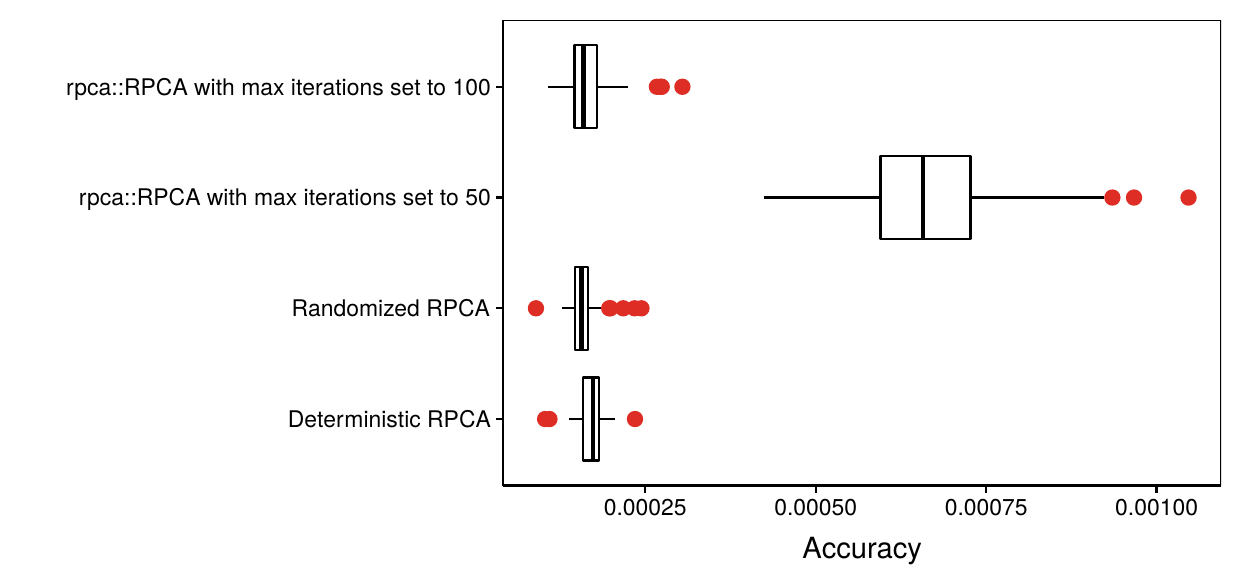}
	\caption{Matrix recovery performance of different RPCA algorithms.}
	\label{fig:rrpca_accurcy}
\end{figure}

It appears that the \code{rpca()} routine of the \pkg{rpca} package is
more accurate, while computationally less attractive (the algorithm
converges slowly). However, using the relative error for comparing the
algorithms might be misleading since we do not know the ground
truth. The relative error is computed using the original data as
baseline, which contains some perturbations itself. Clearly, the
advocated IALM algorithm removes not only the salt and paper noise,
but also shadows, specularities, and saturations from the digits. This
seems to be favorable, yet it leads to a larger relative error.
\begin{table}[t!]
	\centering
	\begin{tabular}{l l l c c c c}
		\hline
		{\bf Package} & {\bf Function} & {\bf Parameters}  & {\bf Time (s)} & {\bf Speedup} & {\bf Error} & {\bf Iterations} \\ 
		\hline
		
		\pkg{rpca}    & \code{rpca()}   & \code{max.iter = 50}  & \centering 11.88 &   *  & {\centering\arraybackslash 0.265 } & 50\\

		\pkg{rsvd}    & \code{rrpca()}   & \code{rand = FALSE}  & \centering 6.33 &   1.8  & {\centering\arraybackslash 0.328 } & 27 \\

		\pkg{rsvd}    & \code{rrpca()}   & \code{rand = TRUE}  & \centering 5.56 &   2.1  & {\centering\arraybackslash 0.329 }   & 27 \\		
		\hline
	\end{tabular}
	\caption{Summary of the computational performance of different RPCA functions.	\label{Tab:digits_noisy}}
\end{table}
Thus, to better compare the algorithms we perform a small simulation study using synthetic data. By superimposing a low-rank matrix with a sparse component, the ground truth is known. First, the low-rank component is generated:
\begin{CodeChunk}
\begin{CodeInput}
R> m <- 300; n <- 300; k <- 5
R> L1 <- matrix(rnorm(m * k), nrow = m, ncol = k) 
R> L2 <- matrix(rnorm(n * k), nrow = k, ncol = n) 
R> L <- L1 %*% L2
\end{CodeInput}
\end{CodeChunk}
The sparse component with about 20\% nonzero i.i.d.~uniform entries in the interval $[-500,500]$ is generated from:
\begin{CodeChunk}
\begin{CodeInput}
R> S <- matrix(runif(m * n, -500, 500), nrow = m, ncol = n)
R> S <- S * matrix(rbinom(m * n, size = 1, prob = 0.2), nrow = m, ncol = n) 
\end{CodeInput}
\end{CodeChunk}
The data matrix is then constructed by superimposing the low-rank and sparse components:
\begin{CodeChunk}
\begin{CodeInput}
R> A <- L + S
\end{CodeInput}
\end{CodeChunk}
Figure~\ref{fig:rrpca_accurcy} shows the performance of the RPCA algorithms over $50$ runs. Both the randomized and deterministic routines provided by the \pkg{rsvd} package show a better performance than the \pkg{rpca} package, when the maximum number of iterations is set to $50$. In addition, we show the performance after $100$ iterations for the RPCA algorithms from the \pkg{rpca} package. 

\section{Additional functionality}\label{sec:rid}

Principal components analysis seeks a set of new components which are formed as weighted linear combinations of the input data. 
This approach allows one to efficiently approximate and summarize the data, however, interpretation of the resulting components can be difficult.
For instance, in a high-dimensional data setting it is cumbersome to interpret the large number of weights (loadings) required to form the components.
While in many applications the eigenvectors have distinct meanings, the orthogonality constraints may not be physical meaningful in other problems.
Thus, it is plausible to look for alternative factorizations which may not provide an optimal rank-$k$ approximation, but which may preserve useful properties of the input matrix, such as sparsity and non-negativity as well as allowing for easier interpretation of its components.
Such properties may be found in the CUR and the interpolative decompositions (ID), which are both tools for computing low-rank approximations.

\subsection{Randomized CUR decomposition}

\begin{figure}[t!]
	\centering
	\DeclareGraphicsExtensions{.pdf}
	\includegraphics[width=0.55\textwidth]{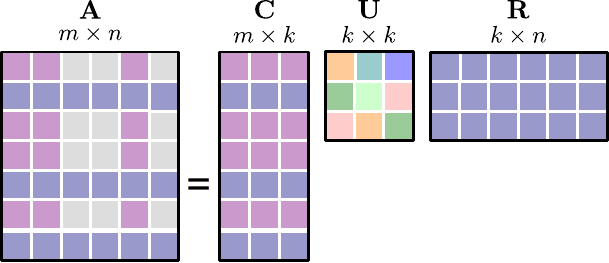}
	\caption{Schematic of the rank-$k$ CUR decomposition of an $m\times n$ matrix. The components  are formed by small subsets of actual columns and rows of the input matrix.}
	\label{Fig:CUR}
\end{figure}

\cite{mahoney2009cur} introduced the CUR matrix decomposition, as an interesting alternative to traditional approximation techniques such as SVD and PCA. The CUR decomposition admits a factorization of the form
\begin{equation*}
\begin{array}{ccccc}
\mathbf{A} & \approx & \mathbf{C} & \mathbf{U} & \mathbf{R}, \\
m\times n &   &  m\times k & k\times k & k\times n
\end{array} 
\end{equation*}
where the components of the matrix $\mathbf{C}\in \mathbb{R}^{m\times k}$ and $\mathbf{R}\in \mathbb{R}^{k\times n}$ are formed by small subsets of actual columns and rows, respectively. The matrix $\mathbf{U}\in \mathbb{R}^{k\times k}$ is formed so that $\|\mathbf{A}- \mathbf{C}\mathbf{U}\mathbf{R}\|_F$ is small. The CUR factorization is illustrated in Figure~\ref{Fig:CUR}.  

The low-rank factor matrices $\mathbf{C}$ and $\mathbf{R}$ are interpretable, since their components maintain the original structure of the data. This allows one to fully leverage any  field-specific knowledge about the data, i.e., experts have often a clear understanding about the actual meaning of certain columns and rows.
However, the CUR decomposition is not unique, and different computational strategies lead to different subsets of columns and rows, for instance, see~\cite{mahoney2009cur} and~\cite{boutsidis2017optimal}. Thus, the practical meaning of the selected rows and columns should always be carefully examined depending on the problem and its objective. 

Note, that the rank-$k$ SVD ($\mathbf{A}_k = \mathbf{U}_k \Sigma_k \mathbf{V}^{\top}_k$) of a general $m\times n$ matrix $\mathbf{A}$ yields an optimal approximation of rank $k$ to $\mathbf{A}$, in the sense that $\|\mathbf{A} - \mathbf{A}_k\| \leq \|\mathbf{A} - \mathbf{M}_k\|$ for any rank $k$ matrix $\mathbf{M}_k$, both in the operator (spectral) and Frobenius norms. However, if $\mathbf{A}$ is a sparse matrix, the $m \times k$ and $n \times k$ factors $\mathbf{U}_k$ and $\mathbf{V}_k$ are typically dense. 
Even though the low-rank SVD is optimal for a given rank $k$, the choice of rank may be limited to relatively low values with respect to $\min(m,n)$ for sparse matrices, in order to achieve any useful compression ratios. Of course, the usefulness of the SVD is not limited to compression; but the utility of a low-rank approximation is greatly reduced once the storage size of the factors exceeds that of the original matrix. The CUR decomposition provides an interesting alternative for compression, since its components preserve sparsity.

The \pkg{rCUR} package provides an implementation in \proglang{R}~\citep{bodor2012rcur}.
The \pkg{rsvd} package implements both the deterministic and randomized CUR decomposition, following the work by~\cite{voronin2017efficient}. Specifically, the interpolative decomposition is used as an algorithmic tool to form the factor matrices $\mathbf{C}$ and $\mathbf{R}$.
Algorithm~\ref{alg:rcur} outlines the computational steps of the \code{rcur()} routine as implemented in the \pkg{rsvd} package. 
		

\subsection[The rcur() function]{The \code{rcur()} function}
The \code{rcur()} function provides the option to compute both the deterministic and the randomized CUR decomposition via Algorithm~\ref{alg:rcur}. The interface of the \code{rcur()} function is as follows:
\begin{Code}
rcur(A, k, p = 10, q = 0, idx_only = FALSE, rand = TRUE)
\end{Code}
The first mandatory argument \code{A} passes the $m\times n$ input data matrix.
The second mandatory argument \code{k} sets the target rank, which is required to be $k < \text{min}\{m,n\}$.
The parameters \code{p} and \code{q} are described in Section~\ref{sec:framework}.
The argument \code{rand} can be used to switch between the deterministic and the randomized algorithm. The latter is used by default, and is more efficient for large-scale matrices. The argument \code{idx_only} can be set to \code{TRUE} in order to return only the column and row index sets which is more memory efficient than returning $\mathbf{C}$ and $\mathbf{R}$.

The resulting model object is a list containing the following components:
\begin{compactitem}  
			\item \code{C}: $m\times k$ matrix containing the column skeleton. 
			\item \code{R}: $k\times n$ matrix containing the row skeleton. 
			\item \code{U}: $k\times k$ matrix. 
			\item \code{C.idx}: $k$-dimensional vector containing the column index set.
			\item \code{R.idx}: $k$-dimensional vector containing the row index set.	
\end{compactitem}

\begin{algorithm}[t!]
	\scalebox{0.9}{\fbox{		
			\begin{minipage}{210mm}
				\begin{tabbing}
					\hspace{2mm} \= \hspace{5mm} \= \hspace{2mm} \= \hspace{50mm} \=\kill
					\textbf{Input:} Input matrix $\mathbf{A}$ with dimensions $m\times n$, and target rank $k<\text{min}\{m,n\}$.\\[1mm]
					\textbf{Optional:} Parameters $p$ and $q$ to control oversampling, and the power scheme.\\[3mm] 
					\> \textbf{function} $\texttt{rcur}(\mathbf{A}, k, p, q)$\\[3mm]
					
					(1)  \> \> $[\mathbf{C},\mathbf{Z},J] = \texttt{rid}(\mathbf{A}, k, p, q)$ \> \> {\color{blue}\textrm{randomized column ID (Algorithm~\ref{alg:rid})}} \\[1mm]
					
					(2)  \> \> $\left[\sim,\mathbf{S}, P\right] = \texttt{qr}(\mathbf{C}^\top)$ \> \> {\color{blue}\textrm{pivoted QR decomposition}}
					\\[1mm]	
					
					(3)  \> \> $I = P(1:k) $ \> \> {\color{blue}\textrm{extract top $k$ row indices}}	\\[1mm]	
					
					(4)  \> \> $\mathbf{R} = \mathbf{A}(I,:) $ \> \> {\color{blue}\textrm{extract $k$ rows from input matrix}}	\\[1mm]

					(5)  \> \> $\mathbf{R^\dagger} = \texttt{pinv}(\mathbf{R})$ \> \> {\color{blue}\textrm{compute pseudoinverse}}	\\[1mm]

					(6)  \> \> $\mathbf{U} = \mathbf{Z}\mathbf{R^\dagger}$ \> \> {\color{blue}{\textrm{compute well-conditioned matrix}}} \\[1mm]

					\textbf{Return:} $\mathbf{C}\in \mathbb{R}^{m\times k}$, $\mathbf{U}\in \mathbb{R}^{k\times k}$ and $\mathbf{R}\in \mathbb{R}^{k\times n}$
				\end{tabbing}
			\end{minipage}}}
			\centering
			\vspace{+.15in}
			\caption{A randomized CUR decomposition algorithm.}
			\label{alg:rcur}
\begin{remark}
The deterministic rank-$k$ CUR decomposition is computed by  replacing the \code{rid()} function with the deterministic \code{id()} function, described in Algorithm~\ref{alg:id}.
\end{remark}			
\end{algorithm}

\subsection{Randomized interpolative decomposition}
The interpolative decomposition yields a low-rank factorization of the form
\begin{equation*}
\begin{array}{ccccc}
\mathbf{A} & \approx & \mathbf{C} & \mathbf{Z}. \\
m\times n &   &  m\times k & k\times n
\end{array} 
\end{equation*}
The factor matrix $\mathbf{C} \in \mathbb{R}^{m\times k}$ is formed by a small number of columns, while $\mathbf{Z} \in \mathbb{R}^{k\times n}$ is a well-conditioned matrix containing the identity. $\mathbf{C}$ is also denoted as a skeleton matrix, and $\mathbf{Z}$ as the interpolation matrix. 

The question is how to choose ``interesting'' columns of $\mathbf{A}$ to form $\mathbf{C}$? 
%
For certain datasets (such as images), one may choose 
the $k$ columns corresponding to highest brightness/contrast, or highest amount of variation or detail. 
As an example, in a photograph of a building structure, the structure portion would be more critical than the ground or sky. 
Of course, a general method is needed to pick $k$ columns from a matrix. In linear algebra, such a method exists: {\em pivoting}. 
Following \cite{halko2011rand}, we advocate the QR factorization with pivoting
\begin{equation*}
\begin{array}{ccccccc}
\mathbf{A} & \mathbf{P} &=& \mathbf{Q} & \mathbf{S},\\
m\times n & n\times n && m\times r & r\times n
\end{array}
\end{equation*}
where $r:=\text{min}\{m,n\}$. $\mathbf{P}$ is the permutation matrix, which simply dictates the re-arrangement of the columns of $\mathbf{A}$. The matrix $\mathbf{Q}$ has orthonormal columns, and $\mathbf{S}$ is upper triangular.\footnote{Here, we denote the upper triangular matrix as $\mathbf{S}$, since $\mathbf{R}$ is occupied by the CUR decomposition.}
Because, the pivoted QR decomposition is an iterative algorithm, it can be stopped after $k$ iterations to obtain only the $k$ dominant pivots. 
Thus, the column subset used to form $\mathbf{C}$ is simply based on the pivoting strategy used in the QR factorization. 
The computational steps required to compute the ID are outlined in Algorithm~\ref{alg:id}.

The procedure can be considerably accelerated by means of randomization. Specifically, we can first compute the randomized QB decomposition via Algorithm~\ref{alg:rqb}. Then, the smaller matrix $\mathbf{B}$ is used to compute the ID decomposition. 
The computational steps are outlined in Algorithm~\ref{alg:rid}. 
For a detailed discussion, and theoretical results we refer to~\cite{voronin2015rsvdpack}, and~\cite{voronin2017efficient}. Therein, it is also described how the factor matrix $\mathbf{Z}$ can be efficiently constructed. Note, however, that our algorithm differs from the implementation by~\cite{voronin2015rsvdpack}. We compute the ID based on the matrix $\mathbf{B}$ (obtained as described in Section~\ref{sec:framework}). In our experiments, this approach shows to be more accurate, while slightly more computational demanding.

\begin{algorithm}[t!]
\scalebox{0.9}{\fbox{		
\begin{minipage}{210mm}
	\begin{tabbing}
					\hspace{2mm} \= \hspace{5mm} \= \hspace{2mm} \= \hspace{50mm} \=\kill
					\textbf{Input:} Input matrix $\mathbf{A}$ with dimensions $m\times n$, and target rank $k<\text{min}\{m,n\}$.\\[3mm]
					
					\textbf{function} $\texttt{id}(\mathbf{A}, k)$\\[3mm]

					(1)  \> \> $\left[\sim,\mathbf{S}, P\right] = \texttt{qr}(\mathbf{A})$ \> \> {\color{blue}\textrm{pivoted QR decomposition}}
					\\[1mm]		
					
					(2)  \> \> $\mathbf{S^\dagger} = \texttt{pinv}(\mathbf{S}(1:k,1:k))$ \> \> {\color{blue}\textrm{compute pseudoinverse}}	\\[1mm]									
					
					(3)  \> \> $\mathbf{T} = \mathbf{S^\dagger} \mathbf{S}(1:k,(k+1):n) $ \> \> {\color{blue}\textrm{compute expansions coefficients}}	\\[1mm]								
					
					(4)  \> \> $\mathbf{Z} = \texttt{matrix}(0,k,n)$ \> \> {\color{blue}\textrm{create empty $k\times n$ matrix}}	\\[1mm]		
					
					(5)  \> \> $\mathbf{Z}(:,P) = \texttt{cbind}(\texttt{diag(k)}, \mathbf{T}) $ \> \> {\color{blue}\textrm{ordered expansions coefficients, using pivots $P$}}	\\[1mm]	
					
					(6)  \> \> $J = P(1:k) $ \> \> {\color{blue}\textrm{extract top $k$ column indices from pivots}}	\\[1mm]	
					
					(7)  \> \> $\mathbf{C} = \mathbf{A}(:,J) $ \> \> {\color{blue}\textrm{extract $k$ columns from input matrix}}	\\[3mm]		
					
					
					\textbf{Return:} $\mathbf{C}\in \mathbb{R}^{m\times k}$, $\mathbf{Z}\in \mathbb{R}^{k\times n}$, and $J\in \mathbb{N}^{k}$
	\end{tabbing}
\end{minipage}}}
\centering
\vspace{+.15in}
\caption{An interpolative decomposition algorithm.}
\label{alg:id}	

\begin{remark}
	The QR decomposition returns the permutation matrix $\mathbf{P}$ in form of a vector $P\in \mathbb{R}^{n}$. This vector contains the indices such that $\mathbf{P}=\mathbf{I}(:,P)$, where $\mathbf{I}\in \mathbb{R}^{n \times n}$ denotes the identity matrix. Thus, $J$ is comprised of the $k$ dominant pivots.
\end{remark}
			
\end{algorithm}

\begin{algorithm}[t!]
\scalebox{0.9}{\fbox{		
\begin{minipage}{210mm}
\begin{tabbing}
\hspace{2mm} \= \hspace{5mm} \= \hspace{2mm} \= \hspace{50mm} \=\kill
\textbf{Input:} Input matrix $\mathbf{A}$ with dimensions $m\times n$, and target rank $k<\text{min}\{m,n\}$.\\[1mm]
							
\textbf{Optional:} Parameters $p$ and $q$ to control oversampling and the power scheme.\\[3mm] 

\> \textbf{function} $\texttt{rid}(\mathbf{A}, k, p, q)$\\[3mm]

(1)  \> \> $\left[\sim, \mathbf{B}\right] = \texttt{rqb}(\mathbf{A}, k, q, p)$ \> \> {\color{blue}\textrm{randomized QB decomposition via Algorithm~\ref{alg:rqb}}} \\[1mm]

(2)  \> \> $\left[\sim,\mathbf{Z}, J\right] = \texttt{id}(\mathbf{B}, k)$ \> \> {\color{blue}\textrm{column ID via Algorithm~\ref{alg:id}}}
							\\[1mm]		
							
(3)  \> \> $\mathbf{C} = \mathbf{A}(:,J) $ \> \> {\color{blue}\textrm{extract $k$ columns from input matrix}}	\\[3mm]

\textbf{Return:} $\mathbf{C}\in \mathbb{R}^{m\times k}$, $\mathbf{V}\in \mathbb{R}^{k\times n}$, and $J\in \mathbb{N}^{k}$
\end{tabbing}
\end{minipage}}}
\centering
\vspace{+.15in}
\caption{A randomized interpolative decomposition algorithm.}
\label{alg:rid}
						
\begin{remark}
	The row ID can be computed by transposing the input matrix, i.e., $\mathbf{A}^\top$.
\end{remark}

\end{algorithm}

\subsection[The rid() function]{The \code{rid()} function}
The \code{rid()} function provides the option to compute both the deterministic and the randomized interpolative decomposition via Algorithms~\ref{alg:id} and~\ref{alg:rid}. 
The interface of the \code{rid()} function takes the following functional form:
\begin{Code}
rid(A, k, mode = "col", p = 10, q = 0, idx_only = FALSE, rand = TRUE)
\end{Code}
The first mandatory argument \code{A} passes the $m\times n$ input data matrix.
The second mandatory argument \code{k} sets the desired target rank, which is required to be $k < \text{min}\{m,n\}$.
The argument \code{mode = c("col", "row")} determines whether the column or row ID should be computed.  
The parameters \code{p} and \code{q} are described in Section~\ref{sec:framework}.
The argument \code{rand} can be used to switch between the deterministic and randomized algorithms. By default the randomized algorithm is selected, and setting this argument \code{rand = FALSE} selects the deterministic algorithm. 

The resulting model object is a list and contains the following components:
\begin{compactitem}  
\item \code{C}: $m\times k$ matrix containing the column skeleton, if \code{mode = "col"}. 
\item \code{R}: $k\times n$ matrix containing the row skeleton, if \code{mode = "row"}. 
\item \code{Z}: $k\times n$ or $m\times k$ matrix (depending on \code{mode}), which is well-conditioned. 
\item \code{idx}: $k$-dimensional vector containing the column or row index set.
\end{compactitem}

\section{Conclusion}\label{sec:conclusion}
\begin{figure}[t!]
  \centering
    \scalebox{0.93}{
			
			\begin{tikzpicture}[auto,node distance = 2cm,>=latex']
			
			\coordinate (O) at (0,0);
			\coordinate (X) at (8,0);
			\coordinate (Y) at (0,4);

			\draw [-latex, black, line width=1.5pt]  (O) -- (X);
			\draw [-latex, black, line width=1.5pt]  (O) -- (Y);
			
			\draw [fill=black] (-0.0,-0.0) circle (0pt) node [left] {Low};
			\draw [fill=black] (0,4) circle (0pt) node [above] {High};
			\draw [fill=black] (8,0) circle (0pt) node [right] {High};
			
			\draw [<->, black, line width=2.5pt]  (-0.2,0.5) -- (-0.2,3.5);
			\draw [fill=black] (-0.2,2) circle (0pt) node [left] {\color{black} Accuracy};
			
			\draw [<->, black, line width=2.5pt]  (1,-0.2) -- (7,-0.2);			
			\draw [fill=black] (4,-0.2) circle (0pt) node [below] {\color{black} Scalability};
			
			\draw [fill=black] (6.0,4.0) circle (0pt) node [below] {\color{black} machine precision};
			
			
			\draw[dotted, darkgreen1, line width=1.0pt] (0.1,2.6) rectangle (8,4);
			
			\node[xshift=2cm, yshift=3.3cm, draw, fill=darkblue1!60, rectangle, minimum height=3em, minimum width=3em, align=center, , rounded corners] 
			{Deterministic\\ algorithms};
			
			\node[xshift=5.5cm, yshift=2.0cm, draw,fill=darkred1!60,rectangle, minimum height=5.5em, minimum width=7.5em, align=center, , rounded corners] 
			{Randomized\\ algorithms};		
			\end{tikzpicture}}
                      \caption{Trade-off between accuracy and
                        scalability. Randomized methods for linear
                        algebra allowing for a scalable architecture
                        for modern ``big data'' applications.}
	\label{fig:portfolio}
\end{figure}
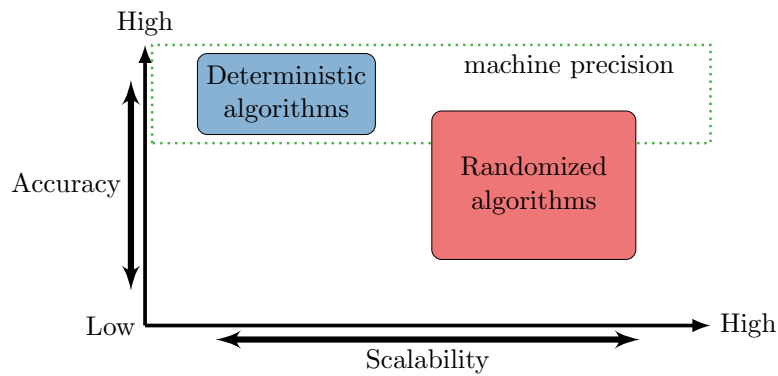

Dimensionality reduction and the related concept of low-rank matrix approximations are fundamental algorithmic tools in machine learning and computational statistics. 
However, high-dimensional data pose a growing computational challenge for traditional matrix algorithms. 
In fact, the exponential growth rate of data is far outstripping advances in computational power, even of modern computational architectures.  
Thus, in the era of ``big data'', the modern computational paradigm of
randomized methods for linear algebra provides an attractive method
for scalable, tractable computations. The price to pay is the
trade-off between the approximation accuracy and computational
costs. In fact, randomized methods are highly scalable, and can be
used to tackle problems which are infeasible otherwise. The different
flavors of both deterministic and randomized methods are illustrated
in Figure~\ref{fig:portfolio}. Thus, randomized algorithms should be
the default choice for applications which involve low-rank matrices
and do not require approximations with full double precision.

Certainly, the randomized singular value decomposition is the most prominent and ubiquitous randomized algorithm. This algorithm comes with strong theoretical error bounds, and the approximation quality can be controlled via oversampling, and power iterations. 
The computational advantage can be substantial compared to other SVD routines in \proglang{R}, provided the target rank $k$ is relatively small. 
While the performance of randomized methods depends on the actual
shape of the matrix, we can state (as a rule of thumb) that
significant computational speedups are achieved if the target rank $k$
is at least $3$--$6$ times smaller than the ambient dimensions of the
measurement space.
The speedup for tall and thin matrices is in general less impressive than for fat matrices.
In addition, the \proglang{R} package \pkg{rsvd} provides several
other randomized matrix decomposition routines, which are all designed
for mid-sized problems, i.e., the input matrix is assumed to fit into
fast memory.
To fully exploit the power of randomized methods, we recommend to use
the enhanced \proglang{R} distribution {Microsoft \proglang{R} Open}
which allows one to use all of the computational resources available.

Future developments of the \pkg{rsvd} package will use randomized
methods to compute linear discriminant analysis, principal component
regression, canonical correlation analysis, and matrix completion
problems. Another important direction is to better integrate the
\pkg{Matrix} package, for instance, to provide efficient routines
which allow to deal better with large-scale sparse
matrices~\citep{Matrixpackage}.

\section*{Acknowledgments}
We would like to express our gratitude to Edzer Pebesma and Julie Josse, and particularly to the two anonymous reviewers. Their helpful feedback allowed us to greatly improve the manuscript. In particular, we would like to thank Gunnar Martinsson for insightful conversations around randomized methods. It is also a pleasure to thank Bryan W. Lewis for illuminating discussions. Further, we thank Thierry Bouwmans, David Harris-Birtill, Aleksandr Aravkin and Lionel Mathelin for many useful comments. 
NBE acknowledges support from the UK Engineering and Physical Sciences Research Council.
JNK acknowledges support from Air Force Office of Scientific Research (AFOSR) grant FA9550-15-1-0385.
SLB acknowledges support from the Army Research Office (ARO) young investigator award grant W911NF-17-1-0422, and from the Air Force Research Laboratory (AFRL) grant FA8651-16-1-0003. 

\bibliography{jss2788}

\end{document}